\documentclass{aa}  

\usepackage{graphicx}
\usepackage{natbib}
\usepackage{array}
\usepackage{rotating}
\usepackage{txfonts}
\usepackage{multirow}
\usepackage{auto-pst-pdf}
\usepackage{longtable}
\usepackage{lscape}
\usepackage{multirow}

\begin{document} 

\title{Chemical abundances of fast-rotating massive stars }
\subtitle{II. Interpretation and comparison with evolutionary models}

   \author{Constantin Cazorla\inst{1}
          \and
          Ya\"el Naz\'e\thanks{Research associate FNRS.}\inst{1}
          \and
          Thierry Morel\inst{1}
          \and
          Cyril Georgy\inst{2}
          \and
          M{\'e}lanie Godart\inst{1}           
          \and
          Norbert Langer\inst{3}
          }

   \institute{Space sciences, Technologies and Astrophysics Research (STAR) Institute, Universit\'e de Li\`ege, Quartier Agora, All\'ee du 6 Ao\^ut 19c, B\^at. B5C, B4000-Li\`ege, Belgium
   \and
   Observatoire de Gen\`eve, Universit\'e de Gen\`eve, Chemin des Maillettes 51, 1290 Versoix, Switzerland
   \and 
   Argelander-Institut f\"ur Astronomie der Universit\"at Bonn, Auf dem H\"ugel 71, 53121 Bonn, Germany
   \newline
              \email{cazorla@astro.ulg.ac.be}
             }

   \date{Received ... ; accepted ...}

 
  \abstract
   {}
   {Past observations of fast-rotating massive stars exhibiting normal nitrogen abundances at their surface have raised questions about the rotational mixing paradigm. We revisit this question thanks to a spectroscopic analysis of a sample of bright fast-rotating OB stars, with the goal of quantifying the efficiency of rotational mixing at high rotation rates.}
   {Our sample consists of 40 fast rotators on the main sequence,  with spectral types comprised between B0.5 and O4. We compare the abundances of some key element indicators of mixing (He, CNO) with the predictions of evolutionary models for single objects and for stars in interacting binary systems.}
   {The properties of half of the sample stars can be reproduced by single evolutionary models, even in the case of probable or confirmed binaries that can therefore be true single stars in a pre-interaction configuration. The main problem for the rest of the sample is a mismatch for the [N/O] abundance ratio (we confirm the existence of fast rotators with a lack of nitrogen enrichment) and/or a high helium abundance that cannot be accounted for by models. Modifying the diffusion coefficient implemented in single-star models does not solve the problem as it cannot simultaneously reproduce the helium abundances and [N/O] abundance ratios of our targets. Since part of them actually are binaries, we also compared their chemical properties with predictions for post-mass transfer systems. We found that these models can explain the abundances measured for a majority of our targets, including some of the most helium-enriched, but fail to reproduce them in other cases. Our study thus reveals that some physical ingredients are still missing in current models.

   }
   {}


   \maketitle
%

\section{Introduction}
Massive stars are generally fast rotators with projected rotational velocities that can amount to up to at least 400 km s$^{-1}$  (e.g. \citealt{how97,duf11}). Such rotation rates can be acquired during their formation or arise later on from interactions with a companion in a binary system (e.g. \citealt{pac81}; \citealt{dem09,dem13}). Stellar rotation has an impact on many facets of stellar physics. In particular, it transports material and angular momentum inside the star, affecting some surface chemical abundances. Observations in the framework of the VLT-FLAMES Survey of Massive Stars \citep{eva08} has suggested that some fast-rotating, evolved B-type stars in the Large Magellanic Cloud (LMC) may exhibit surface abundances that cannot be explained by evolutionary models for single stars incorporating rotational mixing \citep{hun09,bro11b}. A previous exchange of mass and/or angular momentum in an interacting binary system might be able to explain the observations \citep{dem09}, but the binary status of these fast rotators is largely unknown. 

To revisit this question, we have decided to undertake a project combining for the first time a detailed abundance analysis of Galactic fast rotators with a radial velocity (RV) study in order to establish the potential importance of binary effects (Cazorla et al. 2017; hereafter Paper I). Details of our analysis can be found in Paper I, and here we only briefly describe the methods followed to derive the properties of our sample, which is composed of 40 Galactic fast-rotating ($v\sin\,i$ > 200 km s$^{-1}$) OB stars. 

As a first step, we derived the RV associated with  each stellar spectrum thanks to a cross-correlation technique, and calculated the projected rotational velocity through Fourier techniques \citep{gra05,sim07}. The RV measurements were complemented by literature values and were searched for variability to assess the multiplicity status of our targets: (1) if the maximum RV difference was larger than 4\,$\sigma$ and above a threshold of 20 km s$^{-1}$, the star was considered as RV variable (RVVar thereafter), hence a possible binary; (2) period searches were applied to (large) RV datasets, leading to the derivation of SB1 orbital solutions in five cases. In parallel, two different tools, depending on the stellar temperature, were used to derive the atmospheric parameters (effective temperature $T_{\rm{eff}}$ and surface gravity $\log g$), as well as He, C, N, and O abundances: a cooler group of objects (with spectral types in the range B0.5-O9) was analysed with DETAIL/SURFACE \citep{but85,gid81}, while a second group of hotter stars (O4-O9) was analysed with CMFGEN \citep{hil98}. We performed several validation checks to ensure the compatibility of the two methods. First, we  verified that the [N/C] and [N/O] ratios follow the predictions for the CNO cycle, as expected for such massive stars \citep{prz10}. Second, as fast rotation may modify the stellar shape, we used the Code of Massive Binary Spectral Computation \citep[CoMBISpeC;][]{pal12,pal13} to verify that the abundances derived for flattened stars are, within errors, in good agreement with our results assuming no geometrical distortions and gravity darkening. Finally, we  investigated a few stars showing different levels of nitrogen enrichment with both DETAIL/SURFACE and CMFGEN, demonstrating a good agreement, within errors, for the derived parameters. Hence, our dataset is, to first order, homogeneous and all the results can be discussed altogether.  

Paper I provided the individual results for all the stars in our sample. This paper takes a more global view;  the aim is to  compare our observational results with theoretical predictions for single and binary massive stars. To compare our data to expectations for single stars, we employ two independent sets of models,  those of \citet{bro11} (with Z$_{\odot}$ = 0.0088) and  \citet{geo13} (with Z$_{\odot}$ = 0.014). The latter set has been complemented by unpublished calculations having similar physical ingredients as adopted in \citet{geo13}, but extending to higher masses. In the following the two sets of models will be referred to as the Bonn and the Geneva models, respectively. 

This paper is organised as follows: Sect. \ref{sec2} discusses the global characteristics of our sample, Sect. \ref{sec3} compares these properties with predictions of single-star evolutionary models, while Sect. \ref{sec4} considers  these characteristics in the light of the stellar multiplicity status. Finally, Sect. \ref{sec6} summarises our results and presents the conclusions of our project. 

\section{Global properties of our sample}
\label{sec2}

\begin{figure}
\centering
\begin{turn}{90}
\includegraphics[scale=0.49]{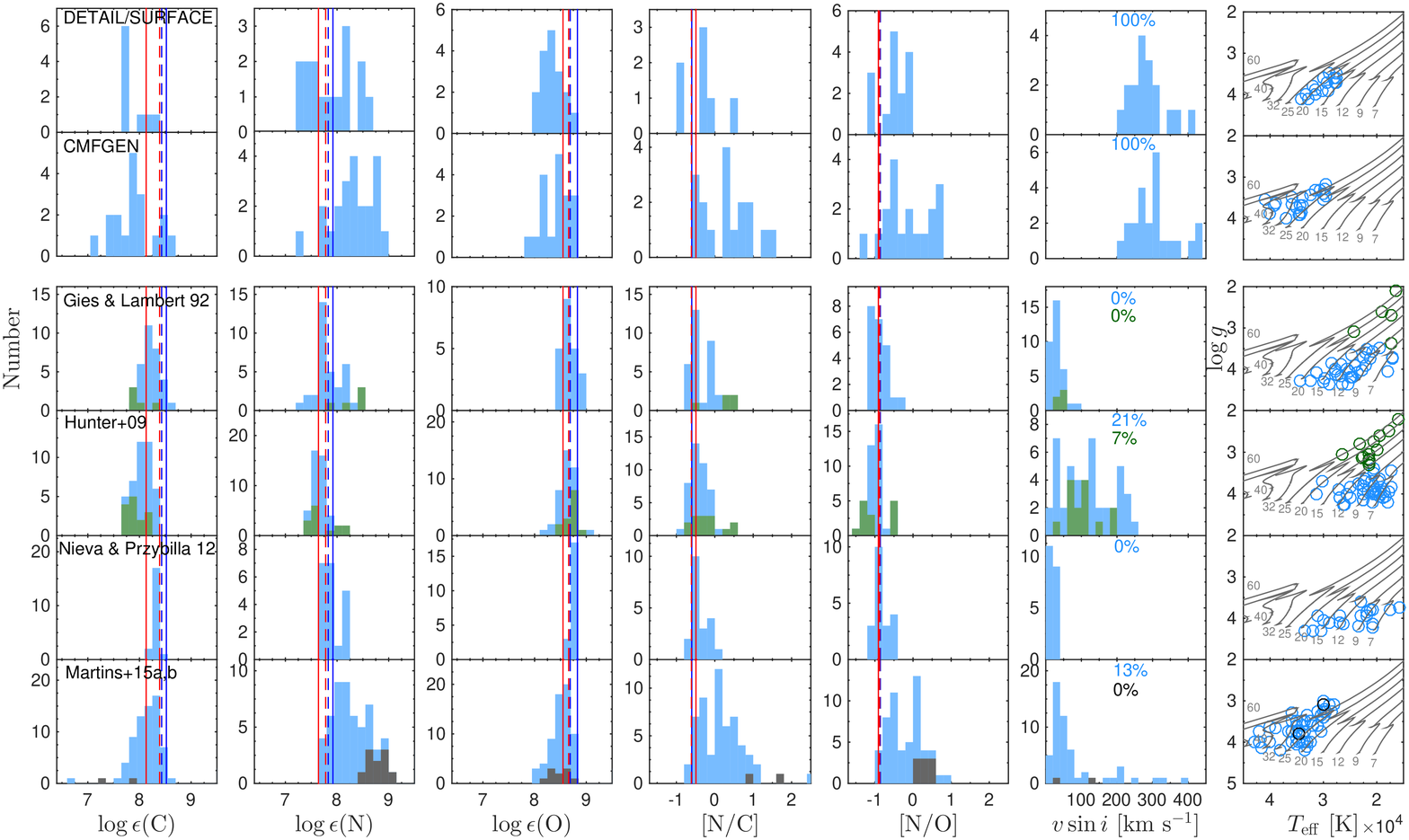} 
\end{turn}
\caption{Comparison between our results (cooler and hotter stars are depicted (from top to bottom) in  Rows 1 and 2) and those of some previous non-LTE studies for nearby OB stars (Rows 3-6 for \citealt{gie92}, \citealt{hun09}, \citealt{nie12}, and \citealt{mar15a,mar15b}). Green histograms/symbols represent supergiant stars. The blue and black histograms/symbols in the bottom panels correspond to the sample of normal O and ON stars analysed by \citet{mar15a} and \citet{mar15b}, respectively. Columns 1, 2, and 3 show the C, N, and O abundances, respectively, and  Cols. 4 and 5 the [N/C] and [N/O] abundance ratios. Lower and upper limits are ignored in all plots. The solar values of \citet{gre98} and \citet{asp09} are shown in these panels as blue solid and blue dashed lines, respectively. The baseline values adopted in the Bonn and Geneva evolutionary models are shown as red solid and red dashed lines, respectively. The two rightmost columns show the breakdown of $v\sin i$ values (the percentage of stars with $v\sin i \ge$ 200 km s$^{-1}$ in each sample is indicated) and the position of targets in the Kiel diagram. In the last column, evolutionary tracks from the Geneva group at solar metallicity and including rotation are overplotted with initial stellar masses (in M$_{\odot}$) indicated. Rotational velocities at the ZAMS for stellar masses higher than 12 M$_{\odot}$ are listed in Table \ref{vZAMSMod}; for 7, 9, and 12 M$_{\odot}$, the initial rotational velocities are 352, 381, and 404 km s$^{-1}$, respectively.}
\label{figCompaLit}
\end{figure}

We compare our CNO abundances with those found in some previous non-LTE studies of nearby OB stars (see Fig. \ref{figCompaLit}) (\citealt{gie92}; \citealt{hun09}; \citealt{nie12}; \citealt{mar15a,mar15b}). These studies sample widely different domains in terms of rotation rate, mass, and evolutionary status. Indeed, the samples of \citet{gie92} and \citet{nie12} are only composed of slow rotators, whereas our stars are all fast rotators. Regarding evolutionary stages, we mostly concentrate on main-sequence (MS) stars with high mass, as in \citet{mar15a,mar15b}, while the Galactic stars studied by \citet{hun09}  typically have lower masses and span various evolutionary stages. 

The three samples of (mostly) slow-rotating MS stars of \citet{gie92}, \citet{nie12}, and \citet{mar15a,mar15b} differ in their chemical properties;  the last sample, which is composed of higher mass stars, exhibits markedly higher $\log \epsilon$(N), [N/C], and [N/O] values. Conversely, the abundance distributions are similar for our stars and those studied by Martins et al., which have comparable masses and evolutionary status but drastically different rotational velocities on average ($\langle v\sin\,i \rangle$ $\sim$ 300 km s$^{-1}$ in our sample versus $\lesssim$ 100 km s$^{-1}$ for the vast majority of the Martins stars). However,  caution must be taken when interpreting these results in the framework of single-star evolutionary models and, in particular, when trying to quantify the relative importance of the various parameters (e.g. mass, rotational velocity) controlling the amount of rotational mixing. For instance, because their masses differ, the stars in the studies of  \citet{gie92} and \citet{mar15a,mar15b} have suffered a different loss of angular momentum because of the stellar winds. As a result, the two samples may have had different rotational velocity distributions on the zero age main sequence (ZAMS) even though the present-day distributions  are quite similar.  Therefore, the different distributions in Fig. \ref{figCompaLit} may not only reflect the dependency of rotational mixing with mass. Also, the proportion of stars in the various samples for which binary effects are important is unknown (see discussion in Sect. \ref{sec4}).

\onecolumn

\begin{figure}
\centering
\begin{turn}{90}
\includegraphics[scale=0.49]{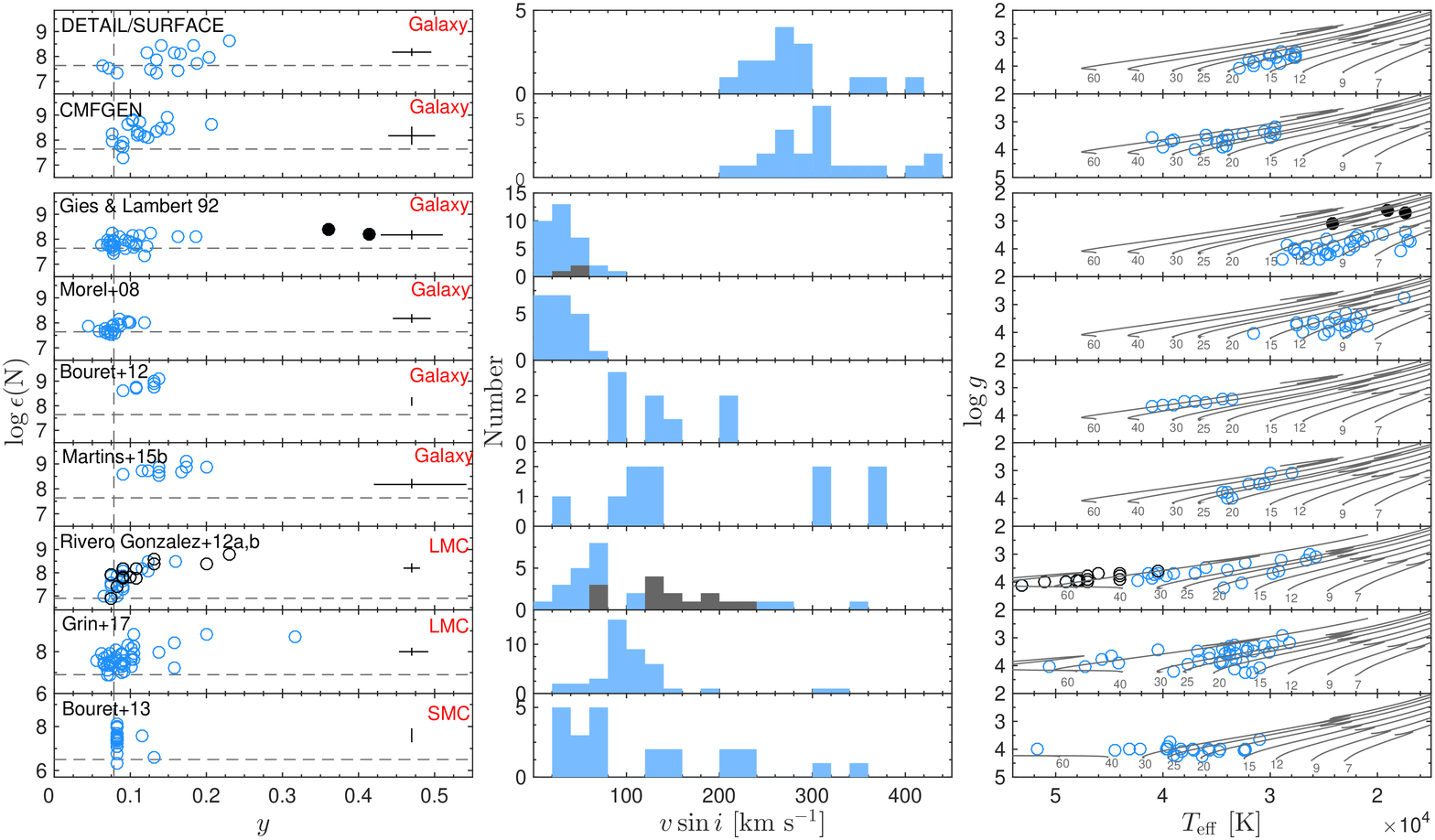} 
\end{turn}
\caption{{\it Left panels:} Nitrogen abundance as a function of the helium abundance ($y=\mathcal{N}(\rm{He})/[\mathcal{N}(\rm{H})\,+\,\mathcal{N}(\rm{He})]$) for this work (with cooler and hotter stars depicted in the first and second rows, respectively) and previous non-LTE studies in the literature (\citealt{gie92}; \citealt{mor08}; \citealt{bou12,bou13}; \citealt{riv12a,riv12b}; \citealt{mar15b}; \citealt{gri17}). The very He-rich stars of \citet{gie92} are highlighted as filled symbols. The helium data of \citet{mor08} are supplemented by results from \citet{mor06}, \citet{bri07a}, \citet{bri07b}, and \citet{hub08}. Sample stars of \citet{riv12a} and \citet{riv12b} are shown in blue and black, respectively. Typical error bars are shown to the right of each panel. The dashed lines show the baseline abundances of \citet{bro11} for the Galaxy and the Magellanic Clouds. {\it Central panels:} Breakdown of $v\sin\,i$ values. {\it Right panels:} Positions of the stars in the Kiel diagram. Evolutionary tracks from \citet{bro11} for the relevant metallicity are overplotted. Initial stellar masses (in solar units) are indicated. Rotational velocities at the ZAMS of Galaxy models for stellar masses higher than 12 M$_{\odot}$ are listed in Table \ref{vZAMSMod}; for 7, 9, and 12 M$_{\odot}$, the initial rotational velocities are chosen to be close to that of the 15 M$_{\odot}$ model, i.e. 339, 333, and 331 km s$^{-1}$, respectively. Assumed rotational velocities at the ZAMS of Magellanic Cloud models are close to the ones of Galaxy models for each mass. Lower and upper limits are ignored in all plots.}
\label{figLoggyNvsiniloggClogT}
\end{figure}

\twocolumn
Figure \ref{figLoggyNvsiniloggClogT} illustrates the dependence between the helium and nitrogen surface abundances for our study and previous non-LTE studies in various environments (Galaxy and Magellanic Clouds). A trend can be seen between the nitrogen and helium abundances of our sample stars;  the two quantities seem to increase in parallel (this aspect is examined in a quantitative way in Sect. \ref{sec3}). This trend is also seen in some literature samples, especially in those of \citet{riv12a,riv12b} and, to a lesser extent, \citet{gri17} which are both composed of LMC stars with on average lower $v\sin\,i$ but higher masses than ours. The nitrogen enrichment in the sample of \citet{gie92}, composed of less massive slow rotators, is low, even if the helium abundance can be high in a few cases. The nitrogen excess is higher in the Galactic sample of \citet{bou12},  which is composed of high-mass stars with moderate $v\sin\,i$. The sample of \citet{mar15b}, partially composed of fast rotators whose mass is $\sim$ 25 M$_{\odot}$, exhibits strong helium and nitrogen enrichments. In contrast, the SMC stars of \citet{bou13} generally show dramatic nitrogen overabundances without a helium abundance enhancement, even if their masses are high.

The aim of this paper is to compare the results of Paper I with models, and to assess whether the models can reproduce these results. The question then arising is the choice of the best observational diagnostics to perform this comparison. For example, are elemental abundances or abundance ratios the best indicators to use?

Figure \ref{figCompaLit} shows that mismatches exist between the derived abundances and the solar values of \citet{gre98} and \citet{asp09}, as well as  the initial values adopted in the Bonn and Geneva evolutionary models. For carbon, there seems to be a systematic underabundance for all observational studies considered, while lower oxygen abundances are detected in our sample and those of \citet{mar15a,mar15b}. The nitrogen abundance displays a more varied behaviour in the samples in Fig. \ref{figCompaLit}:   underabundances and overabundances are both found. 

The CNO abundances are affected by mixing and   a depletion of carbon and, to a lesser extent, oxygen are expected. However, subsolar N abundances are unexpected. The C and O depletions also appear significantly higher than predicted by models. It is therefore likely that shortcomings in the data analysis are at play. It could be deficiencies in the modelling (e.g. \citealt{mor09}) or technical issues (e.g. continuum level set too low in fast rotators because of the lack of continuum windows, which would lead to an underestimation of the strength of the spectral features). In this case, elemental CNO abundances may not be reliable diagnostics. 

In contrast, the lowest [N/C] and [N/O] abundance ratios (which should correspond to unmixed objects) nearly coincide with the solar values and those adopted by evolutionary models on the ZAMS. There are only a few values appearing below (though well within 2$\sigma$ in our case) and they can be explained by statistical fluctuations, considering Gaussian distributions. Abundance ratios thus appear much less affected by systematics.  They are also more sensitive to mixing \citep{mae14}:  [N/C] should increase more rapidly than the nitrogen abundance in the CN cycle because the stellar core should be gradually depleted in carbon; in the ON cycle, the same behaviour is observed for [N/O], due to the depletion of oxygen that occurs in this regime. These abundance ratios are thus better indicators of transport processes. However, numerous [N/C] values in our study are lower limits, mostly because only upper limits could be derived for the carbon abundance in our coolest stars (since the diagnostic \ion{C}{III} lines are then very weak): adopting this ratio would thus lead to the exclusion of many targets. Furthermore, there is a good correlation between [N/C] and [N/O] (see Fig. 7 of Paper I), and the conclusions presented in the following will be unchanged whatever the adopted ratio.  
Therefore, we consider the [N/O] abundance ratio as the main diagnostic of rotational mixing in the rest of this paper\footnote{The [N/O] abundance ratio of \object{HD 150574} is, however, unavailable (see Paper I).}.

\onecolumn

\begin{figure*}
\centering
\begin{turn}{0}
\includegraphics[scale=0.42]{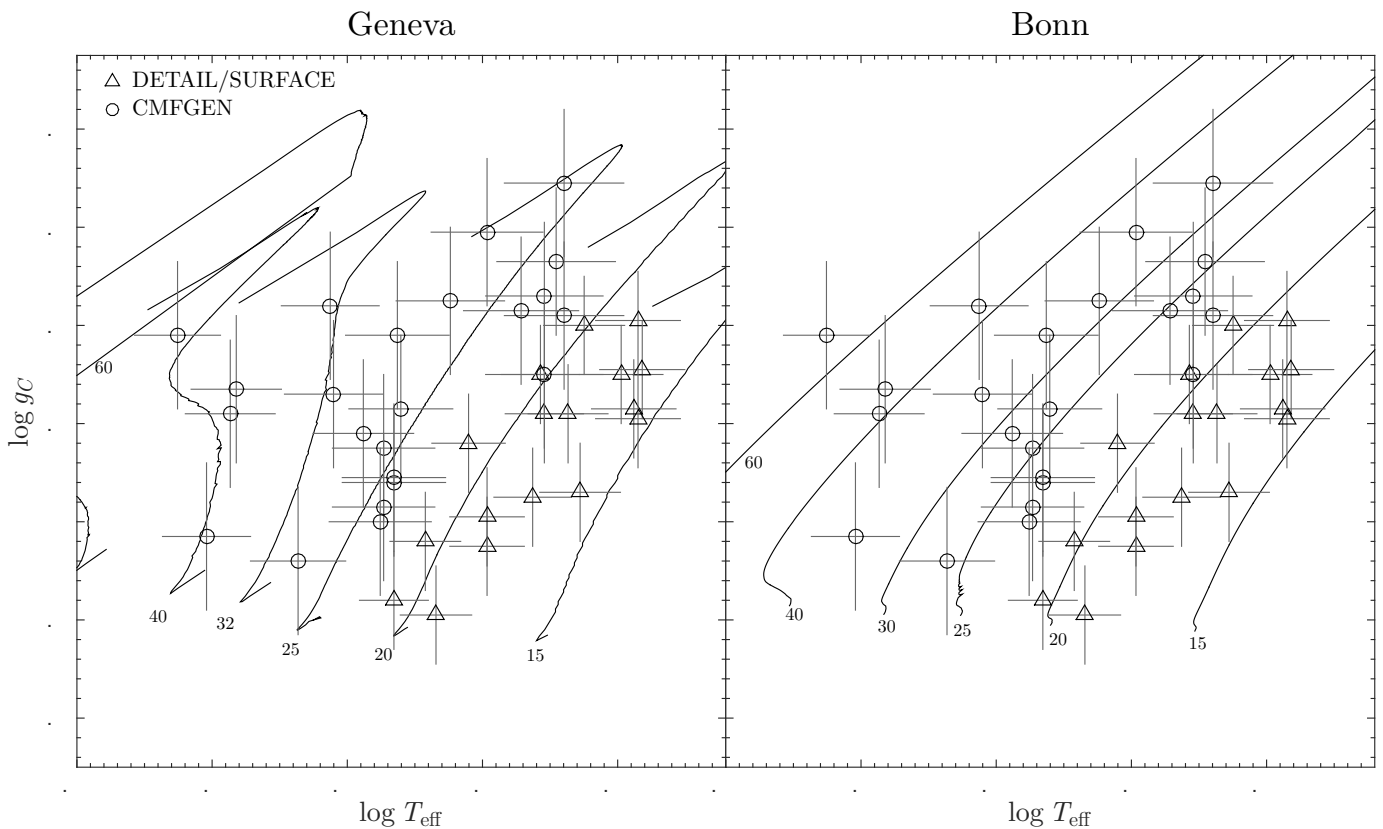}
\end{turn}
\caption{Kiel diagrams comparing the position of our targets with evolutionary tracks (left-hand panel displays Geneva models; right-hand panel displays Bonn models). Only the predictions for the MS phase are illustrated. The initial stellar masses (in solar units) are indicated for each model. Rotational velocities at the ZAMS are listed in Table \ref{vZAMSMod}. Triangle and circle symbols represent the cooler stars studied with DETAIL/SURFACE and the hotter stars studied with CMFGEN, respectively.}
\label{figLoggCTN}
\end{figure*}

\twocolumn
The amount of mixing by rotation inside massive stars depends on several factors: rotational velocity, mass, but also age, metallicity, multiplicity, and possibly magnetic fields. Prior to comparing our results with model predictions, we will thus separate our sample into different subgroups. As the distance and thus the luminosity of our targets are usually not accurately known, their position in a $T_{\rm{eff}}$--$\log g_{\rm{C}}$\footnote{$\log g_{\rm{C}}$ is the surface gravity corrected for the effects of centrifugal forces: $g_{\rm C}$ = $g$ + ($v\sin\,i$)$^2$/$R_{*}$, with $R_{*}$ being the star radius \citep{rep04}.} diagram is used to estimate their evolutionary status. The Kiel diagram is shown in Fig. \ref{figLoggCTN} where the Bonn and Geneva evolutionary models are overplotted. All our targets are on the main sequence. Furthermore, all stars have masses comprised between 15 and $\sim$ 60 M$_{\odot}$; splitting the sample into groups of stars with similar masses may ease the comparison with models. To define such subsamples, we build [N/O] versus $\log g_{\rm{C}}$ diagrams (Fig. \ref{figLoggCNO}) and examine when significant changes in theoretical [N/O] occur. This was done considering Geneva models since there is no monotonic change with mass in the theoretical curves of the Bonn group for the chosen initial rotational velocities (Table \ref{vZAMSMod}). {This slightly non-monotonic behaviour of the Bonn models might occur because mixing through composition barriers can depend (numerically) on spatial resolution (see \citealt{lau14}).} Since the 15, 20, and 25 M$_{\odot}$ Geneva models predict similar [N/O] values during the MS phase, we can thus group all lower-mass targets. The increase of [N/O] for high-mass stars is expected to be significantly greater than for lower-mass stars. We thus define three subsamples: from 15 to 28 M$_{\odot}$, from 29 to 35 M$_{\odot}$, and above 35 M$_{\odot}$. In order to compare the properties of our stars with models of appropriate mass, our data and evolutionary tracks will be colour-coded as a function of the stellar mass in all diagrams of the following sections. As shown by \citet{lan92}, the evolutionary masses of helium-rich stars are overestimated  compared to theoretical tracks computed for a solar He abundance, but neglecting this aspect is not expected to notably affect the breakdown of the stars in the various subsamples. A more significant effect is actually the choice of the evolutionary models; as can be seen, the Geneva and Bonn groups predict very different evolutionary paths in Fig. \ref{figLoggCTN}. For this reason, the population in each subsample is different depending on the chosen family of models.

\begin{table}
\caption{Assumed initial rotational velocities for Geneva and Bonn models.}
\label{vZAMSMod}
\center
\begin{tabular}{ccccccccccccc}
\hline\hline
\multicolumn{2}{c}{\multirow{1}{*}{Geneva}} & \multicolumn{2}{c}{Bonn} \\
$M$ (M$_{\odot}$)&$v_{\rm ZAMS}$ (km s$^{-1}$)&$M$ (M$_{\odot}$)&$v_{\rm ZAMS}$ (km s$^{-1}$)\\ \hline
15 & 400 & 15 & 329\\
20 & 421 & 20 & 324\\
25 & 441 & 25 & 374\\
32 & 614 & 30 & 372\\
40 & 647 & 40 & 417\\
60 & 714 & 60 & 455\\
\hline
\end{tabular}
\tablefoot{Values of the Bonn models were chosen to best represent the typical $v\sin\,i$ of our sample stars, i.e. $\sim$ 300 km s$^{-1}$, at the middle of the MS phase. In this context, we assume $i$ = 70$^{\circ}$ as, being fast rotators, our targets are preferentially seen close to equator on (see Appendix \ref{appB} and discussion in Sect. \ref{sec3} for the inclination values found by BONNSAI; we also note that \citealt{zor02} obtained an average value of 68$\pm$18$^{\circ}$ for a sample of Be stars). We emphasise that another slightly different choice would not affect our conclusions. Geneva models have systematically greater initial rotational velocities than the Bonn models. This is due to the different treatment of rotation within the star: Geneva models assume a less strong coupling between the core and the envelope and the surface spin-down by stellar winds is more efficient than in Bonn models. Higher initial rotational velocities are therefore needed to reach fast rotation during the MS phase.}
\end{table}

Finally, a previous episode of mass transfer may also dramatically alter the chemical properties of components in a massive binary. In Paper I we  derived the multiplicity status of our targets. We found 19 stars to be presumably single (47.5\% of our sample); 9 stars with variable RVs, hence probable binaries (22.5\% of our sample); and only 5 targets with SB1 orbital solution, hence confirmed binaries (12.5\% of our sample). The binary status of  7 stars (17.5\% of our sample)  could not be determined because of a lack of observations. We note  that confirmed double-lined spectroscopic binaries (SB2s) were initially discarded from our sample and that none of our targets were subsequently found to belong to this category. In parallel, a runaway status has been assigned to 10 targets (25\% of the sample). 

Our RV measurements are derived from our own observations and from a large body of spectra retrieved from  public archives (see Paper I for details). In addition, our RV dataset is supplemented by results taken from the literature. Although the sampling of the RV time series varies drastically depending on the target considered, this approach  generally provides a large number of measurements spread over a very long temporal baseline. Of particular interest is the fact that most stars have been monitored over a time span considerably exceeding $\sim$3 years, which implies that we are in principle sensitive to long orbital periods. For most of our targets, we therefore possess all the information necessary to examine the impact of multiplicity on the abundances. However, some caveats do exist and we discuss the consequences of potentially missed binaries 
in Sect. \ref{sec4} .

One can argue that the stars in our sample that we define as being RV variables may actually be pulsating stars. However, the ``presumably single star'' classification that we obtain for stars that have well-characterised pulsations (\object{HD 93521}, \citealt{rau08}; \object{HD 149757}, \citealt{kam97}) suggests that the number of pulsators incorrectly identified as binaries is low and that the criteria, inspired by those of \citet{san13} and used in Paper I to establish whether the measured RVs are variable, generally excludes pulsators. However, we exclude \object{HD 28446A} and \object{HD 41161} from the discussion in Sect. \ref{sec4} as the origin of their RV variations is unclear (see Paper I). 

\section{Comparison with single-star evolution}
\label{sec3}
In this section we confront our results with predictions for single star evolution from Geneva and Bonn models. 

Internal magnetic fields induce the transport of chemical elements and angular momentum inside the star (e.g. \citealt{mae05})\footnote{ The Bonn models do not consider the effect of magnetic fields on the transport of chemical elements;  however,  they do incorporate their effect on the distribution of angular momentum in the interior \citep{bro11}.}. External magnetic fields generate a mechanical coupling between the stellar surface and the winds, taking away some angular momentum from the star \citep{udd02,udd08} and producing a magnetic braking \citep{udd09,mey11}. The presence of magnetic fields is thus predicted to modify the amount of mixing in stellar interiors,  hence to affect abundances at the stellar surface (e.g. \citealt{heg05}; \citealt{mey11}; \citealt{pot12}). However, the abundance analyses of magnetic massive stars did not reveal a clear and systematic difference in their surface CNO properties compared to non-magnetic stars (\citealt{mor12,mar15a}). Furthermore, the incidence of large-scale surface fields with a longitudinal component above $\sim$ 100-200 G is only of the order of 10\% in massive stars (\citealt{fos15,gru17}). This proportion applies to large samples that combine different types of stars, but some categories of objects are known to behave very differently (see e.g. \citealt{wad14}). Are fast rotators also a special group in this respect? Assuming the magnetic fields to be fossil, we would not expect evolved fast rotators to host a field as magnetic braking would have spun them down significantly. On the other hand, considering magnetic field and fast rotation to both arise from a merger event \citep{fer09}, we could instead expect a large fraction of magnetic stars in our sample. It is thus difficult to speculate theoretically on the magnetic field incidence amongst fast rotators. There are, however, some observational constraints, as about a quarter of our targets have been observed in circular spectropolarimetry: \object{HD 66811}, \object{HD 93521}, and \object{HD 149757} were studied by \citet{hub13,hub16}; \object{HD 46056} and \object{HD 93521} were analysed in the framework of the BOB survey \citep{fos15}; while \object{HD 46056}, \object{HD 46485}, \object{HD 66811}, \object{HD 69106}, \object{HD 149757}, \object{HD 192281}, \object{HD 203064}, and \object{HD 210839} were observed as part of the MiMeS survey \citep{gru17}. No significant field detection was reported for any of those stars;  however, this is not surprising as the vast majority of magnetic OB stars are slow rotators. Since there is no convincing evidence for a generalised strong magnetic character in fast rotators, and since evidence for CNO abundance peculiarities in magnetic massive stars is unconvincing, we will not consider the influence of internal or external magnetic fields in our comparison with evolutionary models.

\begin{figure*}
\centering
\begin{turn}{0}
\includegraphics[scale=0.77]{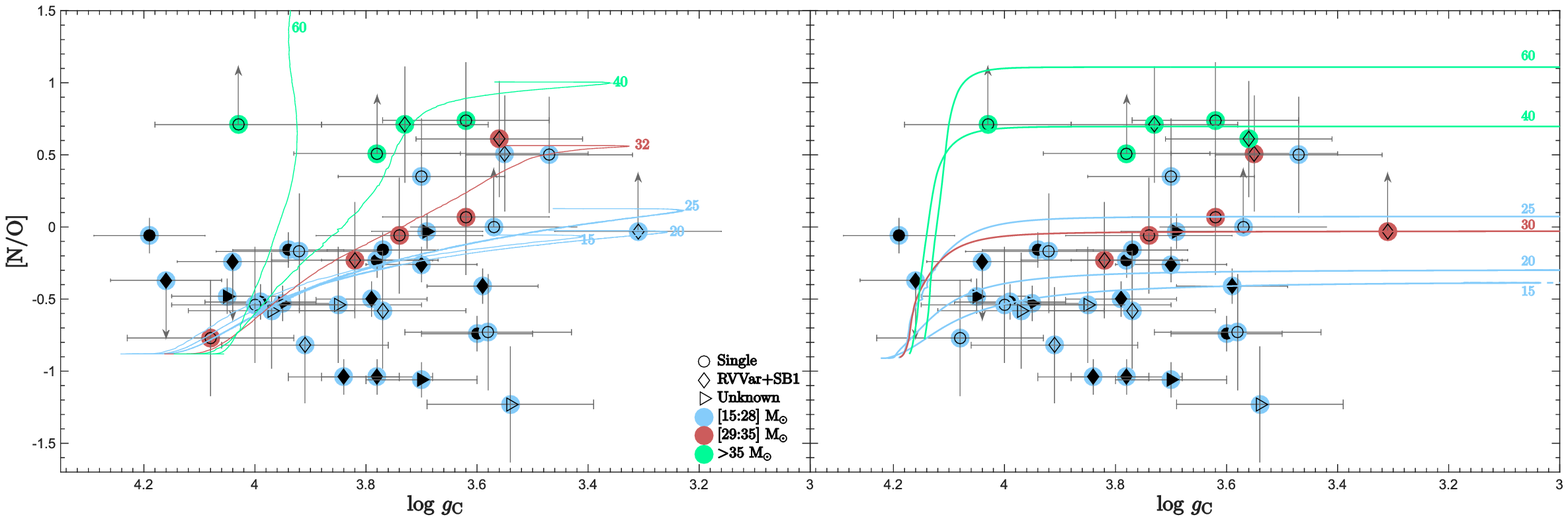}
\end{turn}
\caption{Predicted [N/O] values as a function of $\log g_{\rm{C}}$ by the Geneva (left) and Bonn (right) groups. Initial stellar masses (in solar units) are indicated. Rotational velocities at the ZAMS are listed in Table \ref{vZAMSMod}. Each panel shows the data for the presumably single stars (circles), RV variables and SB1s (diamonds), and stars with unknown multiplicity status (right oriented triangles). Black empty and filled symbols represent the hotter stars studied with CMFGEN and the cooler stars studied with DETAIL/SURFACE, respectively. Blue, brown, and green filled circles represent stars with masses comprised in the ranges from 15 to 28, from 29 to 35, and higher than 35 M$_{\odot}$, respectively.}
\label{figLoggCNO}
\end{figure*}

Figure \ref{figLoggCNO} shows the [N/O] values of our sample stars as a function of their $\log g_{\rm{C}}$, which we use as a proxy for their evolutionary status. In the Geneva models there is no increase of the [N/O] abundance ratios at the very beginning of the MS, but there is a gradual increase afterwards. The evolution of [N/O] is widely different in the Bonn models as they predict a faster increase  of [N/O] when the star evolves off the ZAMS but then no significant change during the rest of the MS. Nevertheless, compared to these models, some of our targets exhibit higher or lower [N/O] abundance ratios than  is predicted for their mass, rotational velocity, and evolutionary status. Of particular importance are the stars in our sample with an apparent lack of CNO-cycled material at their surface (they correspond to the anomalous group 1 of \citealt{hun07,hun09}). In our sample, such objects tend to have a mass in the range 15--28 M$_{\odot}$. We note that these stars are a mixture of objects studied with DETAIL/SURFACE and CMFGEN. Their low [N/O] abundance ratio is therefore very unlikely to be an artefact of the data analysis (see comparison of the two methods for a few illustrative cases in Sect. 6.3 of Paper I). An interesting result is that the higher-mass stars display higher [N/O] abundance ratios, at any given value of the surface gravity, which is a trend predicted by the rotational mixing theory. 

\begin{figure*}
\centering
\begin{turn}{0}
\includegraphics[scale=0.77]{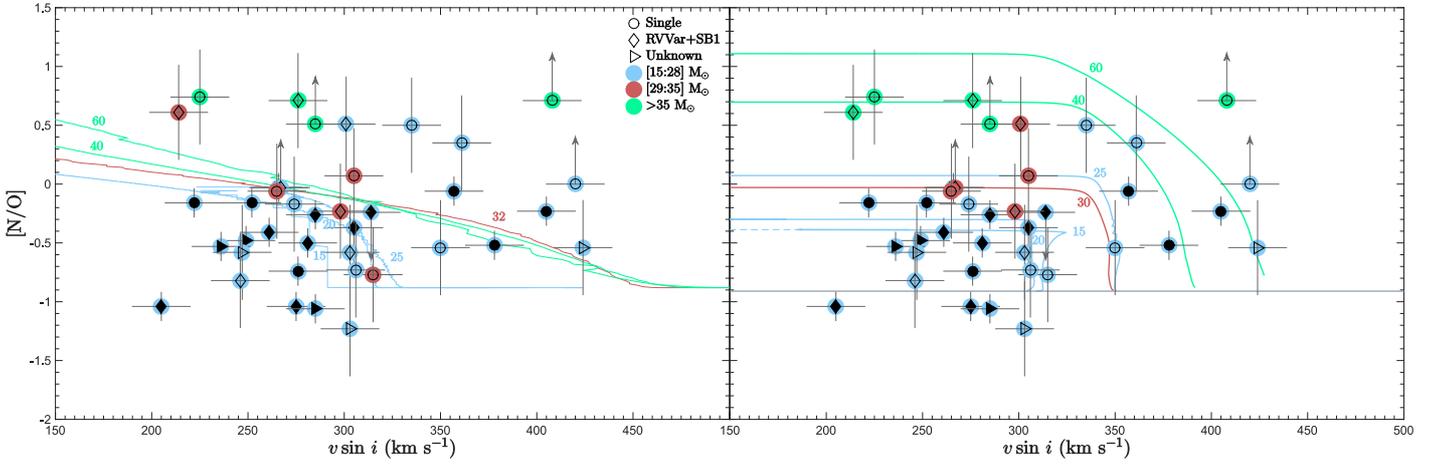}
\end{turn}
\caption{[N/O] abundance ratio as a function of $v\sin\,i$. Theoretical predictions for different masses and rotational velocities are from the Geneva (left) and Bonn (right) groups. Initial stellar masses (in solar units) are indicated. Rotational velocities at the ZAMS are listed in Table \ref{vZAMSMod}. Predicted rotational velocities have been multiplied by $\sin$(70$^\circ$) to take the projection effect into account (see Table \ref{vZAMSMod}); however,  a slightly different choice would not affect these plots in any significant way. Solid lines in both models represent the MS phase, while dashed lines for the Bonn models represent the supergiant phase. Symbols and related colours are the same as in Fig. \ref{figLoggCNO}.}
\label{figvsiniNO}
\end{figure*}

Fig. \ref{figvsiniNO} shows the [N/O] abundance ratios as a function of the projected rotational velocities. The [N/O] of most stars can be explained by single-star models, but some stars show discrepant [N/O] values considering their mass and projected rotational velocity, especially when they are compared to Geneva models. 

\begin{figure*}
\centering
\begin{turn}{0}
\includegraphics[scale=0.77]{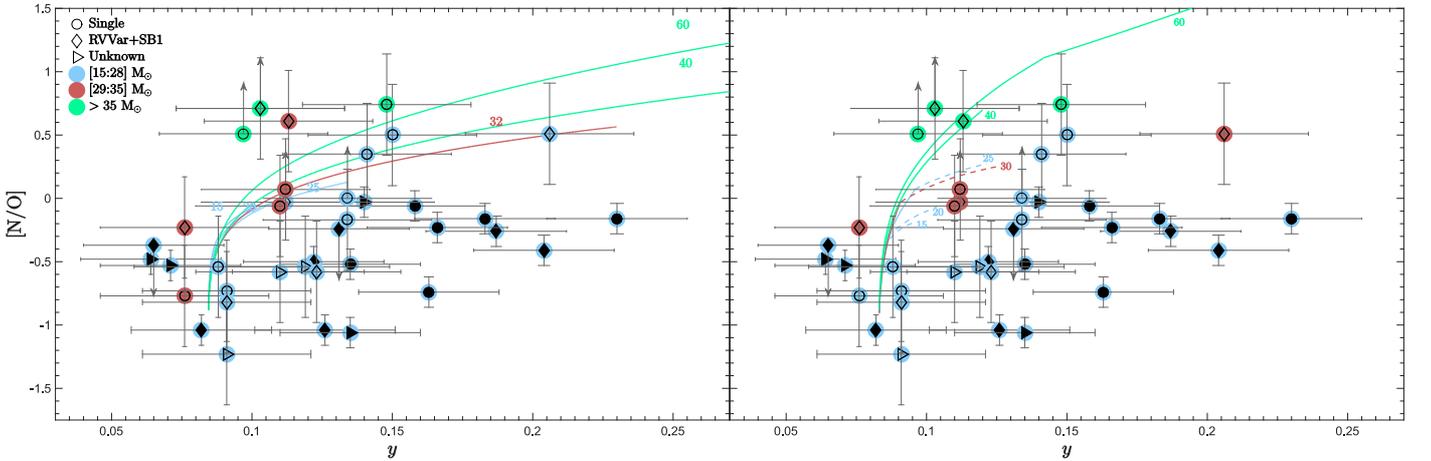}
\end{turn}
\caption{[N/O] as a function of $y$. Graphical conventions are the same as in Figs. \ref{figLoggCNO} and \ref{figvsiniNO}.}
\label{figyNO}
\end{figure*}

Fig. \ref{figyNO} shows the [N/O] abundance ratios as a function of the helium abundances. In order to investigate the likelihood of the presence of a positive correlation between the two quantities, we used the ``bhk'' and ``spearman'' tasks of the stsdas.statistics package in IRAF\footnote{{\tt IRAF} is distributed by the National Optical Astronomy Observatories, operated by the Association of Universities for Research in Astronomy, Inc., under cooperative agreement with the National Science Foundation.} that can take upper/lower limits into account. The highest significance levels are obtained with the ``spearman''  method, but they only amount to 4.4 and 2.6\% when upper and lower limits in [N/O] are taken into account or not, respectively. This suggests that there is no statistically significant correlation between [N/O] and $y$;  we simply note a lack of cases with both high $y$ and low [N/O] ratio. 

 The majority of our targets are located to the right of the curves for MS stars predicted by the Geneva and Bonn models; specifically, they are more enriched in helium than  is predicted by models. It should be noted in this context that the Kiel diagrams of our targets (Fig. \ref{figLoggCTN}) instead indicate that they are  core-hydrogen burning stars. 

Globally, a comparison  with evolutionary tracks reveals general trends, but we also performed a more detailed, object-by-object comparison. To this end, we used the BONN Stellar Astrophysics Interface (BONNSAI; \citealt{sch14})\footnote{The BONNSAI web-service is available at \url{http://www.astro.uni-bonn.de/stars/bonnsai}.}, which relies on the Bonn evolutionary models (unfortunately, a similar tool making use of Geneva models is not available). We considered a Salpeter mass function \citep{sal55} as initial mass prior and a Gaussian initial rotational velocity prior with a mean of 372 km s$^{-1}$ and a full width at half maximum (FWHM) of 57 km s$^{-1}$. {This FWHM value was inferred from a Gaussian fit of the breakdown of the $v\sin i$ values of our sample stars, while the mean initial velocity was chosen to be the one for which the corresponding Bonn model predicts $v\sin i$ $\sim$ 300 km s$^{-1}$ (with $i=70^{\circ}$) at the middle of the MS phase for a 30 M$_{\odot}$ star, which is typical of our sample (Fig. \ref{figLoggCTN}).}

We first tried to use all available parameters ($T_{\rm{eff}}$, $\log g_{\rm C}$, $v\sin\,i$, $y$, C, N, and O abundances) as input parameters. In this case, BONNSAI finds a match between the model predictions and the observed input parameters in 21 cases out of 40. The values provided by BONNSAI are given in Table \ref{resBONN}.  Of these stars, the atmospheric parameters and abundances of \object{HD 66811} cannot be reproduced with our chosen initial rotational velocity prior, but can be explained when a flat distribution is considered. This may be due to the particular properties of this star since it is the highest-mass star in our sample (and notably has a very high mass-loss rate). It is interesting to note that success or failure to get a solution is not linked to  projected rotational velocities  (i.e. the stars for which BONNSAI found a result span the whole range of $v\sin\,i$) or to the multiplicity status (5 RV variables and 3 SB1s are amongst the 21 successes). However, some of the stars for which BONNSAI fails to converge have nitrogen or oxygen abundances that are out of the ranges considered by this tool (7.64 -- 10.12 and 7.19 -- 8.55 dex for $\log \epsilon$(N) and $\log \epsilon$(O), respectively). 

This recalls the elemental abundance problem already mentioned in Sect. \ref{sec2}. However, abundance ratios cannot be entered as input for BONNSAI. To overcome this limitation, we performed a second analysis, ignoring the CNO abundances (i.e. the input parameters are $T_{\rm{eff}}$, $\log g_{\rm C}$, $v\sin\,i$, and $y$). In this case, the atmospheric parameters and abundances of 32 stars can be derived by BONNSAI and are given in Table \ref{resBONN2}; the 8 remaining stars in our sample have a  helium abundance that is too high, which  cannot be reproduced by Bonn models. As a last exercise, we  therefore used BONNSAI without considering any abundance parameters. Convergence is then reached for all targets and the values derived by BONNSAI are given in Table \ref{resBONN3}.

\begin{figure*}
\centering
\begin{turn}{0}
\includegraphics[scale=0.75]{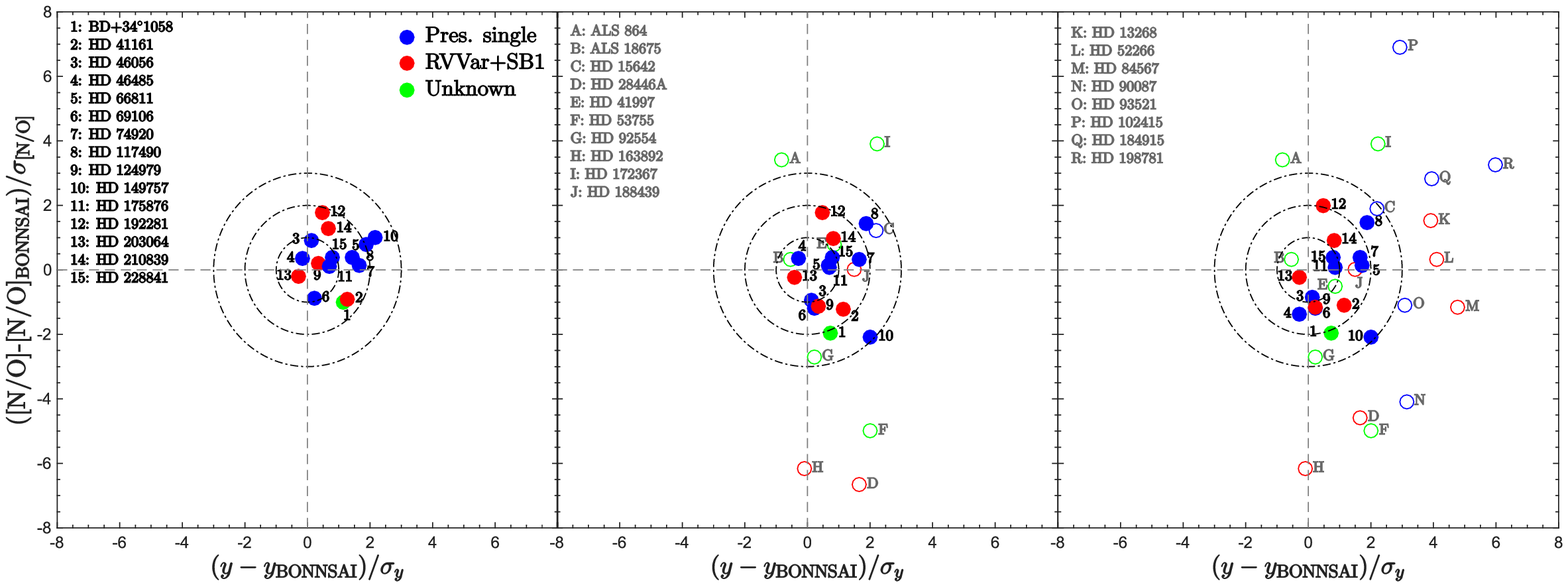}
\end{turn}
\caption{Difference between the observed [N/O] ratio and that predicted by BONNSAI, normalised by the error derived for our objects, as a function of a similar difference for the abundance $y$. Stars with lower/upper limits in [N/O] are not included, explaining why there are fewer stars than indicated in Tables B.1-3. Filled circles represent the stars common to the three panels. The three dash-dotted circles delimitate the areas in which the differences are within 1, 2, and 3\,$\sigma$. Colours represent the multiplicity status of our targets. The input parameters in BONNSAI are different for the three panels -- left panel: full parameter set ($T_{\rm{eff}}$, $\log g_{\rm{C}}$, $v\sin\,i$, He and CNO abundances); middle panel: $T_{\rm{eff}}$, $\log g_{\rm{C}}$, $v\sin\,i$, and $y$; right panel: $T_{\rm{eff}}$, $\log g_{\rm{C}}$, and $v\sin\,i$.}
\label{figyNODiffBONN}
\end{figure*}

When BONNSAI reaches a solution it does not mean that the observed properties (especially the [N/O] ratio and $y$) are well reproduced. We therefore compare the observed and predicted values of the helium abundances and [N/O] ratios (see Fig. \ref{figyNODiffBONN}). Table \ref{rereBOBO} provides the number of stars whose predicted and observed $y$ and [N/O] values both differ by less than 1, 2, and 3$\sigma$. On average, the properties of 50\%\ of our 40 targets can be explained by models within 3$\sigma$ ($\sim$33\%\ for 2$\sigma$ and $\sim$15\% for 1$\sigma$). There is no tendency for our targets to present systematically higher (or lower) [N/O] values than the predictions of the Bonn models. However, as already pointed out above, we note a systematic excess in the observed helium abundance (${y-y_{\rm BONNSAI}} > 0$) for the vast majority of the stars shown in Fig. \ref{figyNODiffBONN} (84\%) and a slight underabundance of helium (${y-y_{\rm BONNSAI}} < 0$) for a minority of them (16\%). More precisely, there is a significant (above 1$\sigma$) excess in helium for 56\% of the stars in the first subsample, the deviation remaining within 1$\sigma$ for 44\% of them. There is no case of significant underabundance of helium.

\begin{table*}
\caption{Number of stars whose helium abundances and [N/O] ratios are both reproduced by Bonn models within 1, 2, and 3$\sigma$. Stars with lower/upper limits in [N/O] are not considered.}
\label{rereBOBO}
\center
\begin{tabular}{cccccccccccccccc}
\hline\hline
BONNSAI input parameters                                                                                                                        & $\le$ 1$\sigma$                         & $\le$ 2$\sigma$                       & $\le$ 3$\sigma$  \\ \hline
$T_{\rm{eff}}$, $\log g_{\rm{C}}$, $v\sin\,i$, $y$, CNO abundances                                                              &\multirow{1}{*}{7}     &\multirow{1}{*}{13}    &\multirow{1}{*}{15}\\
$T_{\rm{eff}}$, $\log g_{\rm{C}}$, $v\sin\,i$, $y$                                                                                              &\multirow{1}{*}{7}     &\multirow{1}{*}{15}    &\multirow{1}{*}{20}\\      
$T_{\rm{eff}}$, $\log g_{\rm{C}}$, $v\sin\,i$                                                                                           &\multirow{1}{*}{6}     &\multirow{1}{*}{14}    &\multirow{1}{*}{20}\\      
\hline \end{tabular}
\end{table*}

\begin{figure*}
\centering
\begin{turn}{0}
\includegraphics[scale=1.5]{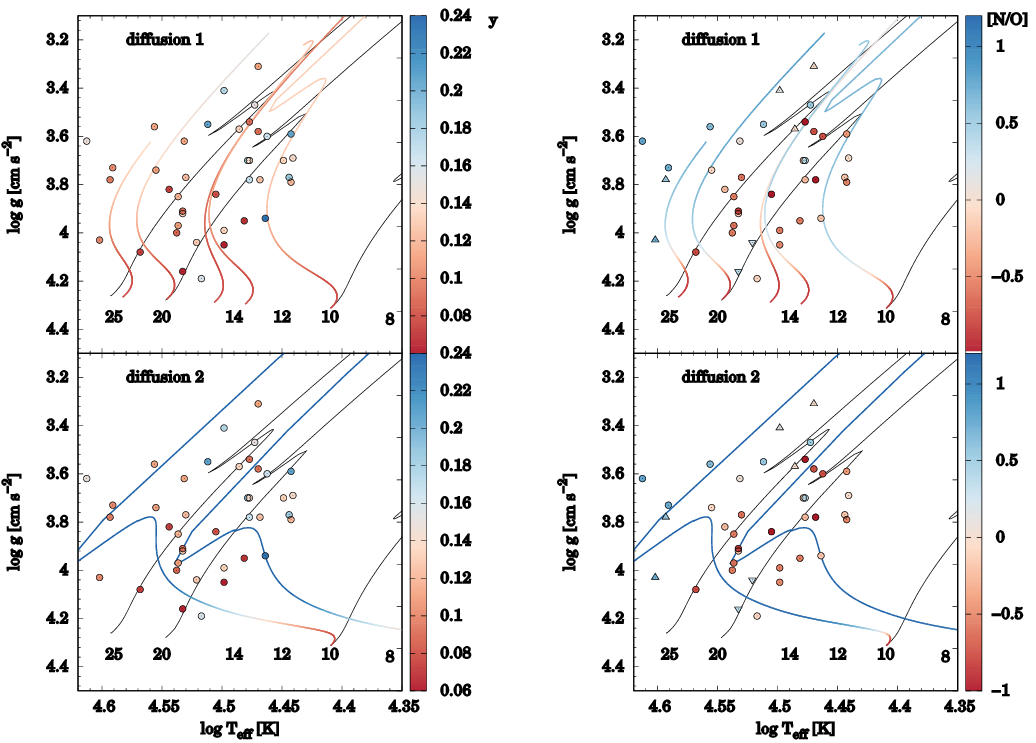}
\end{turn}
\caption{Illustrative impact of the choice of the diffusion coefficients on the CL\'ES evolutionary tracks, as well as the predicted helium abundances and [N/O] ratios. Black tracks stand for models without diffusion, while models with different diffusion coefficients are colour-coded according to helium abundance (left panel) or the [N/O] ratio (right). The first and second rows of panels show the tracks computed with diffusion 1 and 2  (see text), respectively. Circles stand for fixed values, upwards oriented triangles for lower limits, and downwards oriented triangles for upper limits in [N/O].}
\label{figMel}
\end{figure*}

In summary, while [N/O] ratios may or may not be reproduced depending on the object under consideration, single-star models and observations differ in a more systematic way for the helium abundance. 

For very luminous stars, the outer layers can be peeled off because of strong mass loss. Helium-rich material would then be revealed at the surface. However, as shown by \citet{bes14} from LMC observations, it only occurs for $\log(\dot{\rm M}/{\rm M})$ $\gtrsim$ --6.5. Current single-star models for Galactic stars using the mass-loss formalism of \citet{vin01} also do not predict strong helium enrichment for stars with masses below 60 M$_{\odot}$. This conclusion depends on the assumed overshooting parameter as a helium excess due to mass loss is expected in the stellar mass range 40--60 M$_{\odot}$ for high values \citep{cas14}. However, \object{HD 66811} is the only star in our sample falling in this mass range.

Furthermore, it should be noted that in the case of very fast rotation, the mixing timescale becomes shorter than the nuclear timescale \citep{mae87}. No strong chemical gradients can then develop in the stellar interior and it will therefore be completely mixed. Thus, a fast-rotating MS star could exhibit a high helium abundance at its surface; however,  as the opacity is reduced by the large fraction of helium at the surface, such a quasi-chemically homogeneous star will appear overluminous for its mass. This  peculiarity can be revealed with the method of \citet{lan14} for stars exhibiting an excess in helium at their surface and with accurate distance estimates. It is the case for \object{HD 66811} and \object{HD149757}, which have good Hipparcos parallaxes (\citealt{vanl97,mai08}); in the case of {\object{HD 66811}, we assume here that the helium excess at its surface arises from rotational mixing, and is not related to its strong mass loss. We evaluated their luminosities from the distances, reddenings (taken from \citet{bas92} and \citet{mor75} for \object{HD 66811} and \object{HD 149757}, respectively), apparent magnitudes in the $V$ band, and typical bolometric corrections for their spectral types \citep{mar05}. Evidence of an overluminosity is found for these two stars, which -- independent of their evolutionary history -- is consistent with their enhanced helium abundance \citep{lan92}. In the context of single stars, this may indicate a quasi-chemically homogeneous evolution. An assessment of such overluminosities for the full sample must await further {\it Gaia} data releases.

Another way to examine the mismatch between the observations and the theoretical expectations is to investigate whether the efficiency of chemical element transport is different from what is assumed in the models. We  therefore examined the role of turbulent diffusion. To this end, we  used an updated version of the Code Li\'egeois d'\'Evolution Stellaire (CL\'ES; \citealt{scu08}) in which a turbulent diffusion has been implemented for every element following a decreasing law towards the stellar interior, and which is controlled by a constant diffusion coefficient $D_{\rm T}$ as a free input parameter in the models. Predicted helium abundance and [N/O] ratios are illustrated in Kiel diagrams (Fig. \ref{figMel}) for different diffusion rates{, called diffusion 1 and 2, chosen according to their effect on the surface abundances. From these diagrams, it is found that helium and [N/O] enrichments can be explained in most stars by the diffusion 1 configuration, for which the diffusion coefficient is of the order of $D_{\rm T} \sim 10^7$ cm$^2$ s$^{-1}$. This diffusion coefficient, which had to be included in order to fit the observations, is quite large. \citet{mig08} have  shown that in a 6 M$_{\odot}$ model, a diffusion coefficient of $D_{\rm T}$ = 5000 cm$^2$ s$^{-1}$   reproduces the main sequence evolutionary tracks of a 6 M$_{\odot}$ rotating model with an initial velocity of 25 km s$^{-1}$. In order to reproduce the helium abundance of the most helium-enriched star in our sample (\object{HD 198781}), we have to consider an even larger diffusion, which would lower the initial mass of the model to 6 M$_{\odot}$, though (standard) evolutionary tracks (Fig. \ref{figLoggCTN}) hint at a mass of $\sim$15 M$_{\odot}$ for this star. Moreover, we  note that the [N/O] ratio predicted by this model is too high compared to the observed value. Furthermore, we  also tested the effect of combinations of overshooting and mass loss in models computed with the same evolutionary code but without diffusion. The only way to reproduce the observed enrichment in helium (and in [N/O]), as well as the position of our targets in the Kiel diagram, is by considering a very large parameter, such as $\alpha_{\rm oversh}=0.5$, and a mass-loss rate 10 times larger than the predicted values of \citet{vin01} for 40 and 50 M$_{\odot}$ models. In conclusion, such enrichments in helium combined with high [N/O] ratios can only be achieved in models accounting for uncommonly large input parameters.

\section{Comparison with binary star evolution}
\label{sec4}

As shown in the previous section, many aspects of our observations cannot be explained by single-star evolutionary models. In the following, we therefore consider the possible influence of companions on the surface abundances of our targets.

The presence of close companions may modify the surface abundances and the rotation rates of massive stars, because of tidal effects \citep{zah75,hut81,dem09,dem13,son13}, mass accretion \citep{pac81,pol91,pod92,lan03b,pet05a,pet05b,dem09,dem13,der10} or even merging \citep{dem13,tyl11}. For example, \citet{koh12} proposed that fast rotators with low surface nitrogen abundance (which cannot be explained by single-star evolutionary models) may have been slow rotators for most of their lives, and  then experienced a non-conservative mass transfer in a binary system. In parallel, the enrichment in nitrogen and/or helium is considered  a signature of a past Roche-lobe overflow (RLOF) for the \object{Plaskett's Star} \citep{lin08}, \object{$\theta$ Car} \citep{hub08}, \object{HD 149404} \citep{rauc16}, \object{LSS 3074} \citep{rauc17}, as well as the X-ray binaries \object{X Per} \citep{lyu97} and \object{BD +53$^{\circ}$ 2790} \citep{bla06}.

After such a mass-transfer episode, the gainer star may be ``rejuvenated'' (e.g. the blue straggler \object{$\theta$ Car} in \object{IC 2602}; \citealt{hub08}). The stars in clusters should thus appear younger than the other members. Two stars in our sample are confirmed cluster members, so  their ages are  known: \object{HD 46056} and \object{HD 46485} in \object{NGC 2244} \citep{ogu81}. They are both presumably single stars. Using our last BONNSAI run, we estimated the ages of these stars (see Table \ref{resBONN3}) and compared them with the age of their host cluster taken from \citet{hen00}: 2.3 $\pm$ 0.2 Myrs. No significant difference is found. 

While our sample contains several true or probable binaries, it also contains runaway stars; these objects can be the consequence of dynamical interactions in a cluster or the result of a(n) (asymmetric) supernova explosion. In the latter case, surface abundances of the surviving star may be affected. It is thus interesting to examine specifically the results obtained for runaway objects, whatever their multiplicity status. We note that, in dynamical interactions, ejection of a binary occurs in $\sim$ 10\% of the cases \citep{leo90}, while $\sim$ 20 to 40\% of runaways resulting from a supernova explosion remain binary systems \citep{por00}. However, ten of our objects --  five of which are runaways, including the SB1 \object{HD 210839} -- have been searched for pulsed radio emission (\citealt{phi96}; \citealt{say96}), but none was found.

Figures \ref{figLoggCNOBro}, \ref{figvsiniNOBro}, and \ref{figyNO2}  split the results shown in Figs. \ref{figLoggCNO}, \ref{figvsiniNO}, and \ref{figyNO}, respectively, according to the multiplicity status of the targets derived in Paper I. The lack of clear differences between binaries (or runaways) and single stars may at first sight look surprising. However, it should be kept in mind that the classification of targets as presumably single suffers from some unavoidable limitations.  For instance, an obvious observational bias is that several single stars lacking an extensive RV monitoring might have been detected as variable had more data been accumulated. Furthermore, some stars identified as presumably single might actually be the products of mass transfer in a binary. As noted by \citet{dem14}, the RV variation induced by the presence of a companion star will not be large enough to be detected in typically $\sim$ 45\% of all early B- and O-type stars. Indeed, models predict that post-mass exchange systems are relatively long-period binaries with a large mass ratio (e.g. \citealt{wel01}). The companion is thus expected to be much fainter than the actual primary and to reside in an orbit that is  quite wide, which would induce low-amplitude RV variations of the mass gainer (typically $\sim$ 10 km s$^{-1}$ if the unseen companion is a stripped-down remnant; see e.g. \citealt{poe81}, \citealt{pet08,pet13,pet16}). These expected RV variations are comparable to the precision that can be achieved for fast-rotating OB stars. To complicate matters further, stars that survive the supernova explosion of the former companion may not necessarily have a large peculiar velocity \citep{eld11}.

Furthermore, \citet{san12} argued that the evolution of more than 70\,\% of massive stars is affected by binary effects, and \citet{dem11} claimed that these effects mainly result in a single star (after a merging event that occurs, according to \citealt{san12}, for $\sim$25\% of O stars) or a look-alike star (for example one that has gained mass in a post mass-transfer event and which is associated with a very faint companion). Therefore, if we did not detect large differences between our subsamples on the basis of their multiplicity, it may simply be because several presumably single stars are binary products. It is thus worth comparing our results for all stars with those expected from binary evolutionary models.

First, \citet{gle13} calculated the consequences of stellar merging, notably on the surface abundances exhibited by the resulting object. Figure \ref{figNOMrem} compares their results with those obtained for our presumably single  stars. As can be seen, the observed [N/O] ratios agree with expectations for most of the lower-mass subsample, but disagree for the higher-mass objects. The reverse situation is found for the helium abundances. However, it must be noted that the models of \citet{gle13} do not take the rotation of stars into account so their predictions of surface nitrogen and helium abundances should be considered as lower limits. This leads us to speculate that observations and predictions are in fact in better agreement than Fig. \ref{figNOMrem} suggests.

Second, we  consider binary models that include mass and angular momentum transfer \citep{wel99,wel01}. Figure \ref{figLan} compares the predicted values for a mass gainer at the end of mass transfer assuming different initial masses, orbital periods, mass transfer cases, and semi-convective efficiency parameters. We see that binary models computed with a slow semi-convective mixing predict $y \lesssim$ 0.13. Predictions by these models are very similar to those by \citet{gle13} in the [N/O] versus $y$ plane. On the other hand, the model adopting the Schwarzschild condition reaches $y \sim$ 0.19. Binary models with fast semi-convective mixing can therefore   explain the abundances of our most helium-enriched stars, and can  also reproduce the effective temperature and surface gravity of two of them  (\object{HD 13268} and \object{HD 150574}) within the error bars.  There are, however, a few data points in Fig. \ref{figLan} located above the predictions of the binary models investigated here, i.e. stars with little helium enrichment but significant nitrogen excess. Three stars with both high $y$ and low [N/O] values also exist in a region that does not comply with the theory of the CNO-cycle, and the region is thus not expected to be filled. There is no evidence that these abundances are related to artefacts in the data analysis, but a few outliers are expected on statistical grounds, as mentioned in Sect. \ref{sec2},  and they can be reconciled with models at 3$\sigma$.

 \begin{figure*}
\centering
\begin{turn}{0}
\includegraphics[scale=1.5]{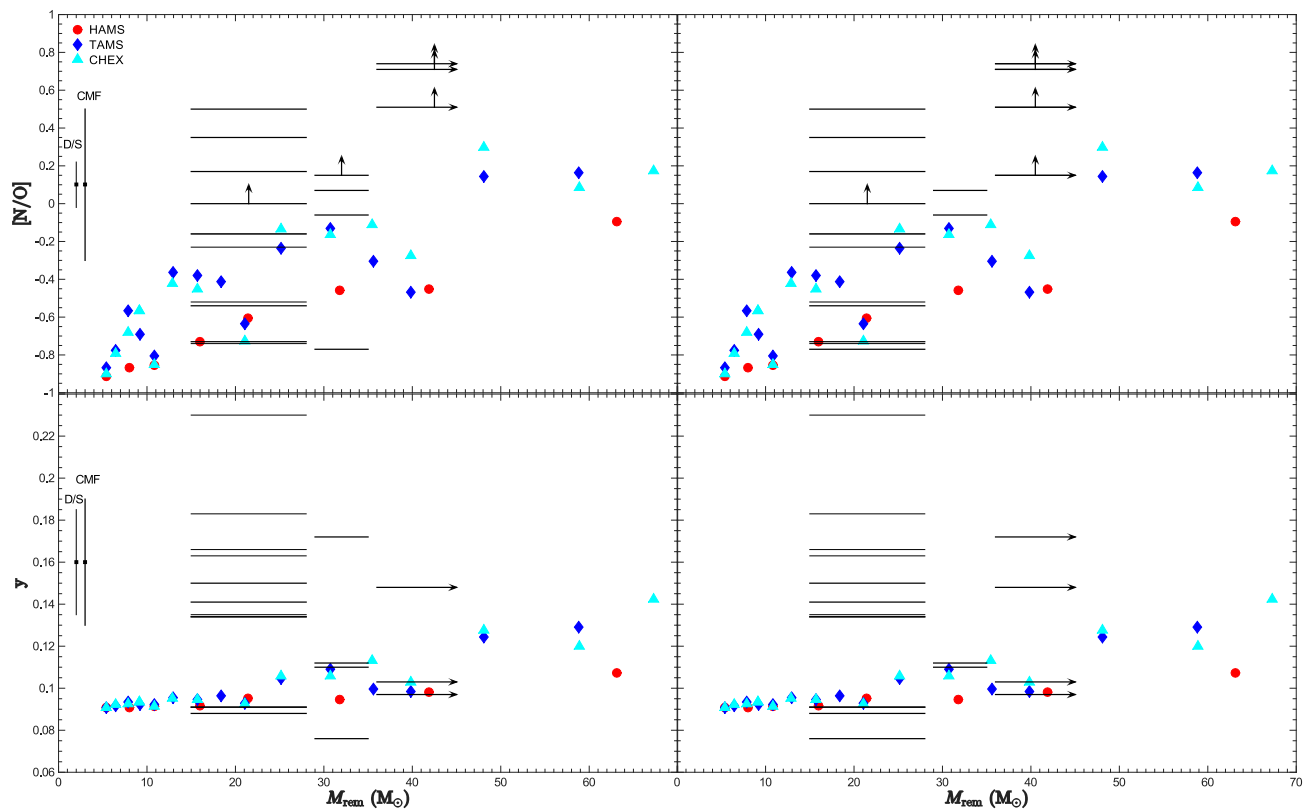}
\end{turn}
\caption{[N/O] abundance ratios (upper panels) and helium abundance (lower panels) as a function of the total mass of merger remnants predicted by the models of \citet{gle13}. Symbols represent the evolutionary stages of the parent stars: solid red circles, blue diamonds, and cyan triangles stand for collisions halfway through the main-sequence lifetime (HAMS), at the terminal-age main sequence (TAMS), and at core hydrogen exhaustion (CHEX), respectively. [N/O] values for our presumably single stars are represented with thin and thick horizontal lines for the cooler and hotter samples, respectively, as a function of their typical masses derived with the Geneva (left panels) and Bonn models (right panels). Typical error bars for the DETAIL/SURFACE and CMFGEN analyses are indicated to the left  of the diagrams.}
\label{figNOMrem}
\end{figure*}

\begin{figure*}
\centering
\begin{turn}{0}
\includegraphics[scale=0.35]{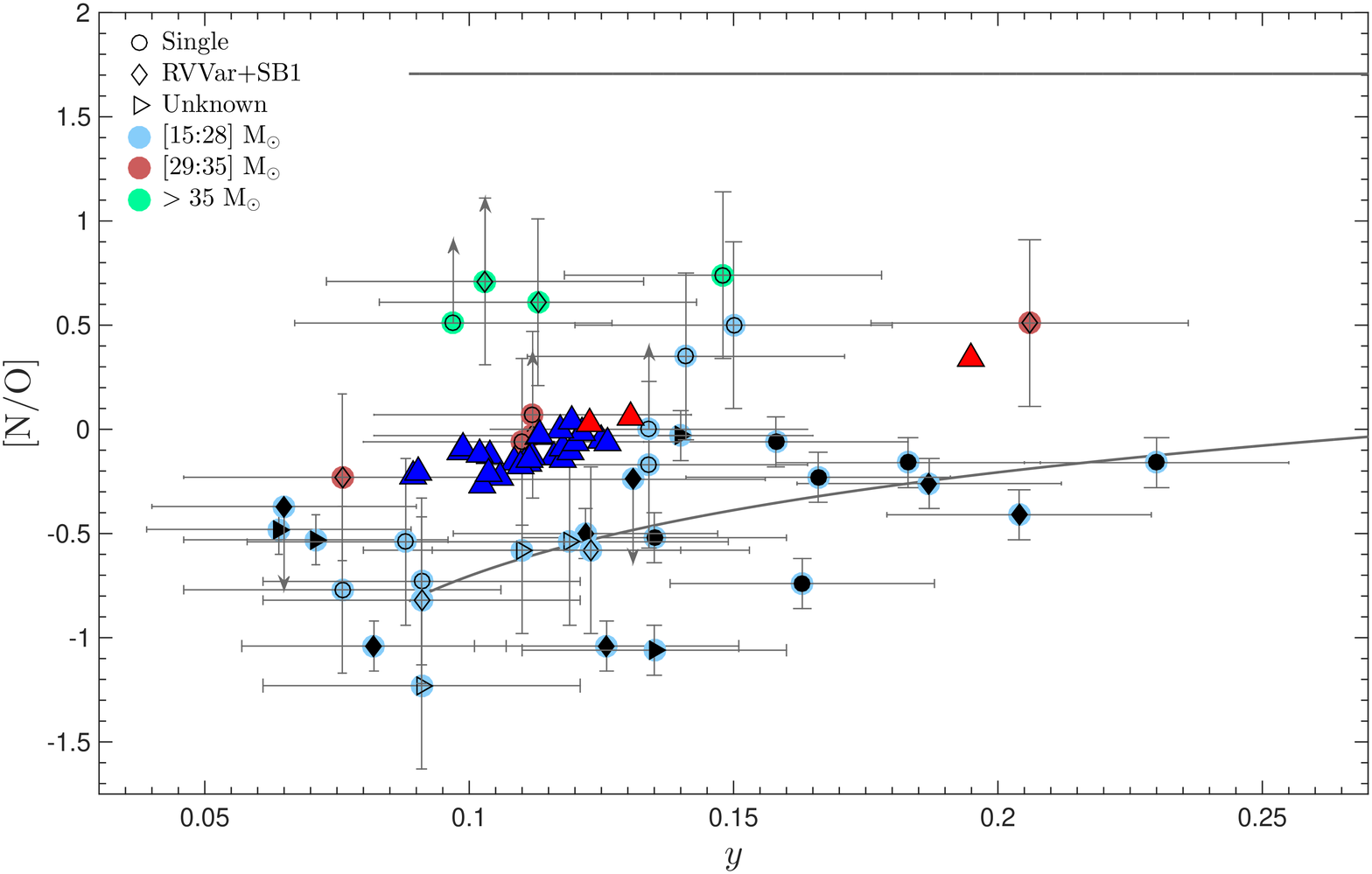}
\end{turn}
\caption{[N/O] abundance ratio as a function of $y$ for our sample stars and comparison with predicted values for mass gainers after a mass transfer, assuming that the mass donor has already exploded (triangles; \citealt{wel99,wel01}). Our data are colour-coded as a function of the stellar mass defined by the Bonn models. Black empty and filled symbols represent the hotter stars studied with CMFGEN and the cooler stars studied with DETAIL/SURFACE, respectively. The mass gainers with initial masses of 8 -- 25 M$_{\odot}$ which grow to 17 -- 40 M$_{\odot}$ are shown in dark blue, while the mass gainers with initial masses of 22 -- 24 M$_{\odot}$ which grow to $\sim$40 M$_{\odot}$ are shown in red (a slow semi-convection mixing, $\alpha_{\rm sc}$ = 0.01, and 0.04, has been considered for the two leftmost red triangles, while a fast semi-convection mixing Schwarzschild criterion, $\alpha_{\rm sc}$ = $\infty$, has been considered for the rightmost one). The upper grey line shows the CNO equilibrium, while the lower grey line assumes that the extra helium at the surface contains the CNO-equilibrium distribution.}
\label{figLan}
\end{figure*}

\section{Conclusion}
\label{sec6}

Following the derivation of the individual stellar parameters and abundances of 40 fast rotators in Paper I, we have analysed the results in a global way.

Using BONNSAI, we found that the \citet{bro11} models can reproduce the atmospheric parameters and abundances of half of our sample. Interestingly, we found that the atmospheric parameters and abundances can be reproduced by single-star evolutionary models whatever the multiplicity status of the targets. Some systems might thus be pre-interaction binaries \citep{dem11}. We found a systematic underprediction of the helium abundance for our targets. Changing the diffusion coefficient in models does not solve this issue as both $y$ and [N/O] ratios cannot be reproduced simultaneously.

As our sample contains known or probable binaries, as well as runaways, and since even presumably single stars may actually be binaries or have suffered from interactions with a companion, we have also compared our results with those from binary evolutionary models. We find that merger models of non-rotating objects \citep{gle13} are not    readily able to  reproduce the [N/O] abundance ratios of our higher-mass single stars and the helium abundances of our lower-mass single stars, but an agreement might be reached after stellar rotation is taken into account. On the other hand, binary models including mass and angular momentum transfer (through RLOF) appear to explain the [N/O] in most cases and can reproduce the helium abundances of some of our most helium-enriched stars, but have difficulties in explaining the properties of some of our stars.

In summary, we confirm the presence of fast massive rotators with no nitrogen enrichment for 10--20\% of our targets (first reported by \citealt{hun09}), but bring to light another unexpected problem: a quite common large abundance of helium at the stellar surface. Such features appear difficult to reproduce by single-star or binary evolutionary models, indicating that some fundamental physics ingredient is missing in (or is not well taken into account by) current models. On the observational side, future work should focus on fast rotators of the SMC and LMC, where the enhancement of the surface nitrogen abundance arising from rotational mixing is expected to be greater than in the Galaxy and which should thus reveal the abundance problems in even greater detail.

\begin{acknowledgements}
{We are very grateful to the referee for providing useful comments.} We thank Dr Fabian Schneider for helping us with BONNSAI, and Drs Nathan Grin and Pablo Marchant for providing useful comments.

We also thank Prof. Arlette Gr{\"o}tsch-Noels and Dr. Richard Scuflaire for very fruitful discussions and for making the necessary modifications in the CL\'ES evolutionary code for our paper.

The Li\`ege team acknowledges support from the Fonds National de la Recherche Scientifique (Belgium), the Communaut\'e Fran\c caise de Belgique, the PRODEX XMM and GAIA-DPAC contracts (Belspo), and an ARC grant for concerted research actions financed by the French community of Belgium (Wallonia-Brussels federation). ADS and CDS were used to prepare this document. 
\end{acknowledgements}


\begin{appendix}
  
\section{Illustration of the results for our sample stars as a function of their multiplicity status}
We present in Figs. \ref{figLoggCNOBro}, \ref{figvsiniNOBro}, and \ref{figyNO2} the results presented in Figs. \ref{figLoggCNO}, \ref{figvsiniNO}, and \ref{figyNO}, respectively, but split according to the multiplicity status derived in Paper I. \object{HD 28446 A} and \object{HD 41161} have been excluded because the origin of their RV variations is unclear (see Paper I).

\begin{figure*}
\centering
\begin{turn}{0}
\includegraphics[scale=1.6]{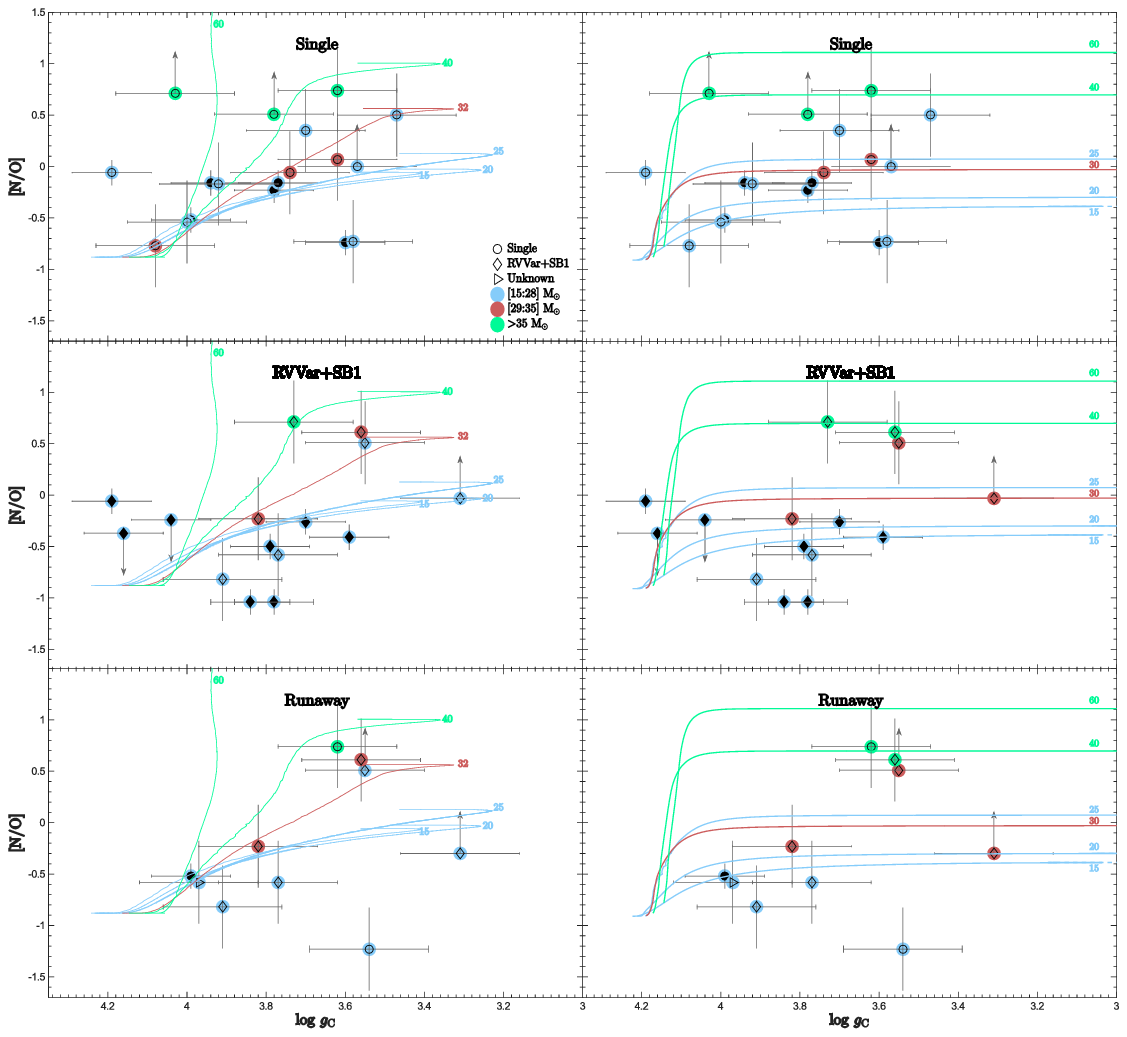}
\end{turn}
\caption{Same as Fig. \ref{figLoggCNO}, but for the different multiplicity status.}
\label{figLoggCNOBro}
\end{figure*}

 \begin{figure*}
\centering
\begin{turn}{0}
\includegraphics[scale=1.6]{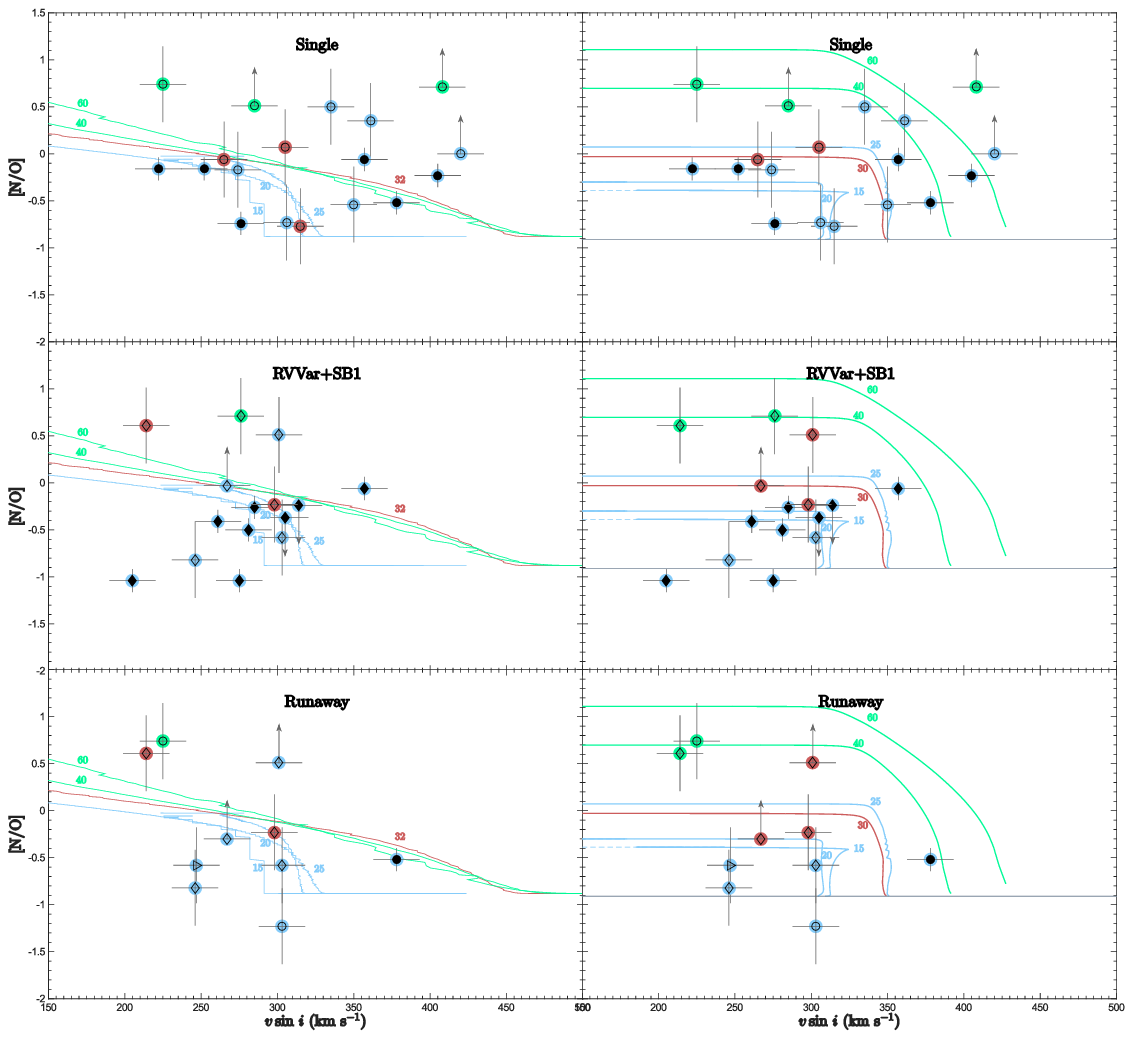}
\end{turn}
\caption{Same as Fig. \ref{figvsiniNO}, but for the different multiplicity status.} 
\label{figvsiniNOBro}
\end{figure*}

\begin{figure*}
\centering
\begin{turn}{0}
\includegraphics[scale=1.6]{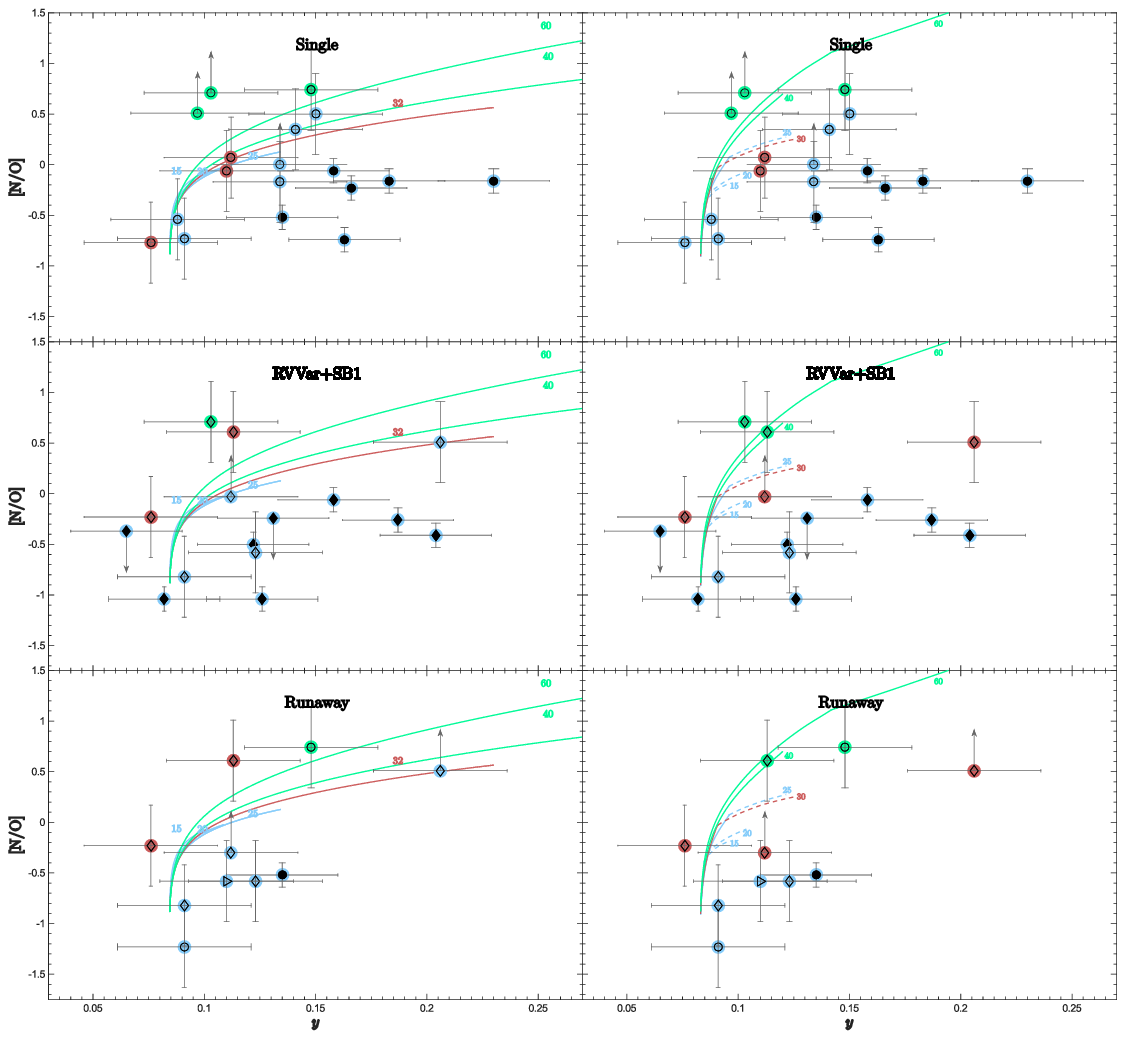}
\end{turn}
\caption{Same as Fig. \ref{figyNO}, but for the different multiplicity status.}
\label{figyNO2}
\end{figure*}

\section{Results obtained with BONNSAI}
\label{appB}
We present in Tables \ref{resBONN}, \ref{resBONN2}, and \ref{resBONN3} the parameters and abundances derived by BONNSAI.

\begin{sidewaystable*} 
\begin{tiny}  
\caption{Results obtained with BONNSAI for stars for which atmospheric parameters and He, C, N, O abundances can be reproduced with Bonn models when $T_{\rm{eff}}$, $\log g_{\rm{C}}$, $v\sin\,i$, $y$, $\log \epsilon$(C), $\log \epsilon$(N), and $\log \epsilon$(O) are given as input to BONNSAI. Observed values of Paper I are indicated between parentheses.}  
\label{resBONN}  
\begin{center}  
\begin{tabular}{ccccccccccccccccccccc}  
\hline\hline  
Star & \object{BD +34$^{\circ}$ 1058} & \object{BD +60$^{\circ}$ 594} & \object{HD 14434} & \object{HD 14442} & \object{HD 15137} & \object{HD 41161} \\ \hline  
$M_{\rm ini}$ (M$_{\odot}$) &  $ 20.0 _{-  1.6 }^{+  2.4 }$ &  
$ 20.6 _{-  1.7 }^{+  1.9 }$ &  
 $ 30.6 _{-  3.3 }^{+  4.0 }$ &  
 $ 39.2 _{-  4.5 }^{+  6.4 }$ &  
 $ 26.2 _{-  3.7 }^{+  7.0 }$ &  
  $ 22.0 _{-  3.0 }^{+  4.6 }$ \\ 
$M_{\rm act}$ (M$_{\odot}$) & $ 20.0 _{-  1.7 }^{+  2.2 }$ &  
$ 20.6 _{-  1.9 }^{+  1.6 }$ &  
$ 30.0 _{-  3.1 }^{+  3.9 }$ &  
$ 36.0 _{-  3.0 }^{+  5.7 }$ &  
$ 25.0 _{-  2.8 }^{+  6.1 }$ &  
$ 21.8 _{-  2.8 }^{+  4.2 }$ \\  
$\log\,L$ (L$_{\odot}$) & $  4.7 _{-  0.1 }^{+  0.1 }$ &  
$  4.7 _{-  0.1 }^{+  0.2 }$ &  
$  5.2 _{-  0.1 }^{+  0.1 }$ &  
$  5.5 _{-  0.1 }^{+  0.1 }$ &  
$  5.4 _{-  0.2 }^{+  0.1 }$ &  
$  4.9 _{-  0.1 }^{+  0.3 }$ \\  
Age (Myr) & $  1.6 _{-  1.0 }^{+  1.9 }$ &  
$  4.0 _{-  0.8 }^{+  0.6 }$ &  
$  1.3 _{-  0.5 }^{+  0.8 }$ &  
$  2.9 _{-  0.5 }^{+  0.4 }$ &  
$  4.6 _{-  0.4 }^{+  1.4 }$ &  
$  4.1 _{-  0.7 }^{+  0.9 }$ \\  
$\tau_{\rm MS}$ & $0.2 _{-  0.1 }^{+  0.2 }$ &  
$  0.5 _{-  0.1 }^{+  0.1 }$ &  
$  0.2 _{-  0.1 }^{+  0.1 }$ &  
$  0.6 _{-  0.1 }^{+  0.1 }$ &  
$  0.8 _{-  0.0 }^{+  0.0 }$ &  
$  0.6 _{-  0.2 }^{+  0.1 }$ \\  
$v_{\rm ini}$ (km s$^{-1}$) & $ 420.0 _{- 19.3 }^{+ 22.2 }$ &  
$ 400.0 _{- 23.2 }^{+ 41.2 }$ &  
$ 420.0 _{- 14.8 }^{+ 26.6 }$ &  
$ 380.0 _{- 33.4 }^{+ 27.1 }$ &  
$ 360.0 _{- 33.3 }^{+ 37.6 }$ &  
$ 330.0 _{- 25.4 }^{+ 34.4 }$ \\  
$v\sin\,i$ (km s$^{-1}$) & $ 420.0 _{- 19.2 }^{+ 12.9 }$ (424$\pm$15) & 
$ 320.0 _{- 19.9 }^{+ 13.1 }$ (314$\pm$15) & 
$ 400.0 _{- 13.8 }^{+ 16.9 }$ (408$\pm$15) & 
$ 280.0 _{- 12.3 }^{+ 18.3 }$ (285$\pm$15) & 
$ 270.0 _{- 17.6 }^{+ 14.1 }$ (267$\pm$15) & 
$ 300.0 _{- 12.7 }^{+ 18.4 }$ (303$\pm$15) \\ 
$i$ ($^{\circ}$) & 90.0  & 
 53.1  & 
77.3  & 
74.9  & 
74.6  & 
75.4  \\ 
$y$ &$ 0.085 _{- 0.000 }^{+ 0.006 }$ (0.119$\pm$0.030) & 
 $ 0.093 _{- 0.003 }^{+ 0.003 }$ (0.131$\pm$0.025) & 
$ 0.093 _{- 0.003 }^{+ 0.003 }$ (0.103$\pm$0.030) & 
$ 0.110 _{- 0.014 }^{+ 0.000 }$$\dagger$ (0.097$\pm$0.030) & 
$ 0.089 _{- 0.003 }^{+ 0.003 }$ (0.112$\pm$0.030) & 
$ 0.085 _{- 0.003 }^{+ 0.003 }$ (0.123$\pm$0.030) \\ 
$\log\,\epsilon$(C) & $ 7.91 _{- 0.21 }^{+ 0.14 }$ (7.90$\pm$0.27) & 
$ 7.72 _{- 0.08 }^{+ 0.12 }$ (7.66$\pm$0.12) & 
$ 7.70 _{- 0.16 }^{+ 0.18 }$ (7.96$\pm$0.27) & 
$ 7.39 _{- 0.03 }^{+ 0.22 }$$\dagger$ (7.10$\pm$0.27) & 
$ 7.80 _{- 0.15 }^{+ 0.12 }$$\dagger$ (7.63$\pm$0.27) & 
$ 7.90 _{- 0.13 }^{+ 0.06 }$ (7.87$\pm$0.27) \\ 
$\log\,\epsilon$(N) & $ 8.36 _{- 0.26 }^{+ 0.13 }$ (8.14$\pm$0.34) & 
$ 8.40 _{- 0.11 }^{+ 0.06 }$ (<8.24) & 
$ 8.50 _{- 0.19 }^{+ 0.06 }$ (8.81$\pm$0.34) & 
$ 8.61 _{- 0.10 }^{+ 0.02 }$ (8.61$\pm$0.34) & 
$ 8.34 _{- 0.08 }^{+ 0.14 }$$\dagger$ (8.27$\pm$0.34) & 
$ 8.26 _{- 0.14 }^{+ 0.10 }$ (8.09$\pm$0.34) \\ 
$\log\,\epsilon$(O) & $ 8.50 _{- 0.09 }^{+ 0.04 }$ (8.68$\pm$0.21) & 
$ 8.40 _{- 0.05 }^{+ 0.05 }$ (8.48$\pm$0.21) & 
$ 8.38 _{- 0.08 }^{+ 0.10 }$ ($\le$8.10) & 
$ 8.13 _{- 0.03 }^{+ 0.19 }$$\dagger$ ($\le$8.10) & 
$ 8.42 _{- 0.05 }^{+ 0.07 }$$\dagger$ ($\le$8.30) & 
$ 8.48 _{- 0.06 }^{+ 0.03 }$ (8.67$\pm$0.21) \\ 
\hline\hline  
Star & \object{HD 46056} & \object{HD 46485} & \object{HD 52533} & \object{HD 66811} & \object{HD 69106} & \object{HD 74920}   \\ \hline  
$M_{\rm ini}$ (M$_{\odot}$) & $ 20.6 _{-  2.1 }^{+  2.4 }$ &  
 $ 24.4 _{-  2.5 }^{+  3.2 }$ &  
 $ 21.4 _{-  1.6 }^{+  1.7 }$ &  
 $ 60.0 _{- 20.7 }^{+ 14.6 }$$\dagger$ &  
 $ 18.0 _{-  2.9 }^{+  3.3 }$ &  
  $ 22.2 _{-  3.0 }^{+  3.7 }$ \\ 
$M_{\rm act}$ (M$_{\odot}$) & $ 20.6 _{-  2.1 }^{+  2.3 }$ &  
$ 24.2 _{-  2.4 }^{+  3.2 }$ &  
$ 21.2 _{-  1.5 }^{+  1.6 }$ &  
$ 53.6 _{- 17.5 }^{+ 10.6 }$$\dagger$ &  
$ 18.8 _{-  3.6 }^{+  2.1 }$ &  
$ 22.0 _{-  2.8 }^{+  3.3 }$ \\  
$\log\,L$ (L$_{\odot}$) & $  4.7 _{-  0.1 }^{+  0.1 }$ &  
$  4.9 _{-  0.1 }^{+  0.1 }$ &  
$  4.8 _{-  0.1 }^{+  0.1 }$ &  
$  5.9 _{-  0.2 }^{+  0.2 }$ &  
$  4.8 _{-  0.2 }^{+  0.2 }$ &  
$  4.9 _{-  0.2 }^{+  0.2 }$ \\  
Age (Myr) & $  0.6 _{-  0.6 }^{+  1.5 }$ &  
$  0.5 _{-  0.5 }^{+  0.9 }$ &  
$  2.8 _{-  0.7 }^{+  0.8 }$ &  
$  2.1 _{-  0.4 }^{+  0.6 }$ &  
$  6.4 _{-  0.8 }^{+  1.3 }$ &  
$  3.8 _{-  0.8 }^{+  0.9 }$ \\  
$\tau_{\rm MS}$ & $  0.1 _{-  0.1 }^{+  0.1 }$ &  
$  0.1 _{-  0.1 }^{+  0.1 }$ &  
$  0.3 _{-  0.1 }^{+  0.1 }$ &  
$  0.6 _{-  0.1 }^{+  0.1 }$ &  
$  0.7 _{-  0.1 }^{+  0.1 }$ &  
$  0.6 _{-  0.2 }^{+  0.1 }$ \\  
$v_{\rm ini}$ (km s$^{-1}$) & $ 350.0 _{- 16.3 }^{+ 32.8 }$ &  
$ 330.0 _{- 26.7 }^{+ 32.9 }$ &  
$ 390.0 _{- 36.4 }^{+ 51.7 }$ &  
$ 380.0 _{- 58.1 }^{+ 29.0 }$ &  
$ 320.0 _{- 23.7 }^{+ 33.6 }$ &  
$ 340.0 _{- 35.0 }^{+ 55.5 }$ \\  
$v\sin\,i$ (km s$^{-1}$) & $ 350.0 _{- 17.8 }^{+ 14.1 }$ (350$\pm$15) & 
$ 320.0 _{- 21.5 }^{+  9.7 }$ (315$\pm$15) & 
$ 310.0 _{- 19.1 }^{+ 13.4 }$ (305$\pm$15) & 
$ 220.0 _{- 10.5 }^{+ 20.8 }$ (225$\pm$15) & 
$ 310.0 _{- 20.4 }^{+ 11.2 }$ (306$\pm$15) & 
$ 280.0 _{- 19.5 }^{+ 12.6 }$ (274$\pm$15) \\ 
$i$ ($^{\circ}$) &  76.5  & 
90.0  & 
52.6  & 
66.4  & 
75.6  & 
58.0  \\ 
$y$ & $ 0.085 _{- 0.000 }^{+ 0.000 }$ (0.088$\pm$0.030) & 
$ 0.081 _{- 0.000 }^{+ 0.000 }$ (0.076$\pm$0.030) & 
$ 0.089 _{- 0.003 }^{+ 0.003 }$ (0.065$\pm$0.025) & 
$ 0.105 _{- 0.010 }^{+ 0.010 }$$\dagger$ (0.148$\pm$0.030) & 
$ 0.085 _{- 0.000 }^{+ 0.003 }$ (0.091$\pm$0.030) & 
$ 0.085 _{- 0.003 }^{+ 0.003 }$ (0.134$\pm$0.030) \\ 
$\log\,\epsilon$(C) & $ 8.12 _{- 0.13 }^{+ 0.02 }$ (8.34$\pm$0.27) & 
$ 8.13 _{- 0.11 }^{+ 0.01 }$ (8.46$\pm$0.27) & 
$ 7.84 _{- 0.12 }^{+ 0.13 }$ (7.76$\pm$0.12) & 
$ 7.22 _{- 0.09 }^{+ 0.21 }$ ($\le$7.00) & 
$ 7.90 _{- 0.08 }^{+ 0.06 }$ (7.88$\pm$0.27) & 
$ 7.90 _{- 0.15 }^{+ 0.10 }$ (7.78$\pm$0.27) \\ 
$\log\,\epsilon$(N) & $ 7.64 _{- 0.01 }^{+ 0.41 }$ (7.78$\pm$0.34) & 
$ 7.64 _{- 0.01 }^{+ 0.35 }$ (7.95$\pm$0.34) & 
$ 8.32 _{- 0.15 }^{+ 0.11 }$ (<7.78) & 
$ 8.65 _{- 0.08 }^{+ 0.09 }$ (8.94$\pm$0.34) & 
$ 8.13 _{- 0.09 }^{+ 0.13 }$ (7.74$\pm$0.34) & 
$ 8.26 _{- 0.14 }^{+ 0.17 }$ (8.34$\pm$0.34) \\ 
$\log\,\epsilon$(O) & $ 8.55 _{- 0.05 }^{+ 0.01 }$ (8.32$\pm$0.21) & 
$ 8.55 _{- 0.04 }^{+ 0.00 }$ (8.72$\pm$0.21) & 
$ 8.45 _{- 0.06 }^{+ 0.05 }$ (8.15$\pm$0.21) & 
$ 8.06 _{- 0.20 }^{+ 0.23 }$$\dagger$ (8.20$\pm$0.21) & 
$ 8.51 _{- 0.03 }^{+ 0.02 }$ (8.47$\pm$0.21) & 
$ 8.49 _{- 0.08 }^{+ 0.04 }$ (8.51$\pm$0.21) \\ 
\hline\hline  
Star & \object{HD 117490} & \object{HD 124979} & \object{HD 149757} & \object{HD 175876} & \object{HD 191423} & \object{HD 192281} \\ \hline  
$M_{\rm ini}$ (M$_{\odot}$) & $ 18.4 _{-  1.9 }^{+  4.5 }$ &  
 $ 20.2 _{-  2.3 }^{+  3.0 }$ &  
 $ 17.2 _{-  1.3 }^{+  1.2 }$ &  
 $ 27.0 _{-  3.4 }^{+  6.8 }$ &  
 $ 18.8 _{-  1.9 }^{+  1.6 }$ &  
  $ 38.0 _{-  5.6 }^{+  6.9 }$ \\ 
$M_{\rm act}$ (M$_{\odot}$) & $ 18.6 _{-  2.0 }^{+  3.8 }$ &  
$ 20.0 _{-  2.1 }^{+  3.0 }$ &  
$ 17.2 _{-  1.3 }^{+  1.2 }$ &  
$ 26.8 _{-  3.3 }^{+  5.5 }$ &  
$ 18.4 _{-  1.7 }^{+  1.6 }$ &  
$ 35.2 _{-  3.9 }^{+  6.2 }$ \\  
$\log\,L$ (L$_{\odot}$) & $  4.9 _{-  0.3 }^{+  0.2 }$ &  
$  4.7 _{-  0.2 }^{+  0.2 }$ &  
$  4.5 _{-  0.1 }^{+  0.1 }$ &  
$  5.3 _{-  0.2 }^{+  0.1 }$ &  
$  4.8 _{-  0.1 }^{+  0.1 }$ &  
$  5.5 _{-  0.2 }^{+  0.1 }$ \\  
Age (Myr) & $  6.1 _{-  0.8 }^{+  1.0 }$ &  
$  3.5 _{-  2.2 }^{+  0.8 }$ &  
$  1.7 _{-  0.9 }^{+  1.0 }$ &  
$  3.4 _{-  0.6 }^{+  0.6 }$ &  
$  6.8 _{-  0.8 }^{+  1.0 }$ &  
$  2.8 _{-  0.5 }^{+  0.4 }$ \\  
$\tau_{\rm MS}$ & $  0.8 _{-  0.1 }^{+  0.1 }$ &  
$  0.2 _{-  0.1 }^{+  0.3 }$ &  
$  0.2 _{-  0.1 }^{+  0.1 }$ &  
$  0.6 _{-  0.1 }^{+  0.1 }$ &  
$  0.7 _{-  0.1 }^{+  0.1 }$ &  
$  0.6 _{-  0.1 }^{+  0.1 }$ \\  
$v_{\rm ini}$ (km s$^{-1}$) & $ 400.0 _{- 34.4 }^{+ 30.8 }$ &  
$ 270.0 _{- 23.2 }^{+ 65.5 }$ &  
$ 370.0 _{- 16.9 }^{+ 30.1 }$ &  
$ 330.0 _{- 36.7 }^{+ 36.6 }$ &  
$ 410.0 _{- 15.4 }^{+ 15.5 }$ &  
$ 360.0 _{- 33.5 }^{+ 28.8 }$ \\  
$v\sin\,i$ (km s$^{-1}$) & $ 360.0 _{- 14.8 }^{+ 16.9 }$ (361$\pm$15) & 
$ 240.0 _{-  7.9 }^{+ 23.8 }$ (246$\pm$15) & 
$ 370.0 _{- 11.3 }^{+ 20.4 }$ (378$\pm$15) & 
$ 270.0 _{- 20.0 }^{+ 12.0 }$ (265$\pm$15) & 
$ 410.0 _{- 11.9 }^{+ 18.3 }$ (420$\pm$15) & 
$ 270.0 _{- 11.3 }^{+ 19.4 }$ (276$\pm$15) \\ 
$i$ ($^{\circ}$) &  76.6  & 
62.7  & 
76.8  & 
74.6  & 
77.5  & 
74.6  \\ 
$y$ & $  0.085 _{- 0.000 }^{+ 0.013  }$$\dagger$ (0.141$\pm$0.030) & 
$  0.081 _{- 0.000 }^{+ 0.000  }$ (0.091$\pm$0.030) & 
$  0.081 _{- 0.000 }^{+ 0.000  }$ (0.135$\pm$0.025) & 
$  0.089 _{- 0.003 }^{+ 0.000  }$ (0.110$\pm$0.030) & 
$  0.093 _{- 0.003 }^{+ 0.003  }$$\dagger$ (0.134$\pm$0.030) & 
$  0.089 _{- 0.003 }^{+ 0.010  }$ (0.103$\pm$0.030) \\ 
$\log\,\epsilon$(C) & $ 7.82 _{- 0.34 }^{+ 0.04 }$$\dagger$ ($\le$7.39) & 
$ 8.12 _{- 0.11 }^{+ 0.02 }$ (8.48$\pm$0.27) & 
$ 8.07 _{- 0.06 }^{+ 0.04 }$ (8.07$\pm$0.12) & 
$ 7.81 _{- 0.04 }^{+ 0.16 }$ (8.04$\pm$0.27) & 
$ 7.61 _{- 0.08 }^{+ 0.08 }$ ($\le$7.24) & 
$ 7.67 _{- 0.24 }^{+ 0.19 }$$\dagger$ (8.00$\pm$0.27) \\ 
$\log\,\epsilon$(N) & $ 8.51 _{- 0.20 }^{+ 0.06 }$$\dagger$ (8.50$\pm$0.34) & 
$ 7.65 _{- 0.01 }^{+ 0.47 }$$\dagger$ (7.92$\pm$0.34) & 
$ 7.90 _{- 0.13 }^{+ 0.15 }$ (7.85$\pm$0.13) & 
$ 8.33 _{- 0.20 }^{+ 0.05 }$ (8.36$\pm$0.34) & 
$ 8.42 _{- 0.07 }^{+ 0.06 }$$\dagger$ (8.33$\pm$0.34) & 
$ 8.43 _{- 0.13 }^{+ 0.19 }$$\dagger$ (8.76$\pm$0.34) \\ 
$\log\,\epsilon$(O) & $ 8.48 _{- 0.17 }^{+ 0.02 }$$\dagger$ (8.15$\pm$0.21) & 
$ 8.55 _{- 0.03 }^{+ 0.00 }$ (8.74$\pm$0.21) & 
$ 8.54 _{- 0.02 }^{+ 0.01 }$ (8.37$\pm$0.21) & 
$ 8.43 _{- 0.02 }^{+ 0.07 }$ (8.42$\pm$0.21) & 
$ 8.40 _{- 0.07 }^{+ 0.02 }$ ($\le$8.33) & 
$ 8.43 _{- 0.18 }^{+ 0.03 }$ (8.05$\pm$0.21) \\ 
\hline  
\end{tabular} 
\end{center}  
\end{tiny} 
\end{sidewaystable*}  
\begin{table*} 
\addtocounter{table}{-1} 
\begin{tiny}  
\caption{Continued.}  
\begin{center}  
\begin{tabular}{ccccccccccccccccccccc}  
\hline\hline  
Star & \object{HD 203064} & \object{HD 210839} & \object{HD 228841}\\ \hline  
$M_{\rm ini}$ (M$_{\odot}$) & $ 23.6 _{-  3.0 }^{+  5.1 }$ &  
 $ 37.8 _{-  6.3 }^{+ 16.7 }$ &  
  $ 25.8 _{-  2.5 }^{+  6.6 }$ \\ 
$M_{\rm act}$ (M$_{\odot}$) & $ 23.2 _{-  2.6 }^{+  4.7 }$ &  
$ 34.6 _{-  4.1 }^{+ 13.5 }$ &  
$ 24.8 _{-  1.9 }^{+  5.8 }$ \\  
$\log\,L$ (L$_{\odot}$) & $  5.1 _{-  0.3 }^{+  0.2 }$ &  
$  5.6 _{-  0.2 }^{+  0.2 }$ &  
$  5.2 _{-  0.1 }^{+  0.2 }$ \\  
Age (Myr) & $  3.6 _{-  0.7 }^{+  0.8 }$ &  
$  3.4 _{-  0.8 }^{+  0.3 }$ &  
$  4.1 _{-  0.4 }^{+  0.9 }$ \\  
$\tau_{\rm MS}$ & $  0.6 _{-  0.2 }^{+  0.1 }$ &  
$  0.7 _{-  0.1 }^{+  0.0 }$ &  
$  0.7 _{-  0.1 }^{+  0.1 }$ \\  
$v_{\rm ini}$ (km s$^{-1}$) & $ 330.0 _{- 25.1 }^{+ 37.9 }$ &  
$ 360.0 _{- 39.2 }^{+ 31.4 }$ &  
$ 380.0 _{- 34.6 }^{+ 30.8 }$ \\  
$v\sin\,i$ (km s$^{-1}$) & $ 300.0 _{- 16.9 }^{+ 14.3 }$ (298$\pm$15) & 
$ 220.0 _{- 20.3 }^{+ 11.0 }$ (214$\pm$15) & 
$ 310.0 _{- 19.5 }^{+ 12.0 }$ (305$\pm$15) \\ 
$i$ ($^{\circ}$) &  75.4  & 
73.0  & 
75.6  \\ 
$y$ & $  0.085 _{- 0.003 }^{+ 0.003  }$ (0.076$\pm$0.030) & 
$  0.093 _{- 0.007 }^{+ 0.007  }$ (0.113$\pm$0.030) & 
$  0.089 _{- 0.000 }^{+ 0.007  }$ (0.112$\pm$0.030) \\ 
$\log\,\epsilon$(C) & $ 7.91 _{- 0.14 }^{+ 0.06 }$ (7.92$\pm$0.27) & 
$ 7.73 _{- 0.37 }^{+ 0.02 }$$\dagger$ (7.83$\pm$0.27) & 
$ 7.78 _{- 0.19 }^{+ 0.07 }$ (7.48$\pm$0.27) \\ 
$\log\,\epsilon$(N) & $ 8.33 _{- 0.19 }^{+ 0.05 }$ (8.23$\pm$0.34) & 
$ 8.43 _{- 0.07 }^{+ 0.20 }$$\dagger$ (8.74$\pm$0.34) & 
$ 8.34 _{- 0.04 }^{+ 0.15 }$ (8.74$\pm$0.34) \\ 
$\log\,\epsilon$(O) & $ 8.48 _{- 0.07 }^{+ 0.02 }$ (8.46$\pm$0.21) & 
$ 8.33 _{- 0.15 }^{+ 0.13 }$$\dagger$ (8.13$\pm$0.21) & 
$ 8.43 _{- 0.11 }^{+ 0.02 }$$\dagger$ (8.67$\pm$0.21) \\ 
\hline  
\end{tabular} 
\end{center}  
\tablefoot{$M_{\rm ini}$ and $M_{\rm act}$ are the initial and current stellar masses, respectively, $L$ is the stellar bolometric luminosity, $\tau_{\rm MS}$ is the fractional MS age, $i$ is the inferred stellar inclination from the actual and projected rotational velocities given by BONNSAI, and $v_{\rm ini}$ is the initial rotational velocity. Errors from BONNSAI correspond to 1\,$\sigma$, except for values flagged with $\dag$ for which the errors are slightly larger than 1\,$\sigma$.} 
\end{tiny} 
\end{table*}

\begin{sidewaystable*} 
\begin{tiny}  
\caption{Same as Fig. \ref{resBONN} but when $T_{\rm{eff}}$, $\log g_{\rm{C}}$, $v\sin\,i$, and $y$ are given as input to BONNSAI. }  
\label{resBONN2}  
\begin{center}  
\begin{tabular}{ccccccccccccccccccccc}  
\hline\hline  
Star & \object{ALS 864} & \object{ALS 18675} & \object{BD +34$^{\circ}$ 1058} & \object{BD +60$^{\circ}$ 594} & \object{HD 14434} & \object{HD 14442} \\ \hline  
$M_{\rm ini}$ (M$_{\odot}$) & $ 17.4 _{-  1.4 }^{+  1.5 }$ &  
 $ 16.4 _{-  1.4 }^{+  1.7 }$ &  
 $ 21.6 _{-  2.1 }^{+  1.6 }$ &  
 $ 19.8 _{-  1.5 }^{+  1.9 }$ &  
 $ 30.2 _{-  3.5 }^{+  4.0 }$ &  
  $ 37.8 _{-  8.2 }^{+  4.0 }$ \\ 
$M_{\rm act}$ (M$_{\odot}$) & $ 17.4 _{-  1.4 }^{+  1.4 }$ &  
$ 16.6 _{-  1.5 }^{+  1.4 }$ &  
$ 21.2 _{-  1.9 }^{+  1.6 }$ &  
$ 20.0 _{-  1.7 }^{+  1.5 }$ &  
$ 29.8 _{-  3.5 }^{+  3.8 }$ &  
$ 35.2 _{-  6.1 }^{+  3.7 }$ \\  
$\log\,L$ (L$_{\odot}$) & $  4.5 _{-  0.1 }^{+  0.1 }$ &  
$  4.5 _{-  0.1 }^{+  0.1 }$ &  
$  4.9 _{-  0.2 }^{+  0.1 }$ &  
$  4.7 _{-  0.1 }^{+  0.1 }$ &  
$  5.2 _{-  0.1 }^{+  0.1 }$ &  
$  5.5 _{-  0.3 }^{+  0.1 }$ \\  
Age (Myr) & $  3.8 _{-  1.9 }^{+  1.1 }$ &  
$  5.3 _{-  1.4 }^{+  1.0 }$ &  
$  5.0 _{-  1.8 }^{+  1.2 }$ &  
$  3.6 _{-  1.6 }^{+  1.0 }$ &  
$  1.2 _{-  1.2 }^{+  0.9 }$ &  
$  2.7 _{-  0.6 }^{+  0.6 }$ \\  
$\tau_{\rm MS}$ & $  0.4 _{-  0.2 }^{+  0.1 }$ &  
$  0.5 _{-  0.2 }^{+  0.1 }$ &  
$  0.6 _{-  0.2 }^{+  0.1 }$ &  
$  0.4 _{-  0.2 }^{+  0.1 }$ &  
$  0.3 _{-  0.2 }^{+  0.1 }$ &  
$  0.6 _{-  0.1 }^{+  0.1 }$ \\  
$v_{\rm ini}$ (km s$^{-1}$) & $ 350.0 _{- 94.3 }^{+ 28.2 }$ &  
$ 330.0 _{- 48.1 }^{+ 73.5 }$ &  
$ 430.0 _{- 19.7 }^{+ 34.4 }$ &  
$ 330.0 _{- 25.5 }^{+ 62.2 }$ &  
$ 430.0 _{- 26.3 }^{+ 28.6 }$ &  
$ 360.0 _{- 48.4 }^{+ 38.5 }$ \\  
$v\sin\,i$ (km s$^{-1}$) & $ 250.0 _{- 15.7 }^{+ 16.3 }$ (249$\pm$15) & 
$ 240.0 _{- 17.5 }^{+ 14.4 }$ (236$\pm$15) & 
$ 420.0 _{- 17.0 }^{+ 14.2 }$ (424$\pm$15) & 
$ 320.0 _{- 20.4 }^{+ 11.2 }$ (314$\pm$15) & 
$ 400.0 _{- 12.8 }^{+ 17.6 }$ (408$\pm$15) & 
$ 280.0 _{- 12.6 }^{+ 18.8 }$ (285$\pm$15) \\ 
$i$ ($^{\circ}$) &  44.0  & 
36.9  & 
90.0  & 
75.9  & 
77.3  & 
74.9  \\ 
$y$ & $ 0.085 _{- 0.000 }^{+ 0.000 }$ (0.064$\pm$0.025) & 
$ 0.085 _{- 0.000 }^{+ 0.003 }$ (0.071$\pm$0.025) & 
$ 0.097 _{- 0.007 }^{+ 0.010 }$$\dagger$ (0.119$\pm$0.030) & 
$ 0.085 _{- 0.000 }^{+ 0.003 }$ (0.131$\pm$0.025) & 
$ 0.085 _{- 0.000 }^{+ 0.013 }$$\dagger$ (0.103$\pm$0.030) & 
$ 0.089 _{- 0.003 }^{+ 0.010 }$$\dagger$ (0.097$\pm$0.030) \\ 
$\log\,\epsilon$(C) & $ 8.12 _{- 0.16 }^{+ 0.01 }$ (<7.86) & 
$ 8.04 _{- 0.14 }^{+ 0.08 }$ (7.78$\pm$0.12) & 
$ 7.47 _{- 0.30 }^{+ 0.23 }$$\dagger$ (7.90$\pm$0.27) & 
$ 8.12 _{- 0.26 }^{+ 0.01 }$ (7.66$\pm$0.12) & 
$ 8.12 _{- 0.67 }^{+ 0.01 }$$\dagger$ (7.96$\pm$0.27) & 
$ 7.81 _{- 0.38 }^{+ 0.15 }$$\dagger$ (7.10$\pm$0.27) \\ 
$\log\,\epsilon$(N) & $ 7.65 _{- 0.02 }^{+ 0.42 }$ (7.64$\pm$0.13) & 
$ 7.97 _{- 0.13 }^{+ 0.30 }$ (7.54$\pm$0.13) & 
$ 8.57 _{- 0.15 }^{+ 0.13 }$ (8.14$\pm$0.34) & 
$ 8.16 _{- 0.27 }^{+ 0.24 }$ (<8.24) & 
$ 8.53 _{- 0.28 }^{+ 0.20 }$ (8.81$\pm$0.34) & 
$ 8.33 _{- 0.10 }^{+ 0.35 }$$\dagger$ (8.61$\pm$0.34) \\ 
$\log\,\epsilon$(O) & $ 8.54 _{- 0.04 }^{+ 0.02 }$ (8.12$\pm$0.21) & 
$ 8.54 _{- 0.05 }^{+ 0.02 }$ (8.07$\pm$0.21) & 
$ 8.32 _{- 0.21 }^{+ 0.12 }$ (8.68$\pm$0.21) & 
$ 8.55 _{- 0.09 }^{+ 0.01 }$ (8.48$\pm$0.21) & 
$ 8.54 _{- 0.28 }^{+ 0.01 }$ ($\le$8.10) & 
$ 8.43 _{- 0.19 }^{+ 0.08 }$ ($\le$8.10) \\ 
\hline\hline  
Star & \object{HD 15137} & \object{HD 15642} & \object{HD 28446A} & \object{HD 41161} & \object{HD 41997} & \object{HD 46056}   \\ \hline  
$M_{\rm ini}$ (M$_{\odot}$) & $ 25.0 _{-  3.2 }^{+  7.8 }$ &  
 $ 24.6 _{-  5.2 }^{+  2.1 }$ &  
 $ 17.2 _{-  1.8 }^{+  2.0 }$ &  
 $ 23.0 _{-  3.2 }^{+  5.0 }$ &  
 $ 21.6 _{-  2.9 }^{+  3.2 }$ &  
  $ 21.6 _{-  2.5 }^{+  2.8 }$ \\ 
$M_{\rm act}$ (M$_{\odot}$) & $ 24.4 _{-  3.0 }^{+  6.1 }$ &  
$ 23.4 _{-  4.3 }^{+  1.9 }$ &  
$ 16.6 _{-  1.2 }^{+  2.4 }$ &  
$ 22.8 _{-  3.1 }^{+  4.2 }$ &  
$ 21.4 _{-  2.7 }^{+  3.0 }$ &  
$ 21.4 _{-  2.4 }^{+  2.7 }$ \\  
$\log\,L$ (L$_{\odot}$) & $  5.3 _{-  0.1 }^{+  0.2 }$ &  
$  5.2 _{-  0.3 }^{+  0.1 }$ &  
$  4.7 _{-  0.1 }^{+  0.2 }$ &  
$  5.1 _{-  0.3 }^{+  0.2 }$ &  
$  4.8 _{-  0.2 }^{+  0.2 }$ &  
$  4.8 _{-  0.2 }^{+  0.2 }$ \\  
Age (Myr) & $  4.5 _{-  0.3 }^{+  1.7 }$ &  
$  6.2 _{-  0.7 }^{+  1.1 }$ &  
$  6.4 _{-  0.7 }^{+  0.9 }$ &  
$  4.2 _{-  0.7 }^{+  1.0 }$ &  
$  3.6 _{-  1.5 }^{+  1.1 }$ &  
$  3.5 _{-  1.9 }^{+  1.0 }$ \\  
$\tau_{\rm MS}$ & $  0.8 _{-  0.0 }^{+  0.0 }$ &  
$  0.8 _{-  0.1 }^{+  0.1 }$ &  
$  0.7 _{-  0.1 }^{+  0.1 }$ &  
$  0.7 _{-  0.2 }^{+  0.1 }$ &  
$  0.5 _{-  0.3 }^{+  0.1 }$$\dagger$ &  
$  0.4 _{-  0.2 }^{+  0.2 }$ \\  
$v_{\rm ini}$ (km s$^{-1}$) & $ 360.0 _{- 50.0 }^{+ 54.2 }$ &  
$ 390.0 _{- 46.8 }^{+ 53.0 }$ &  
$ 290.0 _{- 13.8 }^{+ 96.9 }$ &  
$ 340.0 _{- 31.9 }^{+ 61.9 }$ &  
$ 350.0 _{- 82.4 }^{+ 66.0 }$$\dagger$ &  
$ 360.0 _{- 21.9 }^{+ 40.3 }$ \\  
$v\sin\,i$ (km s$^{-1}$) & $ 270.0 _{- 18.2 }^{+ 13.7 }$ (267$\pm$15) & 
$ 330.0 _{- 11.8 }^{+ 19.6 }$ (335$\pm$15) & 
$ 280.0 _{- 19.6 }^{+ 11.9 }$ (275$\pm$15) & 
$ 300.0 _{- 11.9 }^{+ 19.8 }$ (303$\pm$15) & 
$ 250.0 _{- 17.2 }^{+ 14.9 }$ (247$\pm$15) & 
$ 350.0 _{- 16.6 }^{+ 15.1 }$ (350$\pm$15) \\ 
$i$ ($^{\circ}$) &  74.6  & 
76.1  & 
74.9  & 
69.6  & 
45.6  & 
76.5  \\ 
$y$ & $ 0.089 _{- 0.003 }^{+ 0.007 }$$\dagger$ (0.112$\pm$0.030) & 
$ 0.085 _{- 0.000 }^{+ 0.013 }$$\dagger$ (0.150$\pm$0.030) & 
$ 0.085 _{- 0.003 }^{+ 0.003 }$$\dagger$ (0.126$\pm$0.025) & 
$ 0.089 _{- 0.003 }^{+ 0.003 }$$\dagger$ (0.123$\pm$0.030) & 
$ 0.085 _{- 0.000 }^{+ 0.006 }$ (0.110$\pm$0.030) & 
$ 0.085 _{- 0.000 }^{+ 0.006 }$ (0.088$\pm$0.030) \\ 
$\log\,\epsilon$(C) & $ 7.80 _{- 0.21 }^{+ 0.14 }$$\dagger$ (7.63$\pm$0.27) & 
$ 7.83 _{- 0.41 }^{+ 0.06 }$$\dagger$ ($\le$7.55) & 
$ 7.97 _{- 0.20 }^{+ 0.05 }$$\dagger$ (8.30$\pm$0.12) & 
$ 7.88 _{- 0.28 }^{+ 0.11 }$$\dagger$ (7.87$\pm$0.27) & 
$ 8.13 _{- 0.30 }^{+ 0.01 }$ (8.59$\pm$0.27) & 
$ 8.12 _{- 0.34 }^{+ 0.02 }$ (8.34$\pm$0.27) \\ 
$\log\,\epsilon$(N) & $ 8.34 _{- 0.15 }^{+ 0.17 }$ (8.27$\pm$0.34) & 
$ 8.49 _{- 0.22 }^{+ 0.19 }$$\dagger$ (8.43$\pm$0.34) & 
$ 8.28 _{- 0.33 }^{+ 0.01 }$ (7.48$\pm$0.13) & 
$ 8.33 _{- 0.17 }^{+ 0.19 }$ (8.09$\pm$0.34) & 
$ 7.65 _{- 0.02 }^{+ 0.65 }$ (8.21$\pm$0.34) & 
$ 8.38 _{- 0.35 }^{+ 0.15 }$ (7.78$\pm$0.34) \\ 
$\log\,\epsilon$(O) & $ 8.42 _{- 0.09 }^{+ 0.09 }$ ($\le$8.30) & 
$ 8.48 _{- 0.23 }^{+ 0.03 }$ (7.93$\pm$0.21) & 
$ 8.52 _{- 0.06 }^{+ 0.02 }$ (8.52$\pm$0.21) & 
$ 8.42 _{- 0.07 }^{+ 0.10 }$ (8.67$\pm$0.21) & 
$ 8.54 _{- 0.10 }^{+ 0.02 }$ (8.79$\pm$0.21) & 
$ 8.54 _{- 0.13 }^{+ 0.02 }$ (8.32$\pm$0.21) \\ 
\hline\hline  
Star & \object{HD 46485} & \object{HD 52533} & \object{HD 53755} & \object{HD 66811} & \object{HD 69106} & \object{HD 74920} \\ \hline  
$M_{\rm ini}$ (M$_{\odot}$) & $ 25.0 _{-  2.7 }^{+  3.8 }$ &  
 $ 20.6 _{-  1.5 }^{+  1.6 }$ &  
 $ 16.6 _{-  1.9 }^{+  2.0 }$ &  
 $ 50.0 _{-  8.6 }^{+ 26.5 }$ &  
 $ 18.8 _{-  3.3 }^{+  4.5 }$ &  
  $ 21.6 _{-  2.9 }^{+  3.7 }$ \\ 
$M_{\rm act}$ (M$_{\odot}$) & $ 24.8 _{-  2.5 }^{+  3.6 }$ &  
$ 20.6 _{-  1.5 }^{+  1.6 }$ &  
$ 15.6 _{-  0.9 }^{+  2.8 }$ &  
$ 41.4 _{-  4.2 }^{+ 22.8 }$ &  
$ 18.8 _{-  3.3 }^{+  3.7 }$ &  
$ 21.4 _{-  2.7 }^{+  3.4 }$ \\  
$\log\,L$ (L$_{\odot}$) & $  5.0 _{-  0.1 }^{+  0.2 }$ &  
$  4.7 _{-  0.1 }^{+  0.1 }$ &  
$  4.6 _{-  0.1 }^{+  0.2 }$ &  
$  5.8 _{-  0.1 }^{+  0.3 }$ &  
$  4.9 _{-  0.3 }^{+  0.2 }$ &  
$  4.8 _{-  0.2 }^{+  0.3 }$ \\  
Age (Myr) & $  2.3 _{-  1.6 }^{+  0.7 }$ &  
$  1.2 _{-  1.2 }^{+  1.0 }$ &  
$  7.4 _{-  0.9 }^{+  1.0 }$ &  
$  2.1 _{-  0.3 }^{+  0.7 }$ &  
$  6.2 _{-  0.9 }^{+  1.2 }$ &  
$  3.8 _{-  1.1 }^{+  1.2 }$ \\  
$\tau_{\rm MS}$ & $  0.3 _{-  0.3 }^{+  0.1 }$ &  
$  0.1 _{-  0.1 }^{+  0.1 }$ &  
$  0.7 _{-  0.1 }^{+  0.1 }$ &  
$  0.6 _{-  0.1 }^{+  0.1 }$ &  
$  0.8 _{-  0.1 }^{+  0.1 }$ &  
$  0.6 _{-  0.2 }^{+  0.1 }$ \\  
$v_{\rm ini}$ (km s$^{-1}$) & $ 330.0 _{- 19.8 }^{+ 56.5 }$ &  
$ 320.0 _{- 23.9 }^{+ 57.8 }$ &  
$ 300.0 _{- 17.6 }^{+ 91.2 }$$\dagger$ &  
$ 400.0 _{- 52.8 }^{+ 46.9 }$ &  
$ 320.0 _{- 18.4 }^{+ 65.8 }$ &  
$ 300.0 _{- 15.7 }^{+ 101.7 }$ \\  
$v\sin\,i$ (km s$^{-1}$) & $ 320.0 _{- 20.2 }^{+ 11.4 }$ (315$\pm$15) & 
$ 310.0 _{- 19.1 }^{+ 12.2 }$ (305$\pm$15) & 
$ 290.0 _{- 20.0 }^{+ 11.7 }$ (285$\pm$15) & 
$ 230.0 _{- 18.6 }^{+ 11.4 }$ (225$\pm$15) & 
$ 310.0 _{- 19.2 }^{+ 12.5 }$ (306$\pm$15) & 
$ 280.0 _{- 20.0 }^{+ 12.1 }$ (274$\pm$15) \\ 
$i$ ($^{\circ}$) &  75.9  & 
75.6  & 
69.3  & 
73.4  & 
75.6  & 
74.9  \\ 
$y$ & $  0.085 _{- 0.000 }^{+ 0.003  }$ (0.076$\pm$0.030) & 
$ 0.081 _{- 0.000 }^{+ 0.003 }$ (0.065$\pm$0.025) & 
$ 0.085 _{- 0.000 }^{+ 0.003 }$ (0.135$\pm$0.025) & 
$ 0.128 _{- 0.019 }^{+ 0.022 }$$\dagger$ (0.148$\pm$0.030) & 
$ 0.085 _{- 0.000 }^{+ 0.006 }$$\dagger$ (0.091$\pm$0.030) & 
$ 0.085 _{- 0.003 }^{+ 0.006 }$ (0.134$\pm$0.030) \\ 
$\log\,\epsilon$(C) & $ 8.12 _{- 0.28 }^{+ 0.02 }$ (8.46$\pm$0.27) & 
$ 8.13 _{- 0.14 }^{+ 0.01 }$ (7.76$\pm$0.12) & 
$ 7.95 _{- 0.19 }^{+ 0.05 }$ (...) & 
$ 7.12 _{- 0.03 }^{+ 0.17 }$$\dagger$ ($\le$7.00) & 
$ 7.88 _{- 0.17 }^{+ 0.09 }$$\dagger$ (7.88$\pm$0.27) & 
$ 8.13 _{- 0.37 }^{+ 0.01 }$ (7.78$\pm$0.27) \\ 
$\log\,\epsilon$(N) & $ 7.64 _{- 0.01 }^{+ 0.64 }$ (7.95$\pm$0.34) & 
$ 7.64 _{- 0.01 }^{+ 0.41 }$ (<7.78) & 
$ 8.06 _{- 0.08 }^{+ 0.33 }$$\dagger$ (7.32$\pm$0.13) & 
$ 8.75 _{- 0.11 }^{+ 0.09 }$ (8.94$\pm$0.34) & 
$ 8.24 _{- 0.16 }^{+ 0.16 }$ (7.74$\pm$0.34) & 
$ 8.18 _{- 0.16 }^{+ 0.36 }$ (8.34$\pm$0.34) \\ 
$\log\,\epsilon$(O) & $ 8.55 _{- 0.11 }^{+ 0.01 }$ (8.72$\pm$0.21) & 
$ 8.55 _{- 0.05 }^{+ 0.01 }$ (8.15$\pm$0.21) & 
$ 8.52 _{- 0.06 }^{+ 0.02 }$ (8.38$\pm$0.21) & 
$ 8.06 _{- 0.59 }^{+ 0.01 }$ (8.20$\pm$0.21) & 
$ 8.49 _{- 0.08 }^{+ 0.04 }$ (8.47$\pm$0.21) & 
$ 8.48 _{- 0.08 }^{+ 0.07 }$ (8.51$\pm$0.21) \\ 
\hline  
\end{tabular} 
\end{center}  
\end{tiny} 
\end{sidewaystable*}  
\begin{sidewaystable*}  
\addtocounter{table}{-1} 
\begin{tiny}  
\caption{Continued.}  
\begin{center}  
\begin{tabular}{ccccccccccccccccccccc}  
\hline\hline  
Star & \object{HD 92554} & \object{HD 117490} & \object{HD 124979} & \object{HD 149757} & \object{HD 150574} & \object{HD 163892} \\ \hline  
$M_{\rm ini}$ (M$_{\odot}$) & $ 20.6 _{-  3.7 }^{+  5.0 }$ &  
 $ 18.0 _{-  2.6 }^{+  3.6 }$ &  
 $ 21.4 _{-  3.0 }^{+  3.7 }$ &  
 $ 18.0 _{-  1.4 }^{+  1.6 }$ &  
 $ 39.4 _{- 17.3 }^{+  4.9 }$$\dagger$ &  
  $ 19.8 _{-  1.9 }^{+  2.6 }$ \\ 
$M_{\rm act}$ (M$_{\odot}$) & $ 20.8 _{-  3.8 }^{+  4.0 }$ &  
$ 18.0 _{-  2.5 }^{+  3.2 }$ &  
$ 21.2 _{-  2.8 }^{+  3.4 }$ &  
$ 17.8 _{-  1.2 }^{+  1.6 }$ &  
$ 34.4 _{- 12.8 }^{+  3.5 }$$\dagger$ &  
$ 20.2 _{-  2.4 }^{+  1.8 }$ \\  
$\log\,L$ (L$_{\odot}$) & $  5.0 _{-  0.2 }^{+  0.3 }$ &  
$  4.7 _{-  0.2 }^{+  0.3 }$ &  
$  4.8 _{-  0.2 }^{+  0.3 }$ &  
$  4.6 _{-  0.1 }^{+  0.1 }$ &  
$  5.6 _{-  0.3 }^{+  0.1 }$ &  
$  4.8 _{-  0.2 }^{+  0.1 }$ \\  
Age (Myr) & $  6.0 _{-  1.0 }^{+  1.0 }$ &  
$  6.2 _{-  0.9 }^{+  1.2 }$ &  
$  3.7 _{-  1.1 }^{+  1.1 }$ &  
$  4.7 _{-  1.6 }^{+  1.0 }$ &  
$  4.1 _{-  0.6 }^{+  2.1 }$$\dagger$ &  
$  5.0 _{-  0.7 }^{+  0.7 }$ \\  
$\tau_{\rm MS}$ & $  0.8 _{-  0.1 }^{+  0.1 }$ &  
$  0.7 _{-  0.1 }^{+  0.1 }$ &  
$  0.6 _{-  0.2 }^{+  0.1 }$ &  
$  0.5 _{-  0.2 }^{+  0.1 }$ &  
$  0.8 _{-  0.1 }^{+  0.1 }$ &  
$  0.6 _{-  0.1 }^{+  0.1 }$ \\  
$v_{\rm ini}$ (km s$^{-1}$) & $ 330.0 _{- 24.8 }^{+ 62.1 }$ &  
$ 360.0 _{- 17.1 }^{+ 56.7 }$ &  
$ 350.0 _{- 85.8 }^{+ 36.8 }$ &  
$ 380.0 _{- 22.1 }^{+ 35.7 }$ &  
$ 400.0 _{- 40.5 }^{+ 80.0 }$ &  
$ 350.0 _{- 83.2 }^{+ 42.1 }$ \\  
$v\sin\,i$ (km s$^{-1}$) & $ 300.0 _{- 12.6 }^{+ 19.0 }$ (303$\pm$15) & 
$ 360.0 _{- 15.4 }^{+ 16.3 }$ (361$\pm$15) & 
$ 250.0 _{- 18.1 }^{+ 14.0 }$ (246$\pm$15) & 
$ 380.0 _{- 18.1 }^{+ 13.6 }$ (378$\pm$15) & 
$ 230.0 _{- 15.5 }^{+ 16.7 }$ (233$\pm$15) & 
$ 210.0 _{- 19.3 }^{+ 13.5 }$ (205$\pm$15) \\ 
$i$ ($^{\circ}$) &  75.4  & 
76.6  & 
47.3  & 
77.0  & 
90.0 & 
36.9  \\ 
$y$ & $  0.085 _{- 0.003 }^{+ 0.006  }$ (0.091$\pm$0.030) & 
$  0.085 _{- 0.000 }^{+ 0.006  }$ (0.141$\pm$0.030) & 
$  0.081 _{- 0.000 }^{+ 0.006  }$ (0.091$\pm$0.030) & 
$  0.085 _{- 0.003 }^{+ 0.003  }$ (0.135$\pm$0.025) & 
$  0.089 _{- 0.003 }^{+ 0.029  }$ (0.172$\pm$0.030) & 
$  0.085 _{- 0.003 }^{+ 0.003  }$ (0.082$\pm$0.025) \\ 
$\log\,\epsilon$(C) & $ 7.84 _{- 0.14 }^{+ 0.12 }$ (7.57$\pm$0.27) & 
$ 7.83 _{- 0.28 }^{+ 0.09 }$$\dagger$ ($\le$7.39) & 
$ 8.13 _{- 0.30 }^{+ 0.01 }$ (8.48$\pm$0.27) & 
$ 7.87 _{- 0.19 }^{+ 0.14 }$ (8.07$\pm$0.12) & 
$ 7.11 _{- 0.05 }^{+ 0.58 }$$\dagger$ (7.48$\pm$0.27) & 
$ 7.90 _{- 0.10 }^{+ 0.17 }$ (8.24$\pm$0.12) \\ 
$\log\,\epsilon$(N) & $ 8.33 _{- 0.19 }^{+ 0.13 }$ (7.30$\pm$0.34) & 
$ 8.25 _{- 0.09 }^{+ 0.24 }$ (8.50$\pm$0.34) & 
$ 8.17 _{- 0.23 }^{+ 0.28 }$ (7.92$\pm$0.34) & 
$ 8.22 _{- 0.11 }^{+ 0.24 }$ (7.85$\pm$0.13) & 
$ 8.73 _{- 0.27 }^{+ 0.13 }$$\dagger$ ($\ge$9.08) & 
$ 8.19 _{- 0.25 }^{+ 0.18 }$ (7.34$\pm$0.13) \\ 
$\log\,\epsilon$(O) & $ 8.48 _{- 0.09 }^{+ 0.05 }$ (8.53$\pm$0.21) & 
$ 8.48 _{- 0.12 }^{+ 0.03 }$ (8.15$\pm$0.21) & 
$ 8.54 _{- 0.10 }^{+ 0.02 }$ (8.74$\pm$0.21) & 
$ 8.49 _{- 0.08 }^{+ 0.04 }$ (8.37$\pm$0.21) & 
$ 8.43 _{- 0.75 }^{+ 0.09 }$$\dagger$ ($\ge$8.93) & 
$ 8.49 _{- 0.04 }^{+ 0.06 }$ (8.38$\pm$0.21) \\ 
\hline\hline  
Star & \object{HD 172367} & \object{HD 175876} & \object{HD 188439} & \object{HD 191423} & \object{HD 192281} & \object{HD 203064}\\ \hline  
$M_{\rm ini}$ (M$_{\odot}$) & $ 16.0 _{-  1.8 }^{+  2.1 }$ &  
 $ 26.4 _{-  3.6 }^{+  8.6 }$ &  
 $ 14.8 _{-  1.4 }^{+  1.9 }$ &  
 $ 20.4 _{-  2.5 }^{+  1.5 }$ &  
 $ 39.0 _{-  8.3 }^{+  5.0 }$ &  
  $ 23.6 _{-  3.1 }^{+  5.2 }$ \\ 
$M_{\rm act}$ (M$_{\odot}$) & $ 15.6 _{-  1.3 }^{+  2.3 }$ &  
$ 25.6 _{-  2.6 }^{+  8.7 }$$\dagger$ &  
$ 15.6 _{-  2.1 }^{+  0.9 }$ &  
$ 20.0 _{-  2.3 }^{+  1.2 }$ &  
$ 36.0 _{-  6.1 }^{+  4.3 }$ &  
$ 23.4 _{-  3.0 }^{+  4.5 }$ \\  
$\log\,L$ (L$_{\odot}$) & $  4.6 _{-  0.2 }^{+  0.2 }$ &  
$  5.2 _{-  0.1 }^{+  0.3 }$ &  
$  4.5 _{-  0.2 }^{+  0.2 }$ &  
$  4.9 _{-  0.2 }^{+  0.1 }$ &  
$  5.5 _{-  0.3 }^{+  0.1 }$ &  
$  5.1 _{-  0.3 }^{+  0.2 }$ \\  
Age (Myr) & $  8.0 _{-  0.9 }^{+  1.0 }$ &  
$  3.5 _{-  0.6 }^{+  0.7 }$ &  
$  7.7 _{-  1.0 }^{+  1.1 }$ &  
$  7.0 _{-  0.9 }^{+  0.9 }$ &  
$  2.8 _{-  0.6 }^{+  0.5 }$ &  
$  3.7 _{-  0.8 }^{+  0.9 }$ \\  
$\tau_{\rm MS}$ & $  0.7 _{-  0.1 }^{+  0.1 }$ &  
$  0.7 _{-  0.1 }^{+  0.1 }$ &  
$  0.7 _{-  0.1 }^{+  0.1 }$ &  
$  0.8 _{-  0.1 }^{+  0.1 }$ &  
$  0.6 _{-  0.1 }^{+  0.1 }$ &  
$  0.6 _{-  0.2 }^{+  0.1 }$ \\  
$v_{\rm ini}$ (km s$^{-1}$) & $ 360.0 _{- 95.9 }^{+ 29.1 }$ &  
$ 350.0 _{- 52.3 }^{+ 52.2 }$ &  
$ 290.0 _{- 12.3 }^{+ 94.7 }$$\dagger$ &  
$ 420.0 _{- 22.4 }^{+ 40.6 }$ &  
$ 360.0 _{- 45.4 }^{+ 43.8 }$ &  
$ 330.0 _{- 27.9 }^{+ 53.9 }$ \\  
$v\sin\,i$ (km s$^{-1}$) & $ 270.0 _{- 17.6 }^{+ 13.8 }$ (266$\pm$15) & 
$ 270.0 _{- 20.3 }^{+ 11.5 }$ (265$\pm$15) & 
$ 280.0 _{- 14.0 }^{+ 17.3 }$ (281$\pm$15) & 
$ 420.0 _{- 19.4 }^{+ 12.0 }$ (420$\pm$15) & 
$ 270.0 _{- 11.6 }^{+ 19.6 }$ (276$\pm$15) & 
$ 300.0 _{- 16.9 }^{+ 14.7 }$ (298$\pm$15) \\ 
$i$ ($^{\circ}$) &  45.3  & 
74.6  & 
64.6  & 
90.0  & 
74.6  & 
75.4  \\ 
$y$ & $  0.085 _{- 0.000 }^{+ 0.003  }$ (0.140$\pm$0.025) & 
$  0.089 _{- 0.003 }^{+ 0.007  }$ (0.110$\pm$0.030) & 
$  0.085 _{- 0.000 }^{+ 0.003  }$ (0.122$\pm$0.025) & 
$  0.089 _{- 0.003 }^{+ 0.013  }$ (0.134$\pm$0.030) & 
$  0.089 _{- 0.003 }^{+ 0.013  }$$\dagger$ (0.103$\pm$0.030) & 
$  0.089 _{- 0.003 }^{+ 0.003  }$ (0.076$\pm$0.030) \\ 
$\log\,\epsilon$(C) & $ 7.96 _{- 0.21 }^{+ 0.05 }$ (<8.09) & 
$ 7.81 _{- 0.25 }^{+ 0.19 }$$\dagger$ (8.04$\pm$0.27) & 
$ 7.96 _{- 0.16 }^{+ 0.06 }$ (<8.09) &  
$ 7.61 _{- 0.30 }^{+ 0.12 }$$\dagger$ ($\le$7.24) &  
$ 7.82 _{- 0.40 }^{+ 0.13 }$ (8.00$\pm$0.27) &  
$ 7.91 _{- 0.19 }^{+ 0.10 }$ (7.92$\pm$0.27) \\ 
$\log\,\epsilon$(N) & $ 8.03 _{- 0.09 }^{+ 0.47 }$$\dagger$ (8.44$\pm$0.13) & 
$ 8.34 _{- 0.20 }^{+ 0.22 }$$\dagger$ (8.36$\pm$0.34) & 
$ 8.03 _{- 0.08 }^{+ 0.27 }$$\dagger$ (8.16$\pm$0.13) & 
$ 8.52 _{- 0.14 }^{+ 0.12 }$ (8.33$\pm$0.34) & 
$ 8.43 _{- 0.19 }^{+ 0.20 }$$\dagger$ (8.76$\pm$0.34) & 
$ 8.34 _{- 0.21 }^{+ 0.13 }$ (8.23$\pm$0.34) \\ 
$\log\,\epsilon$(O) & $ 8.53 _{- 0.07 }^{+ 0.01 }$ (8.47$\pm$0.21) & 
$ 8.43 _{- 0.10 }^{+ 0.09 }$ (8.42$\pm$0.21) & 
$ 8.53 _{- 0.05 }^{+ 0.01 }$ (8.66$\pm$0.21) &  
$ 8.38 _{- 0.17 }^{+ 0.07 }$ ($\le$8.33) &  
$ 8.43 _{- 0.25 }^{+ 0.08 }$$\dagger$ (8.05$\pm$0.21) &  
$ 8.48 _{- 0.09 }^{+ 0.04 }$ (8.46$\pm$0.21) \\ 
\hline\hline  
Star & \object{HD 210839} & \object{HD 228841}\\ \hline  
$M_{\rm ini}$ (M$_{\odot}$) & $ 40.0 _{- 12.5 }^{+ 12.3 }$ &  
  $ 25.0 _{-  2.7 }^{+  6.6 }$ \\ 
$M_{\rm act}$ (M$_{\odot}$) & $ 35.2 _{-  7.9 }^{+ 11.1 }$ &  
$ 24.2 _{-  2.1 }^{+  5.8 }$ \\  
$\log\,L$ (L$_{\odot}$) & $  5.6 _{-  0.2 }^{+  0.3 }$ &  
$  5.2 _{-  0.1 }^{+  0.2 }$ \\  
Age (Myr) & $  3.4 _{-  0.9 }^{+  0.4 }$ &  
$  4.1 _{-  0.5 }^{+  0.9 }$ \\  
$\tau_{\rm MS}$ & $  0.7 _{-  0.1 }^{+  0.1 }$ &  
$  0.7 _{-  0.1 }^{+  0.1 }$ \\  
$v_{\rm ini}$ (km s$^{-1}$) & $ 370.0 _{- 60.2 }^{+ 45.2 }$ &  
$ 360.0 _{- 37.1 }^{+ 51.8 }$ \\  
$v\sin\,i$ (km s$^{-1}$) & $ 220.0 _{- 20.1 }^{+ 10.7 }$ (214$\pm$15) & 
$ 310.0 _{- 20.7 }^{+ 10.9 }$ (305$\pm$15) \\ 
$i$ ($^{\circ}$) &  73.0  & 
75.6  \\ 
$y$ & $  0.089 _{- 0.003 }^{+ 0.023  }$$\dagger$ (0.113$\pm$0.030) & 
$  0.089 _{- 0.003 }^{+ 0.007  }$ (0.112$\pm$0.030) \\ 
$\log\,\epsilon$(C) & $ 7.33 _{- 0.01 }^{+ 0.63 }$$\dagger$ (7.83$\pm$0.27) & 
$ 7.78 _{- 0.19 }^{+ 0.15 }$ (7.48$\pm$0.27) \\ 
$\log\,\epsilon$(N) & $ 8.65 _{- 0.41 }^{+ 0.12 }$$\dagger$ (8.74$\pm$0.34) & 
$ 8.34 _{- 0.11 }^{+ 0.18 }$ (8.74$\pm$0.34) \\ 
$\log\,\epsilon$(O) & $ 8.43 _{- 0.33 }^{+ 0.10 }$$\dagger$ (8.13$\pm$0.21) & 
$ 8.42 _{- 0.10 }^{+ 0.07 }$ (8.67$\pm$0.21) \\ 
\hline  
\end{tabular} 
\end{center}  
\end{tiny} 
\end{sidewaystable*}  

\begin{sidewaystable*} 
\begin{tiny}  
\caption{Same as Fig. \ref{resBONN} but when $T_{\rm{eff}}$, $\log g_{\rm{C}}$, and $v\sin\,i$ are given as input to BONNSAI. }  
\label{resBONN3}  
\begin{center}  
\begin{tabular}{ccccccccccccccccccccc}  
\hline\hline  
Star & \object{ALS 864} & \object{ALS 18675} & \object{BD +34$^{\circ}$ 1058} & \object{BD +60$^{\circ}$ 594} & \object{HD 13268} & \object{HD 14434} \\ \hline  
$M_{\rm ini}$ (M$_{\odot}$) & $ 17.4 _{-  1.4 }^{+  1.5 }$ &  
 $ 16.4 _{-  1.4 }^{+  1.7 }$ &  
 $ 21.4 _{-  2.0 }^{+  1.8 }$ &  
 $ 19.8 _{-  1.6 }^{+  1.7 }$ &  
 $ 25.0 _{-  3.9 }^{+  5.6 }$ &  
  $ 30.2 _{-  3.6 }^{+  3.9 }$ \\ 
$M_{\rm act}$ (M$_{\odot}$) & $ 17.4 _{-  1.4 }^{+  1.5 }$ &  
$ 16.6 _{-  1.5 }^{+  1.4 }$ &  
$ 21.0 _{-  1.8 }^{+  1.7 }$ &  
$ 19.6 _{-  1.4 }^{+  1.8 }$ &  
$ 24.0 _{-  3.0 }^{+  5.0 }$ &  
$ 29.8 _{-  3.7 }^{+  3.8 }$ \\  
$\log\,L$ (L$_{\odot}$) & $  4.5 _{-  0.1 }^{+  0.1 }$ &  
$  4.5 _{-  0.1 }^{+  0.2 }$ &  
$  4.8 _{-  0.1 }^{+  0.1 }$ &  
$  4.7 _{-  0.1 }^{+  0.1 }$ &  
$  5.2 _{-  0.2 }^{+  0.2 }$ &  
$  5.2 _{-  0.1 }^{+  0.1 }$ \\  
Age (Myr) & $  3.8 _{-  1.8 }^{+  1.1 }$ &  
$  5.3 _{-  1.4 }^{+  1.0 }$ &  
$  4.9 _{-  2.1 }^{+  1.3 }$ &  
$  3.4 _{-  1.6 }^{+  1.0 }$ &  
$  4.8 _{-  0.7 }^{+  0.8 }$ &  
$  0.8 _{-  0.8 }^{+  1.4 }$ \\  
$\tau_{\rm MS}$ & $  0.4 _{-  0.2 }^{+  0.1 }$ &  
$  0.5 _{-  0.1 }^{+  0.1 }$ &  
$  0.6 _{-  0.3 }^{+  0.1 }$ &  
$  0.4 _{-  0.2 }^{+  0.1 }$ &  
$  0.8 _{-  0.1 }^{+  0.1 }$ &  
$  0.2 _{-  0.1 }^{+  0.2 }$ \\  
$v_{\rm ini}$ (km s$^{-1}$) & $ 350.0 _{- 94.1 }^{+ 72.4 }$$\dagger$ &  
$ 330.0 _{- 47.1 }^{+ 75.8 }$ &  
$ 430.0 _{- 22.3 }^{+ 30.8 }$ &  
$ 330.0 _{- 26.2 }^{+ 53.3 }$ &  
$ 360.0 _{- 42.9 }^{+ 46.4 }$ &  
$ 420.0 _{- 17.8 }^{+ 39.0 }$ \\  
$v\sin\,i$ (km s$^{-1}$) & $ 250.0 _{- 15.7 }^{+ 16.3 }$ (249$\pm$15) & 
$ 240.0 _{- 17.6 }^{+ 14.4 }$ (236$\pm$15) & 
$ 420.0 _{- 17.3 }^{+ 14.0 }$ (424$\pm$15) & 
$ 320.0 _{- 20.6 }^{+ 10.9 }$ (314$\pm$15) & 
$ 300.0 _{- 14.9 }^{+ 16.5 }$ (301$\pm$15) & 
$ 400.0 _{- 12.7 }^{+ 17.7 }$ (408$\pm$15) \\ 
$i$ ($^{\circ}$) &  44.0  & 
36.9  & 
90.0  & 
75.9  & 
75.4  & 
77.3  \\ 
$y$ & $ 0.085 _{- 0.000 }^{+ 0.003 }$ (0.064$\pm$0.025) & 
$ 0.085 _{- 0.000 }^{+ 0.003 }$ (0.071$\pm$0.025) & 
$ 0.097 _{- 0.013 }^{+ 0.007 }$ (0.119$\pm$0.030) & 
$ 0.085 _{- 0.000 }^{+ 0.003 }$ (0.131$\pm$0.025) & 
$ 0.089 _{- 0.003 }^{+ 0.007 }$ (0.206$\pm$0.030) & 
$ 0.085 _{- 0.003 }^{+ 0.013 }$ (0.103$\pm$0.030) \\ 
$\log\,\epsilon$(C) & $ 8.12 _{- 0.17 }^{+ 0.01 }$ (<7.86) & 
$ 8.04 _{- 0.15 }^{+ 0.08 }$ (7.78$\pm$0.12) & 
$ 7.52 _{- 0.30 }^{+ 0.27 }$ (7.90$\pm$0.27) & 
$ 8.12 _{- 0.22 }^{+ 0.01 }$ (7.66$\pm$0.12) & 
$ 7.78 _{- 0.19 }^{+ 0.15 }$$\dagger$ ($\le$7.50) & 
$ 8.11 _{- 0.62 }^{+ 0.02 }$ (7.96$\pm$0.27) \\ 
$\log\,\epsilon$(N) & $ 7.65 _{- 0.02 }^{+ 0.44 }$ (7.64$\pm$0.13) & 
$ 7.97 _{- 0.12 }^{+ 0.31 }$ (7.54$\pm$0.13) & 
$ 8.56 _{- 0.19 }^{+ 0.14 }$ (8.14$\pm$0.34) & 
$ 8.12 _{- 0.27 }^{+ 0.21 }$ (<8.24) & 
$ 8.33 _{- 0.11 }^{+ 0.19 }$ (8.61$\pm$0.34) & 
$ 8.54 _{- 0.31 }^{+ 0.21 }$ (8.61$\pm$0.34) \\ 
$\log\,\epsilon$(O) & $ 8.54 _{- 0.04 }^{+ 0.02 }$ (8.12$\pm$0.21) & 
$ 8.54 _{- 0.05 }^{+ 0.02 }$ (8.07$\pm$0.21) & 
$ 8.32 _{- 0.10 }^{+ 0.24 }$ (8.68$\pm$0.21) & 
$ 8.55 _{- 0.07 }^{+ 0.01 }$ (8.48$\pm$0.21) & 
$ 8.43 _{- 0.09 }^{+ 0.07 }$ (8.10$\pm$0.21) & 
$ 8.54 _{- 0.28 }^{+ 0.01 }$ ($\le$8.10) \\ 
\hline\hline  
Star & \object{HD 14442} & \object{HD 15137} & \object{HD 15642} & \object{HD 28446A} & \object{HD 41161} & \object{HD 41997}    \\ \hline  
$M_{\rm ini}$ (M$_{\odot}$) & $ 38.2 _{-  8.6 }^{+  3.7 }$ &  
 $ 26.0 _{-  4.4 }^{+  6.7 }$ &  
 $ 21.6 _{-  3.7 }^{+  3.8 }$ &  
 $ 17.2 _{-  1.9 }^{+  2.0 }$ &  
 $ 22.6 _{-  3.2 }^{+  5.2 }$ &  
  $ 21.4 _{-  2.8 }^{+  3.3 }$ \\ 
$M_{\rm act}$ (M$_{\odot}$) & $ 35.6 _{-  6.7 }^{+  3.2 }$ &  
$ 24.4 _{-  3.0 }^{+  6.1 }$ &  
$ 20.6 _{-  2.7 }^{+  3.8 }$ &  
$ 16.4 _{-  1.1 }^{+  2.6 }$ &  
$ 22.4 _{-  2.9 }^{+  4.5 }$ &  
$ 21.4 _{-  2.8 }^{+  2.9 }$ \\  
$\log\,L$ (L$_{\odot}$) & $  5.5 _{-  0.3 }^{+  0.1 }$ &  
$  5.3 _{-  0.2 }^{+  0.2 }$ &  
$  5.0 _{-  0.2 }^{+  0.2 }$ &  
$  4.7 _{-  0.2 }^{+  0.2 }$ &  
$  5.1 _{-  0.3 }^{+  0.1 }$ &  
$  4.8 _{-  0.2 }^{+  0.2 }$ \\  
Age (Myr) & $  2.8 _{-  0.7 }^{+  0.6 }$ &  
$  4.6 _{-  0.3 }^{+  1.6 }$ &  
$  6.3 _{-  0.8 }^{+  1.0 }$ &  
$  6.4 _{-  0.7 }^{+  0.9 }$ &  
$  4.1 _{-  0.7 }^{+  1.0 }$ &  
$  3.5 _{-  1.6 }^{+  1.1 }$ \\  
$\tau_{\rm MS}$ & $  0.6 _{-  0.2 }^{+  0.1 }$ &  
$  0.8 _{-  0.0 }^{+  0.0 }$ &  
$  0.8 _{-  0.1 }^{+  0.1 }$ &  
$  0.7 _{-  0.1 }^{+  0.1 }$ &  
$  0.7 _{-  0.1 }^{+  0.1 }$ &  
$  0.5 _{-  0.2 }^{+  0.1 }$ \\  
$v_{\rm ini}$ (km s$^{-1}$) & $ 360.0 _{- 50.5 }^{+ 43.6 }$ &  
$ 350.0 _{- 44.8 }^{+ 57.3 }$ &  
$ 360.0 _{- 29.7 }^{+ 50.6 }$ &  
$ 290.0 _{- 14.7 }^{+ 88.0 }$ &  
$ 330.0 _{- 24.6 }^{+ 60.8 }$ &  
$ 350.0 _{- 89.2 }^{+ 36.4 }$ \\  
$v\sin\,i$ (km s$^{-1}$) & $ 280.0 _{- 12.8 }^{+ 18.7 }$ ($\le$8.10) & 
$ 270.0 _{- 18.3 }^{+ 13.5 }$ (267$\pm$15) & 
$ 330.0 _{- 12.6 }^{+ 19.1 }$ (335$\pm$15) & 
$ 280.0 _{- 19.7 }^{+ 11.9 }$ (275$\pm$15) & 
$ 300.0 _{- 12.1 }^{+ 19.4 }$ (303$\pm$15) & 
$ 250.0 _{- 17.1 }^{+ 14.9 }$ (247$\pm$15) \\ 
$i$ ($^{\circ}$) &  74.9  & 
74.6  & 
76.1  & 
74.9  & 
69.6  & 
47.3  \\ 
$y$ & $ 0.089 _{- 0.003 }^{+ 0.010 }$ (0.097$\pm$0.030) & 
$ 0.089 _{- 0.003 }^{+ 0.003 }$ (0.112$\pm$0.030) & 
$ 0.085 _{- 0.003 }^{+ 0.010 }$ (0.150$\pm$0.030) & 
$ 0.085 _{- 0.000 }^{+ 0.003 }$ (0.126$\pm$0.025) & 
$ 0.089 _{- 0.003 }^{+ 0.003 }$ (0.123$\pm$0.030) & 
$ 0.085 _{- 0.000 }^{+ 0.006 }$ (0.110$\pm$0.030) \\ 
$\log\,\epsilon$(C) & $ 7.82 _{- 0.39 }^{+ 0.15 }$ (7.10$\pm$0.27) & 
$ 7.80 _{- 0.20 }^{+ 0.15 }$$\dagger$ (7.63$\pm$0.27) & 
$ 7.83 _{- 0.24 }^{+ 0.08 }$ ($\le$7.55) & 
$ 7.97 _{- 0.16 }^{+ 0.05 }$ (8.30$\pm$0.12) & 
$ 7.90 _{- 0.23 }^{+ 0.09 }$$\dagger$ (7.87$\pm$0.27) & 
$ 8.12 _{- 0.28 }^{+ 0.01 }$ (8.59$\pm$0.27) \\ 
$\log\,\epsilon$(N) & $ 8.34 _{- 0.09 }^{+ 0.32 }$ (8.61$\pm$0.34) & 
$ 8.35 _{- 0.17 }^{+ 0.16 }$$\dagger$ (8.27$\pm$0.34) & 
$ 8.22 _{- 0.05 }^{+ 0.31 }$$\dagger$ (8.43$\pm$0.34) & 
$ 8.03 _{- 0.07 }^{+ 0.24 }$ (7.48$\pm$0.13) & 
$ 8.34 _{- 0.20 }^{+ 0.14 }$ (8.09$\pm$0.34) & 
$ 8.16 _{- 0.30 }^{+ 0.27 }$ (8.21$\pm$0.34) \\ 
$\log\,\epsilon$(O) & $ 8.43 _{- 0.25 }^{+ 0.09 }$$\dagger$ ($\le$8.10) & 
$ 8.42 _{- 0.07 }^{+ 0.10 }$ ($\le$8.30) & 
$ 8.48 _{- 0.17 }^{+ 0.03 }$$\dagger$ (7.93$\pm$0.21) & 
$ 8.52 _{- 0.05 }^{+ 0.02 }$ (8.52$\pm$0.21) & 
$ 8.48 _{- 0.10 }^{+ 0.05 }$ (8.67$\pm$0.21) & 
$ 8.54 _{- 0.09 }^{+ 0.02 }$ (8.79$\pm$0.21) \\ 
\hline\hline  
Star & \object{HD 46056} & \object{HD 46485} & \object{HD 52266} & \object{HD 52533} & \object{HD 53755} & \object{HD 66811}  \\ \hline  
$M_{\rm ini}$ (M$_{\odot}$) & $ 21.6 _{-  2.5 }^{+  2.8 }$ &  
 $ 25.0 _{-  2.6 }^{+  3.8 }$ &  
 $ 19.2 _{-  2.1 }^{+  2.9 }$ &  
 $ 20.6 _{-  1.5 }^{+  1.7 }$ &  
 $ 16.4 _{-  1.8 }^{+  2.1 }$ &  
  $ 50.0 _{- 10.6 }^{+ 21.5 }$ \\ 
$M_{\rm act}$ (M$_{\odot}$) & $ 21.4 _{-  2.3 }^{+  2.7 }$ &  
$ 24.8 _{-  2.5 }^{+  3.6 }$ &  
$ 19.4 _{-  2.4 }^{+  2.2 }$ &  
$ 20.6 _{-  1.5 }^{+  1.6 }$ &  
$ 15.6 _{-  1.0 }^{+  2.7 }$ &  
$ 42.4 _{-  6.1 }^{+ 18.6 }$ \\  
$\log\,L$ (L$_{\odot}$) & $  4.8 _{-  0.2 }^{+  0.2 }$ &  
$  5.0 _{-  0.1 }^{+  0.2 }$ &  
$  4.9 _{-  0.2 }^{+  0.2 }$ &  
$  4.7 _{-  0.1 }^{+  0.1 }$ &  
$  4.6 _{-  0.1 }^{+  0.2 }$ &  
$  5.8 _{-  0.1 }^{+  0.3 }$ \\  
Age (Myr) & $  3.5 _{-  2.0 }^{+  1.0 }$ &  
$  2.4 _{-  1.6 }^{+  0.8 }$ &  
$  6.0 _{-  0.6 }^{+  0.8 }$ &  
$  1.2 _{-  1.2 }^{+  1.0 }$ &  
$  7.4 _{-  0.8 }^{+  1.0 }$ &  
$  2.1 _{-  0.4 }^{+  0.6 }$ \\  
$\tau_{\rm MS}$ & $  0.4 _{-  0.3 }^{+  0.2 }$ &  
$  0.4 _{-  0.2 }^{+  0.1 }$ &  
$  0.7 _{-  0.1 }^{+  0.1 }$ &  
$  0.1 _{-  0.1 }^{+  0.1 }$ &  
$  0.7 _{-  0.1 }^{+  0.1 }$ &  
$  0.6 _{-  0.1 }^{+  0.1 }$ \\  
$v_{\rm ini}$ (km s$^{-1}$) & $ 360.0 _{- 22.2 }^{+ 41.8 }$ &  
$ 330.0 _{- 19.7 }^{+ 60.6 }$ &  
$ 310.0 _{- 26.4 }^{+ 67.0 }$ &  
$ 320.0 _{- 23.9 }^{+ 60.5 }$ &  
$ 300.0 _{- 20.7 }^{+ 82.3 }$$\dagger$ &  
$ 370.0 _{- 57.3 }^{+ 58.4 }$ \\  
$v\sin\,i$ (km s$^{-1}$) & $ 350.0 _{- 16.5 }^{+ 15.2 }$ (350$\pm$15) & 
$ 320.0 _{- 20.1 }^{+ 11.6 }$ (315$\pm$15) & 
$ 290.0 _{- 19.6 }^{+ 12.4 }$ (285$\pm$15) & 
$ 310.0 _{- 19.1 }^{+ 12.3 }$ (305$\pm$15) & 
$ 290.0 _{- 20.0 }^{+ 11.7 }$ (285$\pm$15) & 
$ 230.0 _{- 19.2 }^{+ 11.2 }$ (225$\pm$15) \\ 
$i$ ($^{\circ}$) &  76.5  & 
75.9  & 
75.2  & 
75.6  & 
69.3  & 
73.4  \\ 
$y$ & $  0.085 _{- 0.003 }^{+ 0.006  }$ (0.088$\pm$0.030) & 
$ 0.085 _{- 0.003 }^{+ 0.003 }$ (0.076$\pm$0.030) & 
$ 0.085 _{- 0.003 }^{+ 0.003 }$ (0.187$\pm$0.025) & 
$ 0.081 _{- 0.000 }^{+ 0.003 }$ (0.065$\pm$0.025) & 
$ 0.085 _{- 0.000 }^{+ 0.003 }$ (0.135$\pm$0.025) & 
$  0.097 _{- 0.010 }^{+ 0.027  }$$\dagger$ (0.148$\pm$0.030) \\ 
$\log\,\epsilon$(C) & $ 8.13 _{- 0.36 }^{+ 0.01 }$ (8.34$\pm$0.27) & 
$ 8.13 _{- 0.31 }^{+ 0.01 }$ (8.46$\pm$0.27) & 
$ 7.93 _{- 0.15 }^{+ 0.06 }$$\dagger$ (7.78$\pm$0.12) & 
$ 8.13 _{- 0.15 }^{+ 0.01 }$ (7.76$\pm$0.12) & 
$ 7.95 _{- 0.15 }^{+ 0.05 }$ (...) & 
$ 7.11 _{- 0.03 }^{+ 0.35 }$ ($\le$7.00) \\ 
$\log\,\epsilon$(N) & $ 8.34 _{- 0.31 }^{+ 0.20 }$ (7.78$\pm$0.34) & 
$ 8.33 _{- 0.39 }^{+ 0.18 }$ (7.95$\pm$0.34) & 
$ 8.22 _{- 0.20 }^{+ 0.12 }$$\dagger$ (7.74$\pm$0.13) & 
$ 7.64 _{- 0.01 }^{+ 0.42 }$ (<7.78) & 
$ 8.06 _{- 0.07 }^{+ 0.21 }$ (7.32$\pm$0.13) & 
$ 8.74 _{- 0.21 }^{+ 0.11 }$ (8.94$\pm$0.34) \\ 
$\log\,\epsilon$(O) & $ 8.54 _{- 0.13 }^{+ 0.02 }$ (8.32$\pm$0.21) & 
$ 8.55 _{- 0.12 }^{+ 0.01 }$ (8.72$\pm$0.21) & 
$ 8.52 _{- 0.08 }^{+ 0.02 }$ (8.00$\pm$0.21) & 
$ 8.55 _{- 0.05 }^{+ 0.01 }$ (8.15$\pm$0.21) & 
$ 8.52 _{- 0.05 }^{+ 0.02 }$ (8.38$\pm$0.21) & 
$ 8.06 _{- 0.44 }^{+ 0.43 }$$\dagger$ (8.20$\pm$0.21) \\ 
\hline  
\end{tabular} 
\end{center}  
\end{tiny} 
\end{sidewaystable*}  
\begin{sidewaystable*}  
\addtocounter{table}{-1} 
\begin{tiny}  
\caption{Continued.}  
\begin{center}  
\begin{tabular}{ccccccccccccccccccccc}  
\hline\hline  
Star & \object{HD 69106} & \object{HD 74920} & \object{HD 84567} & \object{HD 90087} & \object{HD 92554} & \object{HD 93521} \\ \hline  
$M_{\rm ini}$ (M$_{\odot}$) & $ 18.8 _{-  3.3 }^{+  4.5 }$ &  
 $ 21.2 _{-  2.8 }^{+  3.7 }$ &  
 $ 17.4 _{-  2.1 }^{+  2.5 }$ &  
 $ 19.2 _{-  2.3 }^{+  3.2 }$ &  
 $ 20.6 _{-  3.7 }^{+  5.1 }$ &  
  $ 18.0 _{-  1.7 }^{+  1.9 }$ \\ 
$M_{\rm act}$ (M$_{\odot}$) & $ 18.8 _{-  3.3 }^{+  3.8 }$ &  
$ 21.2 _{-  2.7 }^{+  3.3 }$ &  
$ 17.4 _{-  2.3 }^{+  2.0 }$ &  
$ 19.0 _{-  2.1 }^{+  2.9 }$ &  
$ 20.8 _{-  3.8 }^{+  4.0 }$ &  
$ 17.8 _{-  1.6 }^{+  1.7 }$ \\  
$\log\,L$ (L$_{\odot}$) & $  4.9 _{-  0.3 }^{+  0.3 }$$\dagger$ &  
$  4.8 _{-  0.2 }^{+  0.2 }$ &  
$  4.8 _{-  0.2 }^{+  0.2 }$ &  
$  4.9 _{-  0.2 }^{+  0.2 }$ &  
$  5.0 _{-  0.2 }^{+  0.3 }$ &  
$  4.7 _{-  0.1 }^{+  0.2 }$ \\  
Age (Myr) & $  6.2 _{-  0.9 }^{+  1.3 }$ &  
$  3.8 _{-  1.3 }^{+  1.1 }$ &  
$  7.7 _{-  1.1 }^{+  0.9 }$ &  
$  6.5 _{-  0.7 }^{+  0.9 }$ &  
$  6.0 _{-  0.9 }^{+  1.0 }$ &  
$  6.5 _{-  0.7 }^{+  0.8 }$ \\  
$\tau_{\rm MS}$ & $  0.8 _{-  0.1 }^{+  0.1 }$ &  
$  0.6 _{-  0.2 }^{+  0.1 }$ &  
$  0.8 _{-  0.1 }^{+  0.1 }$ &  
$  0.8 _{-  0.1 }^{+  0.0 }$ &  
$  0.8 _{-  0.1 }^{+  0.1 }$ &  
$  0.7 _{-  0.1 }^{+  0.1 }$ \\  
$v_{\rm ini}$ (km s$^{-1}$) & $ 320.0 _{- 19.0 }^{+ 66.7 }$ &  
$ 300.0 _{- 20.9 }^{+ 83.7 }$ &  
$ 280.0 _{- 14.9 }^{+ 95.6 }$ &  
$ 300.0 _{- 19.5 }^{+ 78.8 }$ &  
$ 330.0 _{- 25.5 }^{+ 63.5 }$ &  
$ 400.0 _{- 22.2 }^{+ 26.3 }$ \\  
$v\sin\,i$ (km s$^{-1}$) & $ 310.0 _{- 19.2 }^{+ 12.5 }$ (306$\pm$15) & 
$ 280.0 _{- 19.9 }^{+ 12.2 }$ (274$\pm$15) & 
$ 260.0 _{- 13.3 }^{+ 18.5 }$ (261$\pm$15) & 
$ 280.0 _{- 18.5 }^{+ 13.4 }$ (276$\pm$15) & 
$ 300.0 _{- 12.6 }^{+ 19.0 }$ (303$\pm$15) & 
$ 400.0 _{- 13.9 }^{+ 17.4 }$ (405$\pm$15) \\ 
$i$ ($^{\circ}$) &  75.6  & 
74.9  & 
68.2  & 
74.9  & 
75.4  & 
77.3  \\ 
$y$ & $  0.085 _{- 0.000 }^{+ 0.006  }$ (0.091$\pm$0.030) & 
$ 0.085 _{- 0.000 }^{+ 0.006 }$ (0.134$\pm$0.030) & 
$ 0.085 _{- 0.000 }^{+ 0.003 }$ (0.204$\pm$0.025) & 
$ 0.085 _{- 0.003 }^{+ 0.003 }$ (0.163$\pm$0.025) & 
$ 0.085 _{- 0.003 }^{+ 0.006 }$ (0.091$\pm$0.030) & 
$  0.089 _{- 0.003 }^{+ 0.003  }$ (0.166$\pm$0.025) \\ 
$\log\,\epsilon$(C) & $ 7.88 _{- 0.17 }^{+ 0.09 }$$\dagger$ (7.88$\pm$0.27) & 
$ 7.93 _{- 0.10 }^{+ 0.20 }$ (7.78$\pm$0.27) & 
$ 7.94 _{- 0.15 }^{+ 0.06 }$ (<7.84) & 
$ 7.92 _{- 0.16 }^{+ 0.06 }$$\dagger$ (7.72$\pm$0.12) & 
$ 7.87 _{- 0.22 }^{+ 0.09 }$$\dagger$ (7.57$\pm$0.27) & 
$ 7.74 _{- 0.16 }^{+ 0.09 }$ (7.68$\pm$0.12) \\ 
$\log\,\epsilon$(N) & $ 8.23 _{- 0.16 }^{+ 0.17 }$ (7.74$\pm$0.34) & 
$ 8.18 _{- 0.24 }^{+ 0.26 }$ (8.34$\pm$0.34) & 
$ 8.26 _{- 0.29 }^{+ 0.02 }$ (7.98$\pm$0.13) & 
$ 8.26 _{- 0.24 }^{+ 0.08 }$$\dagger$ (7.42$\pm$0.13) & 
$ 8.33 _{- 0.19 }^{+ 0.14 }$ (7.30$\pm$0.34) & 
$ 8.33 _{- 0.07 }^{+ 0.14 }$ (8.10$\pm$0.13) \\ 
$\log\,\epsilon$(O) & $ 8.49 _{- 0.08 }^{+ 0.04 }$ (8.47$\pm$0.21) & 
$ 8.50 _{- 0.06 }^{+ 0.06 }$ (8.51$\pm$0.21) & 
$ 8.53 _{- 0.06 }^{+ 0.01 }$ (8.39$\pm$0.21) & 
$ 8.51 _{- 0.07 }^{+ 0.03 }$ (8.16$\pm$0.21) & 
$ 8.48 _{- 0.09 }^{+ 0.05 }$ (8.53$\pm$0.21) & 
$ 8.43 _{- 0.07 }^{+ 0.05 }$ (8.33$\pm$0.21) \\ 
\hline\hline  
Star & \object{HD 102415} & \object{HD 117490} & \object{HD 124979} & \object{HD 149757} & \object{HD 150574} & \object{HD 163892} \\ \hline  
$M_{\rm ini}$ (M$_{\odot}$) & $ 19.0 _{-  1.4 }^{+  1.5 }$ &  
 $ 17.8 _{-  2.6 }^{+  3.1 }$ &  
 $ 21.4 _{-  3.0 }^{+  3.7 }$ &  
 $ 18.0 _{-  1.5 }^{+  1.5 }$ &  
 $ 28.4 _{-  5.5 }^{+  9.5 }$ &  
  $ 19.8 _{-  1.9 }^{+  2.6 }$ \\ 
$M_{\rm act}$ (M$_{\odot}$) & $ 19.0 _{-  1.3 }^{+  1.5 }$ &  
$ 17.2 _{-  1.9 }^{+  3.4 }$ &  
$ 21.2 _{-  2.7 }^{+  3.4 }$ &  
$ 17.8 _{-  1.3 }^{+  1.5 }$ &  
$ 28.4 _{-  5.2 }^{+  6.5 }$ &  
$ 20.2 _{-  2.4 }^{+  1.8 }$ \\  
$\log\,L$ (L$_{\odot}$) & $  4.6 _{-  0.1 }^{+  0.1 }$ &  
$  4.7 _{-  0.2 }^{+  0.2 }$ &  
$  4.8 _{-  0.2 }^{+  0.3 }$ &  
$  4.6 _{-  0.1 }^{+  0.1 }$ &  
$  5.4 _{-  0.2 }^{+  0.3 }$ &  
$  4.8 _{-  0.2 }^{+  0.1 }$ \\  
Age (Myr) & $  0.0 _{-  0.0 }^{+  2.2 }$ &  
$  6.2 _{-  0.9 }^{+  1.2 }$ &  
$  3.7 _{-  1.2 }^{+  1.2 }$ &  
$  4.6 _{-  1.7 }^{+  1.0 }$ &  
$  4.1 _{-  0.5 }^{+  1.3 }$ &  
$  5.0 _{-  0.7 }^{+  0.7 }$ \\  
$\tau_{\rm MS}$ & $  0.1 _{-  0.1 }^{+  0.2 }$ &  
$  0.7 _{-  0.1 }^{+  0.1 }$ &  
$  0.6 _{-  0.2 }^{+  0.1 }$ &  
$  0.5 _{-  0.2 }^{+  0.1 }$ &  
$  0.8 _{-  0.1 }^{+  0.0 }$ &  
$  0.6 _{-  0.1 }^{+  0.1 }$ \\  
$v_{\rm ini}$ (km s$^{-1}$) & $ 360.0 _{- 19.9 }^{+ 37.7 }$ &  
$ 360.0 _{- 19.0 }^{+ 41.3 }$ &  
$ 350.0 _{- 87.0 }^{+ 37.4 }$ &  
$ 380.0 _{- 22.6 }^{+ 29.1 }$ &  
$ 350.0 _{- 53.9 }^{+ 56.1 }$ &  
$ 350.0 _{- 84.7 }^{+ 46.8 }$ \\  
$v\sin\,i$ (km s$^{-1}$) & $ 360.0 _{- 20.1 }^{+ 12.4 }$ (357$\pm$15) & 
$ 360.0 _{- 15.8 }^{+ 15.8 }$ (361$\pm$15) & 
$ 250.0 _{- 18.0 }^{+ 14.0 }$ (246$\pm$15) & 
$ 380.0 _{- 18.7 }^{+ 12.8 }$ (378$\pm$15) & 
$ 230.0 _{- 13.4 }^{+ 18.6 }$ (233$\pm$15) & 
$ 210.0 _{- 19.3 }^{+ 13.3 }$ (205$\pm$15) \\ 
$i$ ($^{\circ}$) &  76.6  & 
76.6  & 
47.3  & 
77.0  & 
73.4  & 
36.9  \\ 
$y$ & $  0.085 _{- 0.000 }^{+ 0.000  }$ (0.158$\pm$0.025) & 
$  0.085 _{- 0.003 }^{+ 0.003  }$ (0.141$\pm$0.030) & 
$  0.085 _{- 0.000 }^{+ 0.006  }$ (0.091$\pm$0.030) & 
$  0.085 _{- 0.003 }^{+ 0.003  }$ (0.135$\pm$0.025) & 
$  0.089 _{- 0.003 }^{+ 0.007  }$ (0.172$\pm$0.030) & 
$  0.085 _{- 0.003 }^{+ 0.003  }$ (0.082$\pm$0.025) \\ 
$\log\,\epsilon$(C) & $ 8.13 _{- 0.15 }^{+ 0.01 }$ (<7.54) & 
$ 7.83 _{- 0.17 }^{+ 0.09 }$$\dagger$ ($\le$7.39) & 
$ 8.01 _{- 0.19 }^{+ 0.12 }$ (8.48$\pm$0.27) & 
$ 7.87 _{- 0.14 }^{+ 0.16 }$ (8.07$\pm$0.12) & 
$ 7.81 _{- 0.27 }^{+ 0.20 }$$\dagger$ (7.48$\pm$0.27) & 
$ 7.90 _{- 0.11 }^{+ 0.17 }$ (8.24$\pm$0.12) \\ 
$\log\,\epsilon$(N) & $ 7.66 _{- 0.02 }^{+ 0.41 }$ (8.16$\pm$0.13) & 
$ 8.24 _{- 0.10 }^{+ 0.17 }$ (8.50$\pm$0.34) & 
$ 8.18 _{- 0.25 }^{+ 0.28 }$ (7.92$\pm$0.34) & 
$ 8.22 _{- 0.14 }^{+ 0.20 }$ (7.85$\pm$0.13) & 
$ 8.35 _{- 0.20 }^{+ 0.22 }$$\dagger$ ($\ge$9.08) & 
$ 8.19 _{- 0.25 }^{+ 0.20 }$ (7.34$\pm$0.13) \\ 
$\log\,\epsilon$(O) & $ 8.55 _{- 0.05 }^{+ 0.01 }$ (8.22$\pm$0.21) & 
$ 8.48 _{- 0.07 }^{+ 0.03 }$ (8.15$\pm$0.21) & 
$ 8.54 _{- 0.10 }^{+ 0.02 }$ (8.74$\pm$0.21) & 
$ 8.49 _{- 0.04 }^{+ 0.06 }$ (8.37$\pm$0.21) & 
$ 8.43 _{- 0.13 }^{+ 0.10 }$ ($\ge$8.93) & 
$ 8.49 _{- 0.05 }^{+ 0.07 }$ (8.38$\pm$0.21) \\ 
\hline\hline  
Star & \object{HD 172367} & \object{HD 175876} & \object{HD 184915} & \object{HD 188439} & \object{HD 191423} & \object{HD 192281} \\ \hline  
$M_{\rm ini}$ (M$_{\odot}$) & $ 15.8 _{-  1.7 }^{+  2.1 }$ &  
 $ 26.8 _{-  4.0 }^{+  7.9 }$ &  
 $ 15.2 _{-  1.6 }^{+  1.8 }$ &  
 $ 14.8 _{-  1.4 }^{+  1.8 }$ &  
 $ 19.6 _{-  2.2 }^{+  1.9 }$ &  
  $ 39.2 _{-  8.7 }^{+  4.6 }$ \\ 
$M_{\rm act}$ (M$_{\odot}$) & $ 15.4 _{-  1.2 }^{+  2.3 }$ &  
$ 26.2 _{-  3.2 }^{+  7.4 }$ &  
$ 15.6 _{-  2.0 }^{+  1.2 }$ &  
$ 15.6 _{-  2.2 }^{+  0.8 }$ &  
$ 19.8 _{-  2.5 }^{+  1.2 }$ &  
$ 35.6 _{-  6.2 }^{+  4.4 }$ \\  
$\log\,L$ (L$_{\odot}$) & $  4.6 _{-  0.2 }^{+  0.2 }$ &  
$  5.2 _{-  0.2 }^{+  0.2 }$ &  
$  4.5 _{-  0.2 }^{+  0.2 }$ &  
$  4.5 _{-  0.2 }^{+  0.2 }$ &  
$  4.9 _{-  0.2 }^{+  0.1 }$ &  
$  5.5 _{-  0.3 }^{+  0.1 }$ \\  
Age (Myr) & $  7.9 _{-  0.9 }^{+  1.0 }$ &  
$  3.5 _{-  0.6 }^{+  0.7 }$ &  
$  7.7 _{-  0.9 }^{+  1.0 }$ &  
$  7.7 _{-  0.9 }^{+  1.1 }$ &  
$  7.0 _{-  0.9 }^{+  0.9 }$ &  
$  2.8 _{-  0.6 }^{+  0.6 }$ \\  
$\tau_{\rm MS}$ & $  0.7 _{-  0.1 }^{+  0.1 }$ &  
$  0.7 _{-  0.1 }^{+  0.1 }$ &  
$  0.7 _{-  0.1 }^{+  0.1 }$ &  
$  0.7 _{-  0.1 }^{+  0.1 }$ &  
$  0.7 _{-  0.1 }^{+  0.1 }$ &  
$  0.6 _{-  0.1 }^{+  0.1 }$ \\  
$v_{\rm ini}$ (km s$^{-1}$) & $ 280.0 _{- 13.4 }^{+ 94.2 }$ &  
$ 340.0 _{- 47.8 }^{+ 56.1 }$ &  
$ 340.0 _{- 84.5 }^{+ 35.7 }$ &  
$ 290.0 _{- 14.1 }^{+ 92.1 }$$\dagger$ &  
$ 420.0 _{- 26.5 }^{+ 29.6 }$ &  
$ 360.0 _{- 49.8 }^{+ 46.9 }$ \\  
$v\sin\,i$ (km s$^{-1}$) & $ 270.0 _{- 17.9 }^{+ 13.7 }$ (266$\pm$15) & 
$ 270.0 _{- 20.3 }^{+ 11.4 }$ (265$\pm$15) & 
$ 260.0 _{- 21.9 }^{+ 10.3 }$ (252$\pm$15) & 
$ 280.0 _{- 14.1 }^{+ 17.3 }$ (281$\pm$15) & 
$ 420.0 _{- 20.3 }^{+ 11.2 }$ (420$\pm$15) & 
$ 270.0 _{- 11.8 }^{+ 19.4 }$ (276$\pm$15) \\ 
$i$ ($^{\circ}$) &  54.9  & 
74.6  & 
49.9  & 
64.6  & 
90.0  & 
74.6  \\ 
$y$ & $  0.085 _{- 0.000 }^{+ 0.003  }$ (0.140$\pm$0.025) & 
$  0.085 _{- 0.003 }^{+ 0.006  }$ (0.110$\pm$0.030) & 
$  0.085 _{- 0.000 }^{+ 0.003  }$ (0.183$\pm$0.025) & 
$  0.085 _{- 0.000 }^{+ 0.003  }$ (0.122$\pm$0.025) & 
$  0.089 _{- 0.003 }^{+ 0.010  }$ (0.134$\pm$0.030) & 
$  0.089 _{- 0.003 }^{+ 0.013  }$ (0.103$\pm$0.030) \\ 
$\log\,\epsilon$(C) & $ 7.96 _{- 0.16 }^{+ 0.04 }$ (<8.09) & 
$ 7.81 _{- 0.19 }^{+ 0.20 }$ (8.04$\pm$0.27) & 
$ 7.98 _{- 0.17 }^{+ 0.05 }$ (<8.18) &  
$ 7.96 _{- 0.14 }^{+ 0.06 }$ (<8.09) &  
$ 7.67 _{- 0.28 }^{+ 0.07 }$ ($\le$7.24) &  
$ 7.81 _{- 0.45 }^{+ 0.14 }$$\dagger$ (8.00$\pm$0.27) \\  
$\log\,\epsilon$(N) & $ 8.03 _{- 0.07 }^{+ 0.23 }$$\dagger$ (8.44$\pm$0.13) & 
$ 8.34 _{- 0.22 }^{+ 0.18 }$ (8.36$\pm$0.34) & 
$ 8.03 _{- 0.11 }^{+ 0.23 }$ (8.46$\pm$0.13) & 
$ 8.03 _{- 0.08 }^{+ 0.23 }$ (8.16$\pm$0.13) & 
$ 8.41 _{- 0.09 }^{+ 0.18 }$$\dagger$ (8.33$\pm$0.34) & 
$ 8.34 _{- 0.08 }^{+ 0.33 }$ (8.76$\pm$0.34) \\ 
$\log\,\epsilon$(O) & $ 8.53 _{- 0.06 }^{+ 0.01 }$ (8.47$\pm$0.21) & 
$ 8.43 _{- 0.10 }^{+ 0.10 }$$\dagger$ (8.42$\pm$0.21) & 
$ 8.53 _{- 0.05 }^{+ 0.02 }$ (8.62$\pm$0.21) &  
$ 8.53 _{- 0.05 }^{+ 0.01 }$ (8.66$\pm$0.21) &  
$ 8.40 _{- 0.14 }^{+ 0.06 }$ ($\le$8.33) &  
$ 8.43 _{- 0.25 }^{+ 0.08 }$ (8.05$\pm$0.21) \\  
\hline  
\end{tabular} 
\end{center}  
\end{tiny} 
\end{sidewaystable*}  
\begin{table*}  
\addtocounter{table}{-1} 
\begin{tiny}  
\caption{Continued.}  
\begin{center}  
\begin{tabular}{ccccccccccccccccccccc}  
\hline\hline  
Star & \object{HD 198781} & \object{HD 203064} & \object{HD 210839} & \object{HD 228841} \\ \hline  
$M_{\rm ini}$ (M$_{\odot}$) & $ 15.2 _{-  1.4 }^{+  1.5 }$ &  
 $ 23.8 _{-  3.2 }^{+  5.1 }$ &  
 $ 40.0 _{- 12.4 }^{+ 11.4 }$ &  
  $ 25.4 _{-  3.2 }^{+  6.2 }$ \\ 
$M_{\rm act}$ (M$_{\odot}$) & $ 15.2 _{-  1.3 }^{+  1.5 }$ &  
$ 23.4 _{-  2.9 }^{+  4.6 }$ &  
$ 35.2 _{-  8.2 }^{+ 10.2 }$ &  
$ 24.8 _{-  2.8 }^{+  5.2 }$ \\  
$\log\,L$ (L$_{\odot}$) & $  4.4 _{-  0.1 }^{+  0.1 }$ &  
$  5.1 _{-  0.3 }^{+  0.2 }$ &  
$  5.6 _{-  0.2 }^{+  0.3 }$ &  
$  5.2 _{-  0.2 }^{+  0.2 }$ \\  
Age (Myr) & $  6.1 _{-  1.5 }^{+  1.2 }$ &  
$  3.7 _{-  0.7 }^{+  1.0 }$ &  
$  3.4 _{-  0.9 }^{+  0.4 }$ &  
$  4.1 _{-  0.5 }^{+  0.9 }$ \\  
$\tau_{\rm MS}$ & $  0.5 _{-  0.1 }^{+  0.1 }$ &  
$  0.6 _{-  0.2 }^{+  0.1 }$ &  
$  0.7 _{-  0.1 }^{+  0.1 }$ &  
$  0.7 _{-  0.1 }^{+  0.1 }$ \\  
$v_{\rm ini}$ (km s$^{-1}$) & $ 370.0 _{- 97.3 }^{+ 24.6 }$ &  
$ 330.0 _{- 28.0 }^{+ 59.8 }$ &  
$ 350.0 _{- 50.9 }^{+ 65.5 }$ &  
$ 360.0 _{- 40.3 }^{+ 46.2 }$ \\  
$v\sin\,i$ (km s$^{-1}$) & $ 220.0 _{- 12.1 }^{+ 20.5 }$ (222$\pm$15) & 
$ 300.0 _{- 16.8 }^{+ 14.8 }$ (298$\pm$15) & 
$ 220.0 _{- 20.3 }^{+ 10.6 }$ (214$\pm$15) & 
$ 300.0 _{- 11.0 }^{+ 20.6 }$ (305$\pm$15) \\ 
$i$ ($^{\circ}$) &  36.5  & 
75.4  & 
73.0  & 
69.6  \\ 
$y$ & $  0.081 _{- 0.000 }^{+ 0.003  }$ (0.230$\pm$0.025) & 
$  0.085 _{- 0.003 }^{+ 0.006  }$ (0.076$\pm$0.030) & 
$  0.089 _{- 0.003 }^{+ 0.020  }$ (0.113$\pm$0.030) & 
$  0.089 _{- 0.003 }^{+ 0.003  }$ (0.112$\pm$0.030) \\ 
$\log\,\epsilon$(C) & $ 8.03 _{- 0.15 }^{+ 0.08 }$ ($\le$8.09) & 
$ 7.91 _{- 0.24 }^{+ 0.09 }$ (7.92$\pm$0.27) & 
$ 7.12 _{- 0.00 }^{+ 0.00 }$$\dagger$ (7.83$\pm$0.27) & 
$ 7.78 _{- 0.18 }^{+ 0.15 }$$\dagger$ (7.48$\pm$0.27) \\ 
$\log\,\epsilon$(N) & $ 7.98 _{- 0.15 }^{+ 0.27 }$$\dagger$ (8.62$\pm$0.34) & 
$ 8.34 _{- 0.21 }^{+ 0.15 }$ (8.23$\pm$0.34) & 
$ 8.69 _{- 0.47 }^{+ 0.08 }$$\dagger$ (8.74$\pm$0.34) & 
$ 8.34 _{- 0.12 }^{+ 0.17 }$ (8.74$\pm$0.34) \\ 
$\log\,\epsilon$(O) & $ 8.53 _{- 0.04 }^{+ 0.02 }$ (8.78$\pm$0.21) & 
$ 8.48 _{- 0.10 }^{+ 0.05 }$ (8.46$\pm$0.21) & 
$ 8.45 _{- 0.35 }^{+ 0.08 }$$\dagger$ (8.13$\pm$0.21) & 
$ 8.42 _{- 0.09 }^{+ 0.07 }$$\dagger$ (8.67$\pm$0.21) \\  
\hline  
\end{tabular} 
\end{center}  
\end{tiny} 
\end{table*}

\end{appendix}


\begin{thebibliography}{}

\bibitem[Asplund et al.(2009)]{asp09} Asplund, M., Grevesse, N., Sauval, A.~J., \& Scott, P.\ 2009, \araa, 47, 481
\bibitem[Bastiaansen(1992)]{bas92} Bastiaansen, P.~A.\ 1992, \aaps, 93, 449 
\bibitem[Bestenlehner et al.(2014)]{bes14} Bestenlehner, J.~M., Gr{\"a}fener, G., Vink, J.~S., et al.\ 2014, \aap, 570, A38 
\bibitem[Blay et al.(2006)]{bla06} Blay, P., Negueruela, I., Reig, P., et al.\ 2006, \aap, 446, 1095 
\bibitem[Bouret et al.(2012)]{bou12} Bouret, J.-C., Hillier, D.~J., Lanz, T., \& Fullerton, A.~W.\ 2012, \aap, 544, A67
\bibitem[Bouret et al.(2013)]{bou13} Bouret, J.-C., Lanz, T., Martins, F., et al.\ 2013, \aap, 555, A1
\bibitem[Briquet \& Morel(2007)]{bri07a} Briquet, M., \& Morel, T.\ 2007, Communications in Asteroseismology, 150, 183 
\bibitem[Briquet et al.(2007)]{bri07b} Briquet, M., Morel, T., Thoul, A., et al.\ 2007, \mnras, 381, 1482 
\bibitem[Brott et al.(2011a)]{bro11} Brott, I., de Mink, S.~E., Cantiello, M., et al.\ 2011a, \aap, 530, A115
\bibitem[Brott et al.(2011b)]{bro11b} Brott, I., Evans, C.~J., Hunter, I., et al.\ 2011b, \aap, 530, A116
\bibitem[Butler \& Giddings(1985)]{but85} Butler, K., \& Giddings, J.~R.\ 1985, in Newsletter of Analysis of Astronomical Spectra, No. 9 (Univ. London)
\bibitem[Castro et al.(2014)]{cas14} Castro, N., Fossati, L., Langer, N., et al.\ 2014, \aap, 570, L13 
\bibitem[Cazorla et al.(2017)]{caz17} Cazorla, C., Morel, T., Naz{\'e}, Y., et al.\ 2017, A\&A, in press, arXiv:1703.05592 (Paper I)
\bibitem[de Mink et al.(2009)]{dem09} de Mink, S.~E., Cantiello, M., Langer, N., et al.\ 2009, \aap, 497, 243
\bibitem[de Mink et al.(2011)]{dem11} de Mink, S.~E., Langer, N., \& Izzard, R.~G.\ 2011, Bulletin de la Soci{\'e}t{\'e} Royale des Sciences de Li{\`e}ge, 80, 543
\bibitem[de Mink et al.(2013)]{dem13} de Mink, S.~E., Langer, N., Izzard, R.~G., Sana, H., \& de Koter, A.\ 2013, \apj, 764, 166
\bibitem[de Mink et al.(2014)]{dem14} de Mink, S.~E., Sana, H., Langer, N., Izzard, R.~G., \& Schneider, F.~R.~N.\ 2014, \apj, 782, 7
\bibitem[Dervi{\c s}o{\v g}lu et al.(2010)]{der10} Dervi{\c s}o{\v g}lu, A., Tout, C.~A., \& Ibano{\v g}lu, C.\ 2010, \mnras, 406, 1071
\bibitem[Dufton et al.(2011)]{duf11} Dufton, P.~L., Dunstall, P.~R., Evans, C.~J., et al.\ 2011, \apjl, 743, L22 
\bibitem[Eldridge et al.(2011)]{eld11} Eldridge, J.~J., Langer, N., \& Tout, C.~A.\ 2011, \mnras, 414, 3501 
\bibitem[Evans et al.(2008)]{eva08} Evans, C., Hunter, I., Smartt, S., et al.\ 2008, The Messenger, 131, 25 
\bibitem[Ferrario et al.(2009)]{fer09} Ferrario, L., Pringle, J.~E., Tout, C.~A., \& Wickramasinghe, D.~T. 2009, MNRAS, 400, L71 
\bibitem[Fossati et al.(2015)]{fos15} Fossati, L., Castro, N., Sch{\"o}ller, M., et al.\ 2015, \aap, 582, A45 
\bibitem[Georgy et al.(2013)]{geo13} Georgy, C., Ekstr{\"o}m, S., Granada, A., et al.\ 2013, \aap, 553, A24 
\bibitem[Gies \& Lambert(1992)]{gie92} Gies, D.~R., \& Lambert, D.~L.\ 1992, \apj, 387, 673 
\bibitem[Giddings(1981)]{gid81} Giddings, J.~R.\ 1981, Ph.D.~Thesis
\bibitem[Glebbeek et al.(2013)]{gle13} Glebbeek, E., Gaburov, E., Portegies Zwart, S., \& Pols, O.~R.\ 2013, \mnras, 434, 3497 
\bibitem[Gray(2005)]{gra05} Gray, D.~F.\ 2005, ''The Observation and Analysis of Stellar Photospheres", 3rd Edition, ISBN 0521851866, Cambridge University Press
\bibitem[Grin et al.(2017)]{gri17} Grin, N.~J., Ram{\'{\i}}rez-Agudelo, O.~H., de Koter, A., et al.\ 2017, \aap, 600, A82 
\bibitem[Grunhut et al.(2017)]{gru17} Grunhut, J.~H., Wade, G.~A., Neiner, C., et al.\ 2017, \mnras, 465, 2432 
\bibitem[Grevesse \& Sauval(1998)]{gre98} Grevesse, N., \& Sauval, A.~J.\ 1998, \ssr, 85, 161 
\bibitem[Heger et al.(2005)]{heg05} Heger, A., Woosley, S. E., \& Spruit, H. C. 2005, \apj, 626, 350
\bibitem[Hensberge et al.(2000)]{hen00} Hensberge, H., Pavlovski, K., \& Verschueren, W.\ 2000, \aap, 358, 553
\bibitem[Hillier \& Miller(1998)]{hil98} Hillier, D.~J., \& Miller, D.~L.\ 1998, \apj, 496, 407
\bibitem[Howarth et al.(1997)]{how97} Howarth, I.~D., Siebert, K.~W., Hussain, G.~A.~J., \& Prinja, R.~K.\ 1997, \mnras, 284, 265
\bibitem[Hubrig et al.(2008)]{hub08} Hubrig, S., Briquet, M., Morel, T., et al.\ 2008, \aap, 488, 287 
\bibitem[Hubrig et al.(2013)]{hub13} Hubrig, S., Sch{\"o}ller, M., Ilyin, I., et al.\ 2013, \aap, 551, A33 
\bibitem[Hubrig et al.(2016)]{hub16} Hubrig, S., Kholtygin, A., Ilyin, I., Sch{\"o}ller, M., \& Oskinova, L.~M.\ 2016, \apj, 822, 104 
\bibitem[Hunter et al.(2007)]{hun07} Hunter, I., Dufton, P.~L., Smartt, S.~J., et al.\ 2007, \aap, 466, 277
\bibitem[Hunter et al.(2009)]{hun09} Hunter, I., Brott, I., Langer, N., et al.\ 2009, \aap, 496, 841
\bibitem[Hut(1981)]{hut81} Hut, P.\ 1981, \aap, 99, 126
\bibitem[Kambe et al.(1997)]{kam97} Kambe, E., Hirata, R., Ando, H., et al.\ 1997, \apj, 481, 406 
\bibitem[K{\"o}hler et al.(2012)]{koh12} K{\"o}hler, K., Borzyszkowski, M., Brott, I., Langer, N., \& de Koter, A.\ 2012, \aap, 544, A76
\bibitem[Langer(1992)]{lan92} Langer, N.\ 1992, \aap, 265, L17 
\bibitem[Langer et al.(2003)]{lan03b} Langer, N., Wellstein, S., \& Petrovic, J.\ 2003, A Massive Star Odyssey: From Main Sequence to Supernova, 212, 275 
\bibitem[Langer \& Kudritzki(2014)]{lan14} Langer, N., \& Kudritzki, R.~P.\ 2014, \aap, 564, A52 
\bibitem[Lau et al.(2014)]{lau14} Lau, H.~H.~B., Izzard, R.~G., \& Schneider, F.~R.~N.\ 2014, \aap, 570, A125 
\bibitem[Leonard \& Duncan(1990)]{leo90} Leonard, P.~J.~T., \& Duncan, M.~J.\ 1990, \aj, 99, 608
\bibitem[Linder et al.(2008)]{lin08} Linder, N., Rauw, G., Martins, F., et al.\ 2008, \aap, 489, 713 
\bibitem[Lyubimkov et al.(1997)]{lyu97} Lyubimkov, L.~S., Rostopchin, S.~I., Roche, P., \& Tarasov, A.~E.\ 1997, \mnras, 286, 549 
\bibitem[Maeder(1987)]{mae87} Maeder, A.\ 1987, \aap, 178, 159 
\bibitem[Maeder \& Meynet(2005)]{mae05} Maeder, A., \& Meynet, G.\ 2005, \aap, 440, 1041 
\bibitem[Maeder et al.(2014)]{mae14} Maeder, A., Przybilla, N., Nieva, M.-F., et al.\ 2014, \aap, 565, A39
\bibitem[Mahy et al.(2011)]{mah11} Mahy, L., Martins, F., Machado, C., Donati, J.-F., \& Bouret, J.-C.\ 2011, \aap, 533, A9 
\bibitem[Ma{\'{\i}}z Apell{\'a}niz et al.(2008)]{mai08} Ma{\'{\i}}z Apell{\'a}niz, J., Alfaro, E.~J., \& Sota, A.\ 2008, arXiv:0804.2553 
\bibitem[Martins et al.(2005)]{mar05} Martins, F., Schaerer, D., \& Hillier, D.~J.\ 2005, \aap, 436, 1049 
\bibitem[Martins et al.(2015a)]{mar15a} Martins, F., Herv{\'e}, A., Bouret, J.-C., et al.\ 2015a, \aap, 575, A34
\bibitem[Martins et al.(2015b)]{mar15b} Martins, F., Sim{\'o}n-D{\'{\i}}az, S., Palacios, A., et al.\ 2015b, \aap, 578, A109
\bibitem[Meynet et al.(2011)]{mey11} Meynet, G., Eggenberger, P., \& Maeder, A.\ 2011, \aap, 525, L11
\bibitem[Miglio et al.(2008)]{mig08} Miglio, A., Montalb{\'a}n, J., Noels, A., \& Eggenberger, P.\ 2008, \mnras, 386, 1487 
\bibitem[Morel et al.(2006)]{mor06} Morel, T., Butler, K., Aerts, C., Neiner, C., \& Briquet, M.\ 2006, \aap, 457, 651 
\bibitem[Morel et al.(2008)]{mor08} Morel, T., Hubrig, S., \& Briquet, M.\ 2008, \aap, 481, 453
\bibitem[Morel(2009)]{mor09} Morel, T.\ 2009, Communications in Asteroseismology, 158, 122 
\bibitem[Morel(2012)]{mor12} Morel, T.\ 2012, Proceedings of a Scientific Meeting in Honor of Anthony F.~J.~Moffat, 465, 54 
\bibitem[Morton(1975)]{mor75} Morton, D.~C.\ 1975, \apj, 197, 85 
\bibitem[Nieva \& Przybilla(2012)]{nie12} Nieva, M.-F., \& Przybilla, N.\ 2012, \aap, 539, A143 
\bibitem[Ogura \& Ishida(1981)]{ogu81} Ogura, K., \& Ishida, K.\ 1981, \pasj, 33, 149 
\bibitem[Packet(1981)]{pac81} Packet, W.\ 1981, \aap, 102, 17
\bibitem[Palate \& Rauw(2012)]{pal12} Palate, M., \& Rauw, G.\ 2012, \aap, 537, A119
\bibitem[Palate et al.(2013)]{pal13} Palate, M., Rauw, G., Koenigsberger, G., \& Moreno, E.\ 2013, \aap, 552, A39
\bibitem[Peters et al.(2008)]{pet08} Peters, G.~J., Gies, D.~R., Grundstrom, E.~D., \& McSwain, M.~V.\ 2008, \apj, 686, 1280-1291 
\bibitem[Peters et al.(2013)]{pet13} Peters, G.~J., Pewett, T.~D., Gies, D.~R., Touhami, Y.~N., \& Grundstrom, E.~D.\ 2013, \apj, 765, 2 
\bibitem[Peters et al.(2016)]{pet16} Peters, G.~J., Wang, L., Gies, D.~R., \& Grundstrom, E.~D.\ 2016, \apj, 828, 47 
\bibitem[Petrovic et al.(2005a)]{pet05a} Petrovic, J., Langer, N., Yoon, S.-C., \& Heger, A.\ 2005a, \aap, 435, 247
\bibitem[Petrovic et al.(2005b)]{pet05b} Petrovic, J., Langer, N., \& van der Hucht, K.~A.\ 2005b, \aap, 435, 1013
\bibitem[Philp et al.(1996)]{phi96} Philp, C.~J., Evans, C.~R., Leonard, P.~J.~T., \& Frail, D.~A.\ 1996, \aj, 111, 1220 
\bibitem[Podsiadlowski et al.(1992)]{pod92} Podsiadlowski,P., Joss, P.~C., \& Hsu, J.~J.~L.\ 1992, \apj, 391, 246
\bibitem[Poeckert(1981)]{poe81} Poeckert, R.\ 1981, \pasp, 93, 297 
\bibitem[Pols et al.(1991)]{pol91} Pols, O.~R., Cote, J., Waters, L.~B.~F.~M., \& Heise, J.\ 1991, \aap, 241, 419
\bibitem[Portegies Zwart(2000)]{por00} Portegies Zwart, S.~F.\ 2000, \apj, 544, 437
\bibitem[Potter at al.(2012)]{pot12} Potter, A.~T., Chitre, S.~M., \& Tout, C.~A. 2012, \mnras, 424, 2358 
\bibitem[Przybilla et al.(2010)]{prz10} Przybilla, N., Firnstein, M., Nieva, M.-F., Meynet, G., \& Maeder, A.\ 2010, \aap, 517, A38
\bibitem[Raucq et al.(2016)]{rauc16} Raucq, F., Rauw, G., Gosset, E., et al.\ 2016, \aap, 588, A10 
\bibitem[Raucq et al.(2017)]{rauc17} Raucq, F., Gosset, E., Rauw, G., et al.\ 2017, \aap, in press, arXiv:1703.03247 
\bibitem[Rauw et al.(2008)]{rau08} Rauw, G., De Becker, M., van Winckel, H., et al.\ 2008, \aap, 487, 659 
\bibitem[Repolust et al.(2004)]{rep04} Repolust, T., Puls, J., \& Herrero, A.\ 2004, \aap, 415, 349 
\bibitem[Rivero Gonz{\'a}lez et al.(2012a)]{riv12a} Rivero Gonz{\'a}lez, J.~G., Puls, J., Najarro, F., \& Brott, I.\ 2012a, \aap, 537, A79 
\bibitem[Rivero Gonz{\'a}lez et al.(2012b)]{riv12b} Rivero Gonz{\'a}lez, J.~G., Puls, J., Massey, P., \& Najarro, F.\ 2012b, \aap, 543, A95 
\bibitem[Salpeter(1955)]{sal55} Salpeter, E.~E.\ 1955, \apj, 121, 161 
\bibitem[Sana et al.(2012)]{san12} Sana, H., de Mink, S.~E., de Koter, A., et al.\ 2012, Science, 337, 444
\bibitem[Sana et al.(2013)]{san13} Sana, H., de Koter, A., de Mink, S.~E., et al.\ 2013, \aap, 550, A107 
\bibitem[Sayer et al.(1996)]{say96} Sayer, R.~W., Nice, D.~J., \& Kaspi, V.~M.\ 1996, \apj, 461, 357 
\bibitem[Schneider et al.(2014)]{sch14} Schneider, F.~R.~N., Langer, N., de Koter, A., et al.\ 2014, \aap, 570, A66 
\bibitem[Scuflaire et al.(2008)]{scu08} Scuflaire, R., Th{\'e}ado, S., Montalb{\'a}n, J., et al.\ 2008, \apss, 316, 83 
\bibitem[Sim{\'o}n-D{\'{\i}}az \& Herrero(2007)]{sim07} Sim{\'o}n-D{\'{\i}}az, S., \& Herrero, A.\ 2007, \aap, 468, 1063
\bibitem[Song et al.(2013)]{son13} Song, H.~F., Maeder, A., Meynet, G., et al.\ 2013, \aap, 556, A100
\bibitem[Tylenda et al.(2011)]{tyl11} Tylenda, R., Hajduk, M., Kami{\'n}ski, T., et al.\ 2011, \aap, 528, A114
\bibitem[ud-Doula \& Owocki(2002)]{udd02} ud-Doula, A., \& Owocki, S.~P.\ 2002, \apj, 576, 413 
\bibitem[ud-Doula et al.(2008)]{udd08} ud-Doula, A., Owocki, S.~P., \& Townsend, R.~H.~D.\ 2008, \mnras, 385, 97 
\bibitem[ud-Doula et al.(2009)]{udd09} ud-Doula, A., Owocki, S.~P., \& Townsend, R.~H.~D.\ 2009, \mnras, 392, 1022 
\bibitem[van Leeuwen et al.(1997)]{vanl97} van Leeuwen, F., Evans, D.~W., Grenon, M., et al.\ 1997, \aap, 323, L61 
\bibitem[Vink et al.(2001)]{vin01} Vink, J.~S., de Koter, A., \& Lamers, H.~J.~G.~L.~M.\ 2001, \aap, 369, 574
\bibitem[Wade et al.(2014)]{wad14} Wade, G.~A., et al., 2014, IAUS, 302, 265 
\bibitem[Wellstein \& Langer(1999)]{wel99} Wellstein, S., \& Langer, N.\ 1999, \aap, 350, 148 
\bibitem[Wellstein et al.(2001)]{wel01} Wellstein, S., Langer, N., \& Braun, H.\ 2001, \aap, 369, 939 
\bibitem[Zahn(1975)]{zah75} Zahn, J.-P.\ 1975, \aap, 41, 329
\bibitem[Zorec et al.(2002)]{zor02} Zorec, J., Fr{\'e}mat, Y., Hubert, A.~M., \& Floquet, M.\ 2002, IAU Colloq.~185: Radial and Nonradial Pulsations as Probes of Stellar Physics, 259, 244 

\end{thebibliography}
\end{document}